%% file: SOK-llmsecurity.tex
\documentclass[a4paper,fleqn]{cas-dc}
\usepackage{lipsum}
 
\usepackage{shortcuts}
\usepackage{fontawesome}
\usepackage[labelsep=newline,justification=centering]{caption} 
\usepackage{mwe}
\usepackage{wasysym}

\usepackage[numbers]{natbib}
\def\tsc#1{\csdef{#1}{\textsc{\lowercase{#1}}\xspace}}
\tsc{WGM}
\tsc{QE}
\tsc{EP}
\tsc{PMS}
\tsc{BEC}
\tsc{DE}
\usepackage[most]{tcolorbox}
\usepackage{bookmark}

\begin{document}
\let\WriteBookmarks\relax
\def\floatpagepagefraction{1}
\def\textpagefraction{.001}

\hypersetup{
    colorlinks = true,
    linkcolor = blue,
    anchorcolor = blue,
    citecolor = blue,
    filecolor = blue,
    urlcolor = blue,
    bookmarksopen=true
}

\input{macro}

\shorttitle{A Survey on Large Language Model (LLM) Security and Privacy: The
Good, the Bad, and Ugly}

\shortauthors{Yifan Yao et~al.}
 
\title [mode = title]{\centering A Survey on Large Language Model (LLM) Security and Privacy: \\ The Good, the Bad, and the Ugly}                      

 
\newtcolorbox[%
auto counter]{mybox}[2][]{%
	enhanced jigsaw,
        colback=blue!12,
	breakable,
	#1}

%
\author{Yifan Yao}

\ead{yy566@drexel.edu}

\credit{Conceptualization of this study, Methodology, Software}

\affiliation[1]{organization={Drexel University},
    addressline={3675 Market St.}, 
    city={Philadelphia},
    state={PA},
    postcode={19104}, 
    country={USA}}

\author{Jinhao Duan} 
\ead{jd3734@drexel.edu}
\author{Kaidi Xu} 
\ead{kx46@drexel.edu}
\author{Yuanfang Cai} 
\ead{yfcai@cs.drexel.edu}
\author{Zhibo Sun} 
\ead{zs384@drexel.edu}
\author{Yue Zhang} 
\ead{yz899@drexel.edu}

\begin{abstract}
Large Language Models (LLMs), such as ChatGPT and Bard, have revolutionized natural language understanding and generation. They possess deep language comprehension, human-like text generation capabilities, contextual awareness, and robust problem-solving skills, making them invaluable in various domains (e.g., search engines, customer support, translation).  In the meantime, LLMs have also gained traction in the security community, revealing security vulnerabilities and showcasing their potential in security-related tasks.
This paper explores the intersection of LLMs with security and privacy. Specifically, we investigate how LLMs positively impact security and privacy, potential risks and threats associated with their use, and inherent vulnerabilities within LLMs. Through a comprehensive literature review, the paper categorizes the papers into ``The Good'' (beneficial LLM applications), ``The Bad'' (offensive applications), and ``The Ugly'' (vulnerabilities of LLMs and their defenses). We have some interesting findings. For example, LLMs have proven to enhance code security (code vulnerability detection) and data privacy (data confidentiality protection), outperforming traditional methods. However, they can also be harnessed for various attacks (particularly user-level attacks) due to their human-like reasoning abilities. We have identified areas that require further research efforts. For example, 
Research on model and parameter extraction attacks is limited and often theoretical, hindered by LLM parameter scale and confidentiality.  Safe instruction tuning, a recent development, requires more exploration.
We hope that our work can shed light on the LLMs' potential to both bolster and jeopardize cybersecurity.
\end{abstract}

\begin{keywords}

Large Language Model (LLM), LLM Security, LLM Privacy, ChatGPT, LLM Attacks, LLM Vulnerabilities  
  
\end{keywords}

\maketitle

\input{sections/sec1-introduction}
\input{sections/sec2-background}
\input{sections/sec3-researchproblems}
\input{sections/sec4-thegood}
\input{sections/sec5-thebad}

\input{sections/sec6-theugly}
\input{sections/sec7-discussion}

\input{sections/sec8-releatedwork}
\input{sections/sec9-conclusion}

\bibliographystyle{IEEEtranS}
\bibliography{cas-refs}

\end{document}

%% file: macro.tex
\newcommand{\totalPaper}{281}
\newcommand{\goodPaper}{83}
\newcommand{\badPaper}{54}
\newcommand{\uglyPaper}{144}

\newcommand{\userLevelPaper}{32}
\newcommand{\prePaper}{82}

%% file: sections/sec1-introduction.tex
\section{Introduction}

\noindent
A large language model is the language model with massive parameters that undergoes pretraining tasks (e.g., masked language modeling and autoregressive prediction) to understand and process human language, by modeling the contextualized text semantics and probabilities from large amounts of text data. A capable LLM should have four key features~\cite{yang2023harnessing}: (i) profound comprehension of natural language context; (ii) ability to generate human-like text; (iii) contextual awareness, especially in knowledge-intensive domains; (iv) strong instruction-following ability which is useful for problem-solving and decision-making.

There are a number of LLMs that were developed and released in 2023, gaining significant popularity. Notable examples include OpenAI's \textit{ChatGPT}~\cite{openai2023gpt4},  Meta AI's \textit{LLaMA}~\cite{meta_ai2023llama}, and Databricks' Dolly 2.0~\cite{databricks2023free}. For instance, ChatGPT alone boasts a user base of over 180 million~\cite{chatgptusers}. LLMs now offer a wide range of versatile applications across various domains. Specifically, they not only provide technical support to domains directly related to language processing (e.g., search engines~\cite{ziems2023large,arcila2023platform}, customer support~\cite{spatharioti2023comparing}, translation~\cite{yao2023empowering,karpinska2023large}) but also find utility in more general scenarios such as code generation~\cite{jain2023code}, healthcare~\cite{thirunavukarasu2023large}, finance~\cite{wu2023bloomberggpt}, and education~\cite{mbakwe2023chatgpt}. This showcases their adaptability and potential to streamline language-related tasks across diverse industries and contexts.

LLMs are gaining popularity within the security community. As of February 2023, a research study reported that GPT-3 uncovered 213 security vulnerabilities (only 4 turned out to be false positives)~\cite{gpt213bugs} in a code repository. In contrast, one of the leading commercial tools in the market detected only 99 vulnerabilities. More recently, several LLM-powered security papers have emerged in prestigious conferences. For instance, in IEEE S\&P 2023, Hammond Pearce et al.~\cite{10179324} conducted a comprehensive investigation employing various commercially available LLMs,  evaluating them across synthetic, hand-crafted, and real-world security bug scenarios. The results are promising, as LLMs successfully addressed all synthetic and hand-crafted scenarios. In NDSS 2024,  a tool named Fuzz4All~\cite{xia2023universal} showcased the use of LLMs for input generation and mutation, accompanied by an innovative autoprompting technique and fuzzing loop.

These remarkable initial attempts prompt us to delve into three crucial security-related research questions: 
\begin{itemize}
    \item \textit{\textbf{RQ1}. How do LLMs make a positive impact on security and privacy across diverse domains, and what advantages do they offer to the security community?}
    \item \textit{\textbf{RQ2}. What potential risks and threats emerge from the utilization of LLMs within the realm of cybersecurity?}
    \item \textit{\textbf{RQ3}. What vulnerabilities and weaknesses within LLMs, and how to defend against those threats?}
\end{itemize}

\paragraph{Findings} To comprehensively address these questions, we conducted a meticulous literature review and assembled a collection of \totalPaper~papers pertaining to the intersection of LLMs with security and privacy. We categorized these papers into three distinct groups: those highlighting security-beneficial applications (i.e., the good), those exploring applications that could potentially exert adverse impacts on security (i.e., the bad), and those focusing on the discussion of security vulnerabilities (alongside potential defense mechanisms) within LLMs (i.e., the ugly). To be more specific: 

\begin{itemize}
       \item \textbf{The Good (\S\ref{sec:good}):} LLMs have a predominantly positive impact on the security community, as indicated by the most significant number of papers dedicated to enhancing security. Specifically, LLMs have made contributions to both code security and data security and privacy. In the context of code security, LLMs have been used for the whole life cycle of the code (e.g., secure coding, test case generation, vulnerable code detection, malicious code detection, and code fixing). In data security and privacy, LLMs have been applied to ensure data integrity, data confidentiality, data reliability, and data traceability. Meanwhile, 
Compared to state-of-the-art methods, most researchers found LLM-based methods to outperform traditional approaches.
           \item \textbf{The Bad (\S\ref{sec:bad}):} LLMs also have offensive applications against security and privacy. We categorized the attacks into five groups: hardware-level attacks (e.g., side-channel attacks), OS-level attacks (e.g.,  analyzing information from operating systems), software-level attacks (e.g., creating malware), network-level attacks (e.g., network phishing), and user-level attacks (e.g., misinformation, social engineering, scientific misconduct).  User-level attacks, with \userLevelPaper~ papers, are the most prevalent due to LLMs' human-like reasoning abilities. Those attacks threaten both security (e.g., malware attacks) and privacy (e.g., social engineering). Nowadays, LLMs lack direct access to OS and hardware-level functions. The potential threats of LLMs could escalate if they gain such access.
            \item \textbf{The Ugly (\S\ref{sec:ugly}):} We explore the vulnerabilities and defenses in LLMs, categorizing vulnerabilities into two main groups: AI Model Inherent Vulnerabilities (e.g., data poisoning, backdoor attacks, training data extraction) and Non-AI Model Inherent Vulnerabilities (e.g., remote code execution, prompt injection, side channels). 
            These attacks pose a dual threat, encompassing both security concerns (e.g., remote code execution attacks) and privacy issues (e.g., data extraction).   Defenses for LLMs are divided into strategies placed in the architecture, and applied during the training and inference phases. Training phase defenses involve  corpora cleaning, and optimization methods, while inference phase defenses include instruction pre-processing, malicious detection, and generation post-processing. These defenses collectively aim to enhance the security, robustness, and ethical alignment of LLMs. We found that model extraction, parameter extraction, and similar attacks have received limited research attention, remaining primarily theoretical with minimal practical exploration. The vast scale of LLM parameters makes traditional approaches less effective, and the confidentiality of powerful LLMs further shields them from conventional attacks. Strict censorship of LLM outputs challenges even black-box ML attacks. Meanwhile, research on the impact of model architecture on LLM safety is scarce, partly due to high computational costs. Safe instruction tuning, a recent development, requires further investigation.
\end{itemize}

\paragraph{Contributions} Our work makes a dual contribution. First, we are pioneers summarizing the role of LLMs in security and privacy. We delve deeply into the positive impacts of LLMs on security, their potential risks and threats, vulnerabilities in LLMs, and the corresponding defense mechanisms. 
Other surveys may focus on one or two specific aspects, such as beneficial applications, offensive applications, vulnerabilities, or defenses. To the best of our knowledge, our survey is the first to cover all three key aspects related to security and privacy for the first time.
Second, we have made several interesting discoveries. For instance, our research reveals that LLMs contribute more positively than negatively to security and privacy. Moreover, we observe that most researchers concur that LLMs outperform state-of-the-art methods when employed for securing code or data. Concurrently, it becomes evident that user-level attacks are the most prevalent, largely owing to the human-like reasoning abilities exhibited by LLMs.

\paragraph{Roadmap}
The rest of the paper is organized as follows. We begin with a brief introduction to LLM in \S\ref{sec:background}. \S\ref{sec:overview} presents the overview of our work.  In \S\ref{sec:good}, we explore the beneficial impacts of employing LLMs. \S\ref{sec:bad} discusses the negative impacts on security and privacy.  In \S\ref{sec:ugly}, we discuss the prevalent threats, vulnerabilities associated with LLMs as well as the countermeasures to mitigate these risks.
\S\ref{sec:discuss} discuss LLMs in other security related topics and possible directions. We conclude the paper in \S\ref{sec:conclusion}.

%% file: sections/sec2-background.tex
\section{Background}
\label{sec:background}

\subsection{Large Language Models (LLMs)}

\smallskip

    \noindent Large Language Models (LLMs)~\cite{zhao2023survey} represents an evolution from language models. Initially, language models were statistical in nature and laid the groundwork for computational linguistics. The advent of transformers has significantly increased their scale. This expansion, along with the use of extensive training corpora and advanced pre-training techniques is pivotal in areas such as AI for science, logical reasoning, and embodied AI. These models undergo extensive training on vast datasets to comprehend and produce text that closely mimics human language. Typically, LLMs are endowed with hundreds of billions, or even more, parameters, honed through the processing of massive textual data. They have spearheaded substantial advancements in the realm of Natural Language Processing (NLP)~\cite{feng2021survey} and find applications in a multitude of fields (e.g., risk assessment~\cite{novelli2023taking}, programming~\cite{cai2023low}, vulnerability detection~\cite{jain2023code}, medical text analysis~\cite{thirunavukarasu2023large}, and search engine optimization~\cite{arcila2023platform}).

Based on Yang's study~\cite{yang2023harnessing}, an LLM should have at least four key features. First, an LLM should demonstrate a deep understanding and interpretation of natural language text, enabling it to extract information and perform various language-related tasks (e.g., translation). Second, it should have the capacity to generate human-like text (e.g., completing sentences, composing paragraphs, and even writing articles) when prompted. Third, LLMs should exhibit contextual awareness by considering factors such as domain expertise, a quality referred to as ``Knowledge-intensive''. Fourth, these models should excel in problem-solving and decision-making, leveraging information within text passages to make them invaluable for tasks such as information retrieval and question-answering systems.\looseness=-1

\subsection{Comparison of Popular LLMs}

\smallskip

\noindent
As shown in \autoref{tab:llmscomparison}~\cite{llmcomparision1,llmcomparision2}, there is a diversity of providers for language models, including industry leaders such as OpenAI, Google, Meta AI, and emerging players such as Anthropic and Cohere. The release dates span from 2018 to 2023, showcasing the rapid development and evolution of language models in recent years. Newer models such as ``gpt-4'' have emerged in 2023, highlighting the ongoing innovation in this field. While most of the models are not open-source, it is interesting to note that models like BERT, T5, PaLM, LLaMA, and CTRL are open-source, which can facilitate community-driven development and applications. Larger models tend to have more parameters, potentially indicating increased capabilities but also greater computational demands. For example, ``PaLM'' stands out with a massive 540 billion parameters. It can also be observed that LLMs tend to have more parameters, potentially indicating increased capabilities but also greater computational demands. The ``\textit{Tunability}'' column suggests whether these models can be fine-tuned for specific tasks. In other words, it is possible to take a large, pre-trained language model and adjust its parameters and training on a smaller, domain-specific dataset to make it perform better on a particular task. For instance, with tunability, one can fine-tune BERT on a dataset of movie reviews to make it highly effective at sentiment analysis.

\begin{table} 
\centering
\setlength\tabcolsep{2pt}
\scriptsize
\caption{Comparison of Popular LLMs}
\label{tab:llmscomparison}
\begin{tabular}{lllclr}
\hline
\textbf{Model} & \textbf{Date} & \textbf{Provider} & \textbf{Open-Source} & \textbf{Params} & \textbf{Tunability} \\
\hline
gpt-4~\cite{Ding_2023} &  2023.03 & OpenAI & \tickNo & 1.7T & \tickNo \\
gpt-3.5-turbo & 2021.09 & OpenAI & \tickNo & 175B & \tickNo \\
gpt-3 \cite{brown2020language}& 2020.06 & OpenAI & \tickNo & 175B & \tickNo \\
cohere-medium \cite{liang2023holistic} & 2022.07 & Cohere & \tickNo & 6B & \tickYes \\
cohere-large \cite{liang2023holistic} & 2022.07 & Cohere & \tickNo & 13B & \tickYes \\
cohere-xlarge \cite{liang2023holistic} & 2022.06 & Cohere & \tickNo & 52B & \tickYes \\
BERT \cite{devlin2019bert} & 2018.08 & Google & \tickYes & 340M & \tickYes \\
T5 \cite{raffel2023exploring} & 2019 & Google & \tickYes & 11B & \tickYes \\
PaLM \cite{narang_chowdhery2022palm} & 2022.04 & Google & \tickYes & 540B & \tickYes \\
LLaMA \cite{meta_ai2023llama} & 2023.02 & Meta AI & \tickYes & 65B & \tickYes \\
CTRL \cite{salesforce2023conditional} & 2019 & Salesforce & \tickYes & 1.6B & \tickYes \\
Dolly 2.0 \cite{databricks2023free} & 2023.04 & Databricks & \tickYes & 12B & \tickYes \\
\hline
\end{tabular}

\end{table}

%% file: sections/sec3-researchproblems.tex
\section{Overview}
\label{sec:overview}
\subsection{Scope}

\smallskip

\noindent Our paper endeavors to conduct a thorough literature review, with the objective of collating and scrutinizing existing research and studies about the realms of security and privacy in the context of LLMs. The effort is geared towards both establishing the current state of the art in this domain and pinpointing gaps in our collective knowledge. While it is true that LLMs wield multifaceted applications extending beyond security considerations (e.g., social and financial impacts), our primary focus remains steadfastly on matters of security and privacy. 
Moreover, it is noteworthy that GPT models have attained significant prominence within this landscape. Consequently, when delving into specific content and examples, we aim to employ GPT models as illustrative benchmarks.

\subsection{The Research Questions}
\smallskip

\noindent LLMs have carried profound implications across diverse domains. However, it is essential to recognize that, as with any powerful technology, LLMs bear a significant responsibility. Our paper delves deeply into the multifaceted role of LLMs in the context of security and privacy. We intend to scrutinize their positive contributions to these domains, explore the potential threats they may engender, and uncover the vulnerabilities that could compromise their integrity.  To accomplish this, our study will conduct a thorough literature review centered around three pivotal research questions:

\begin{itemize}
    \item \textbf{The Good (\S\ref{sec:good}):} How do LLMs positively contribute to security and privacy in various domains, and what are the potential benefits they bring to the security community? 

    \item \textbf{The Bad (\S\ref{sec:bad}):}  What are the potential risks and threats associated with the use of LLMs in the context of cybersecurity? Specifically, how can LLMs be used for malicious purposes, and what types of cyber attacks can be facilitated or amplified using LLMs? 

    \item \textbf{The Ugly (\S\ref{sec:ugly}):} What vulnerabilities and weaknesses exist within LLMs, and how do these vulnerabilities pose a threat to security and privacy?
\end{itemize}

\begin{figure}
 \includegraphics[width= .45\textwidth]{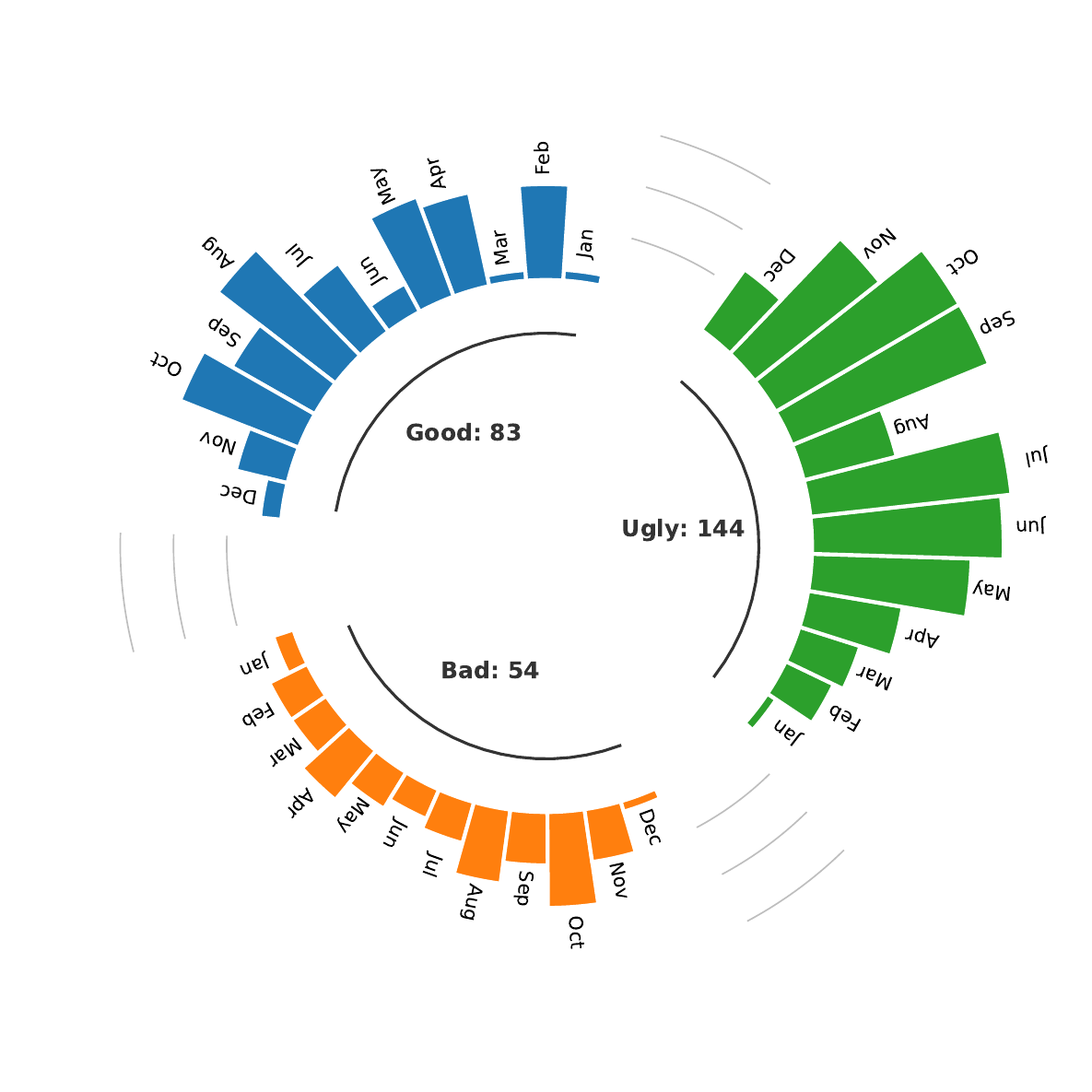}
 \caption{An overview of our collected papers.} 
\label{fig:overview}
\end{figure} 

Motivated by these questions, we conducted a search on Google Scholar and compiled papers related to security and privacy involving LLMs. As shown in \autoref{fig:overview}, we gathered a total of \goodPaper~``good'' papers that highlight the positive contributions of LLMs to security and privacy. Additionally, we identified \badPaper~``bad'' papers, in which attackers exploited LLMs to target users, and \uglyPaper~``ugly'' papers, in which authors discovered vulnerabilities within LLMs. Most of the papers were published in 2023, with only \prePaper~of them released in between 2007 and 2022. Notably, there is a consistent upward trend in the number of papers released each month, with October reaching its peak, boasting the highest number of papers published (38 papers in total, accounting for 15.97\% of all the collected papers). It is conceivable that more security-related LLM papers will be published in the near future.

\begin{mybox}[boxsep=0pt,
	boxrule=1pt,
	left=4pt,
	right=4pt,
 	top=4pt,
 	bottom=4pt,
	]

 \textbf{Finding I.} In terms of security-related applications (i.e., the ``good'' and the ``bad'' parts), it is evident that the majority of researchers are inclined towards using LLMs to bolster the security community, such as in vulnerability detection and security test generation, despite the presence of some vulnerabilities in LLMs at this stage. There are relatively few researchers who employ LLMs as tools for conducting attacks. In summary, LLMs contribute more positively than negatively to the security community.  \looseness=-1
 \end{mybox}

%% file: sections/sec4-thegood.tex
\section{Positive Impacts on Security and Privacy}
\label{sec:good}

\smallskip
 \noindent
 In this section, we explore the beneficial impacts of employing LLMs. In the context of code or data privacy, we have opted to use the term ``privacy'' to characterize scenarios in which LLMs are utilized to ensure the confidentiality of either code or data. However, given that we did not come across any papers specifically addressing code privacy, our discussion focuses on code security (\S\ref{code:good}) as well as both data security and privacy (\S\ref{data:good}).

\vspace{2mm}

\begin{table*}[]
\scriptsize
\setlength\tabcolsep{1.5pt}
\caption{LLMs for Code Security and Privacy}
\label{tab:llmforcodesecurity}
\begin{tabular}{@{}lllccccccllr@{}}
\toprule[1.5pt]
\multirow{5}{*}{\textbf{Work}} &  & \multicolumn{6}{c}{\textbf{Life Cycle}}                                                                                                                                                                                                                                                                                                                                                                                                                       & \multirow{5}{*}{\textbf{LLM(s)}} & \multirow{5}{*}{\textbf{Domain}} & \multirow{5}{*}{\textbf{\begin{tabular}[c]{@{}c@{}} When compared to \\ SOTA ways?\end{tabular}}} \\ \cmidrule(lr){3-3} \cmidrule(lr){4-4} \cmidrule(lr){5-8}
                               &                                & \multirow{2}{*}{\textbf{ Coding (C)}} & \multirow{2}{*}{\textbf{\begin{tabular}[c]{@{}c@{}}Test Case\\ Generating\\ (TCG) \end{tabular}}} & \multicolumn{4}{c}{\textbf{Running and Executing (RE)}}                                                                                                                                                                                                                                                                                                                                                &                                 &                                       &                                                                                                      \\
                               \cmidrule(lr){5-5} \cmidrule(lr){6-6}
                                \cmidrule(lr){7-7} \cmidrule(lr){8-8}
                               &                                &                                      &                                   & \textbf{\begin{tabular}[c]{@{}c@{}}Bug\\ Detecting\end{tabular}} & \multicolumn{1}{c}{\textbf{\begin{tabular}[c]{@{}c@{}}Malicious \\ Code Detecting\end{tabular}}} & \multicolumn{1}{c}{\textbf{\begin{tabular}[c]{@{}c@{}}Vulnerability \\  Detecting\end{tabular}}} & \multicolumn{1}{c}{\textbf{\begin{tabular}[c]{@{}c@{}} Fixing\end{tabular}}} &                                 &                                       &                                                                                                      \\    \midrule

\multicolumn{1}{l}{Sandoval et al.~\cite{sandoval2023lost}}           & \multicolumn{1}{l}{}           & \multicolumn{1}{c}{\faCircle}                 & \multicolumn{1}{c}{\faCircleO}              & \multicolumn{1}{c}{\faCircleO}                                                    &    \faCircleO                                                                                              &      \faCircleO                                                                                             &     \faCircleO                                                                                              & \multicolumn{1}{c}{Codex}            & \multicolumn{1}{c}{-}                  & \multicolumn{1}{r}{\tickUp Negligible risks}                                                                                 \\
\multicolumn{1}{l}{\textsf{SVEN}~\cite{he2023large}}           & \multicolumn{1}{l}{}           & \multicolumn{1}{c}{\faCircle}                 & \multicolumn{1}{c}{\faCircleO}              & \multicolumn{1}{c}{\faCircleO}                                                    &    \faCircleO                                                                                              &      \faCircleO                                                                                             &     \faCircleO                                                                                              & \multicolumn{1}{c}{CodeGen}            & \multicolumn{1}{c}{-}                  & \multicolumn{1}{r}{\tickUp More faster/secure}

\\
\multicolumn{1}{l}{\textsf{SALLM}~\cite{siddiq2023generate}}           & \multicolumn{1}{l}{}           & \multicolumn{1}{c}{\faCircle}                 & \multicolumn{1}{c}{\faCircleO}              & \multicolumn{1}{c}{\faCircleO}                                                    &    \faCircleO                                                                                              &      \faCircleO                                                                                             &     \faCircleO                                                                                              & \multicolumn{1}{c}{ChatGPT etc.}            & \multicolumn{1}{c}{-}                  & \multicolumn{1}{r}{-}                                                                                 \\
\multicolumn{1}{l}{Madhav et al.~\cite{cryptoeprint:2023/212}}           & \multicolumn{1}{l}{}           & \multicolumn{1}{c}{\faCircle}                 & \multicolumn{1}{c}{\faCircleO}              & \multicolumn{1}{c}{\faCircleO}                                                    &    \faCircleO                                                                                              &      \faCircleO                                                                                             &     \faCircleO                                                                                              & \multicolumn{1}{c}{ChatGPT}            & \multicolumn{1}{c}{Hardware}                  & \multicolumn{1}{r}{-}                                                                                 \\

\multicolumn{1}{l}{Zhang et al.~\cite{zhang2023well}}           & \multicolumn{1}{l}{}           & \multicolumn{1}{c}{\faCircleO}                 & \multicolumn{1}{c}{\faCircle}              & \multicolumn{1}{c}{\faCircleO}                                                    &    \faCircleO                                                                                              &      \faCircle                                                                                             &     \faCircleO                                                                                              & \multicolumn{1}{c}{ChatGPT}            & \multicolumn{1}{c}{Supply chain}                  & \multicolumn{1}{r}{\tickUp More valid cases}                                                                                 \\

\multicolumn{1}{l}{\textsf{Libro}~\cite{kang2023llm}}           & \multicolumn{1}{l}{}           & \multicolumn{1}{c}{\faCircleO}                 & \multicolumn{1}{c}{\faCircle}              & \multicolumn{1}{c}{\faCircleO}                                                    &    \faCircleO                                                                                              &      \faCircle                                                                                             &     \faCircleO                                                                                              & \multicolumn{1}{c}{LLaMA}            & \multicolumn{1}{c}{-}                  & \multicolumn{1}{r}{\tickDown Higher FP/FN}                                                                                 \\

\multicolumn{1}{l}{\textsf{TitanFuzz}~\cite{deng2022fuzzing}}           & \multicolumn{1}{l}{}           & \multicolumn{1}{c}{\faCircleO}                 & \multicolumn{1}{c}{\faCircle}              & \multicolumn{1}{c}{\faCircle}                                                    &    \faCircleO                                                                                              &      \faCircle                                                                                             &     \faCircleO                                                                                              & \multicolumn{1}{c}{Codex}            & \multicolumn{1}{c}{DL libs}                  & \multicolumn{1}{r}{\tickUp Higher coverage}                                                                                 \\

\multicolumn{1}{l}{\textsf{FuzzGPT}~\cite{deng2023largefuzzinggpt}}           & \multicolumn{1}{l}{}           & \multicolumn{1}{c}{\faCircleO}                 & \multicolumn{1}{c}{\faCircle}              & \multicolumn{1}{c}{\faCircle}                                                    &    \faCircleO                                                                                              &      \faCircle                                                                                             &     \faCircleO                                                                                              & \multicolumn{1}{c}{ChatGPT}            & \multicolumn{1}{c}{DL libs}                  & \multicolumn{1}{r}{\tickUp Higher coverage}                                                                                 \\
\multicolumn{1}{l}{\textsf{Fuzz4All}~\cite{xia2023universal}}           & \multicolumn{1}{l}{}           & \multicolumn{1}{c}{\faCircleO}                 & \multicolumn{1}{c}{\faCircle}              & \multicolumn{1}{c}{\faCircle}                                                    &    \faCircleO                                                                                              &      \faCircle                                                                                            &     \faCircleO                                                                                              & \multicolumn{1}{c}{ChatGPT}            & \multicolumn{1}{c}{Languages}                  & \multicolumn{1}{r}{\tickUp Higher coverage}                                                                                 \\
\multicolumn{1}{l}{\textsf{WhiteFox}~\cite{yang2023whitebox}}           & \multicolumn{1}{l}{}           & \multicolumn{1}{c}{\faCircleO}                 & \multicolumn{1}{c}{\faCircle}              & \multicolumn{1}{c}{\faCircle}                                                    &    \faCircleO                                                                                              &      \faCircle                                                                                             &     \faCircleO                                                                                              & \multicolumn{1}{c}{GPT4}            & \multicolumn{1}{c}{Compiler}                  & \multicolumn{1}{r}{\tickUp High-quality tests}                                                                                 \\
\multicolumn{1}{l}{Zhang et al.~\cite{zhang2023understanding}}           & \multicolumn{1}{l}{}           & \multicolumn{1}{c}{\faCircleO}                 & \multicolumn{1}{c}{\faCircle}              & \multicolumn{1}{c}{\faCircle}                                                    &    \faCircleO                                                                                              &      \faCircle                                                                                             &     \faCircleO                                                                                              & \multicolumn{1}{c}{ChatGPT}            & \multicolumn{1}{c}{API}                  & \multicolumn{1}{r}{-}                                                                                 \\

\multicolumn{1}{l}{\textsf{CHATAFL}~\cite{zhou2012hey}}           & \multicolumn{1}{l}{}           & \multicolumn{1}{c}{\faCircleO}                 & \multicolumn{1}{c}{\faCircle}              & \multicolumn{1}{c}{\faCircle}                                                    &    \faCircleO                                                                                              &      \faCircle                                                                                             &     \faCircleO                                                                                              & \multicolumn{1}{c}{ChatGPT}            & \multicolumn{1}{c}{Protocol}                  & \multicolumn{1}{r}{\tickUp Higher coverage}                                                                                 \\
\multicolumn{1}{l}{Henrik~\cite{gpt3malware2023}}           & \multicolumn{1}{l}{}           & \multicolumn{1}{c}{\faCircleO}                 & \multicolumn{1}{c}{\faCircleO}              & \multicolumn{1}{c}{\faCircleO}                                                    &    \faCircle                                                                                              &      \faCircleO                                                                                             &     \faCircleO                                                                                              & \multicolumn{1}{c}{ChatGPT}            & \multicolumn{1}{c}{-}                  & \multicolumn{1}{r}{\tickDown Higer FP/FN}                                                                                 \\
\multicolumn{1}{l}{\textsf{Apiiro}~\cite{apiiro2023}}           & \multicolumn{1}{l}{}           & \multicolumn{1}{c}{\faCircleO}                 & \multicolumn{1}{c}{\faCircleO}              & \multicolumn{1}{c}{\faCircleO}                                                    &    \faCircle                                                                                              &      \faCircleO                                                                                             &     \faCircleO                                                                                              & \multicolumn{1}{c}{N/A}            & \multicolumn{1}{c}{-}                  & \multicolumn{1}{r}{-}                                                                                 \\
\multicolumn{1}{l}{Noever~\cite{noever2023large}}           & \multicolumn{1}{l}{}           & \multicolumn{1}{c}{\faCircleO}                 & \multicolumn{1}{c}{\faCircleO}              & \multicolumn{1}{c}{\faCircleO}                                                    &    \faCircleO                                                                                              &      \faCircle                                                                                             &     \faCircle                                                                                              & \multicolumn{1}{c}{ChatGPT}            & \multicolumn{1}{c}{-}                  & \multicolumn{1}{r}{\tickUp 4X faster}                                                                                 \\
\multicolumn{1}{l}{Bakhshandeh et al.~\cite{bakhshandeh2023using}}           & \multicolumn{1}{l}{}           & \multicolumn{1}{c}{\faCircleO}                 & \multicolumn{1}{c}{\faCircleO}              & \multicolumn{1}{c}{\faCircleO}                                                    &    \faCircleO                                                                                              &      \faCircle                                                                                             &     \faCircleO                                                                                              & \multicolumn{1}{c}{ChatGPT}            & \multicolumn{1}{c}{-}                  & \multicolumn{1}{r}{\tickUp Low FP/FN}                                                                                 \\
\multicolumn{1}{l}{Moumita et al.~\cite{10301302}}           & \multicolumn{1}{l}{}           & \multicolumn{1}{c}{\faCircleO}                 & \multicolumn{1}{c}{\faCircleO}              & \multicolumn{1}{c}{\faCircleO}                                                    &    \faCircleO                                                                                              &      \faCircle                                                                                             &     \faCircleO                                                                                              & \multicolumn{1}{c}{ChatGPT}            & \multicolumn{1}{c}{-}                  & \multicolumn{1}{r}{\tickDown Higher FP/FN}                                                                                 \\
\multicolumn{1}{l}{Cheshkov et al.~\cite{cheshkov2023evaluation}}           & \multicolumn{1}{l}{}           & \multicolumn{1}{c}{\faCircleO}                 & \multicolumn{1}{c}{\faCircleO}              & \multicolumn{1}{c}{\faCircleO}                                                    &    \faCircleO                                                                                              &      \faCircle                                                                                             &     \faCircleO                                                                                              & \multicolumn{1}{c}{ChatGPT}            & \multicolumn{1}{c}{-}                  & \multicolumn{1}{r}{\tickDown No better}                                                                                 \\
%
\multicolumn{1}{l}{\textsf{LATTE}~\cite{liu2023harnessing}}           & \multicolumn{1}{l}{}           & \multicolumn{1}{c}{\faCircleO}                 & \multicolumn{1}{c}{\faCircleO}              & \multicolumn{1}{c}{\faCircleO}                                                    &    \faCircleO                                                                                              &      \faCircle                                                                                             &     \faCircleO                                                                                              & \multicolumn{1}{c}{GPT}            & \multicolumn{1}{c}{-}                  & \multicolumn{1}{r}{\tickUp Cost effective}                                                                                 \\
\multicolumn{1}{l}{\textsf{DefectHunter}~\cite{wang2023defecthunter}}           & \multicolumn{1}{l}{}           & \multicolumn{1}{c}{\faCircleO}                 & \multicolumn{1}{c}{\faCircleO}              & \multicolumn{1}{c}{\faCircleO}                                                    &    \faCircleO                                                                                              &      \faCircle                                                                                             &     \faCircleO                                                                                              & \multicolumn{1}{c}{Codex}            & \multicolumn{1}{c}{-}                  & \multicolumn{1}{r}{-}                                                                                 \\
\multicolumn{1}{l}{Chen et al.~\cite{chen2023when}}           & \multicolumn{1}{l}{}           & \multicolumn{1}{c}{\faCircleO}                 & \multicolumn{1}{c}{\faCircleO}              & \multicolumn{1}{c}{\faCircleO}                                                    &    \faCircleO                                                                                              &      \faCircle                                                                                             &     \faCircleO                                                                                              & \multicolumn{1}{c}{ChatGPT}            & \multicolumn{1}{c}{Blockchain}                  & \multicolumn{1}{r}{-}                                                                                 \\
\multicolumn{1}{l}{Hu et al.~\cite{hu2023large}}           & \multicolumn{1}{l}{}           & \multicolumn{1}{c}{\faCircleO}                 & \multicolumn{1}{c}{\faCircleO}              & \multicolumn{1}{c}{\faCircleO}                                                    &    \faCircleO                                                                                              &      \faCircle                                                                                             &     \faCircleO                                                                                              & \multicolumn{1}{c}{ChatGPT}            & \multicolumn{1}{c}{Blockchain}                  & \multicolumn{1}{r}{-}                                                                                 \\

\multicolumn{1}{l}{KARTAL~\cite{sakaoglu2023kartal}}           & \multicolumn{1}{l}{}           & \multicolumn{1}{c}{\faCircleO}                 & \multicolumn{1}{c}{\faCircleO}              & \multicolumn{1}{c}{\faCircleO}                                                    &    \faCircleO                                                                                              &      \faCircle                                                                                             &     \faCircleO                                                                                              & \multicolumn{1}{c}{ChatGPT}            & \multicolumn{1}{c}{Web apps}                  & \multicolumn{1}{r}{\tickUp Less manual}                                                                                 \\

\multicolumn{1}{l}{\textsf{VulLibGen}~\cite{chen2023vullibgen}}           & \multicolumn{1}{l}{}           & \multicolumn{1}{c}{\faCircleO}                 & \multicolumn{1}{c}{\faCircleO}              & \multicolumn{1}{c}{\faCircleO}                                                    &    \faCircleO                                                                                              &      \faCircle                                                                                             &     \faCircleO                                                                                              & \multicolumn{1}{c}{LLaMa}            & \multicolumn{1}{c}{Libs}                  & \multicolumn{1}{r}{\tickUp Higher accuracy/speed}                                                                                 \\
\multicolumn{1}{l}{Ahmad et al.~\cite{ahmad2023fixing}}           & \multicolumn{1}{l}{}           & \multicolumn{1}{c}{\faCircleO}                 & \multicolumn{1}{c}{\faCircleO}              & \multicolumn{1}{c}{\faCircleO}                                                    &    \faCircleO                                                                                              &      \faCircle                                                                                             &     \faCircle                                                                                              & \multicolumn{1}{c}{Codex}            & \multicolumn{1}{c}{Hardware}                  & \multicolumn{1}{r}{\tickUp Fix more bugs}                                                                                 \\

\multicolumn{1}{l}{InferFix~\cite{jin2023inferfix}}           & \multicolumn{1}{l}{}           & \multicolumn{1}{c}{\faCircleO}                 & \multicolumn{1}{c}{\faCircleO}              & \multicolumn{1}{c}{\faCircle}                                                    &    \faCircleO                                                                                              &      \faCircle                                                                                             &     \faCircle                                                                                              & \multicolumn{1}{c}{Codex}            & \multicolumn{1}{c}{-}                  & \multicolumn{1}{r}{\tickUp CI Pipeline} \\

\multicolumn{1}{l}{Pearce et al.~\cite{10179324}}           & \multicolumn{1}{l}{}           & \multicolumn{1}{c}{\faCircleO}                 & \multicolumn{1}{c}{\faCircleO}              & \multicolumn{1}{c}{\faCircle}                                                    &    \faCircleO                                                                                              &      \faCircle                                                                                             &     \faCircleO                                                                                              & \multicolumn{1}{c}{Codex etc.}            & \multicolumn{1}{c}{-}                  & \multicolumn{1}{r}{\tickUp Zero-shot} \\
\multicolumn{1}{l}{Fu et al.~\cite{fu2023chatgpt}}           & \multicolumn{1}{l}{}           & \multicolumn{1}{c}{\faCircleO}                 & \multicolumn{1}{c}{\faCircleO}              & \multicolumn{1}{c}{\faCircle}                                                    &    \faCircleO                                                                                              &      \faCircle                                                                                             &     \faCircle                                                                                              & \multicolumn{1}{c}{ChatGPT}            & \multicolumn{1}{c}{APR}                  & \multicolumn{1}{r}{\tickUp Higher accuracy} \\
\multicolumn{1}{l}{Sobania et al.~\cite{sobania2023analysis}}           & \multicolumn{1}{l}{}           & \multicolumn{1}{c}{\faCircleO}                 & \multicolumn{1}{c}{\faCircleO}              & \multicolumn{1}{c}{\faCircleO}                                                    &    \faCircleO                                                                                              &      \faCircleO                                                                                             &     \faCircle                                                                                              & \multicolumn{1}{c}{ChatGPT etc.}            & \multicolumn{1}{c}{APR}                  & \multicolumn{1}{r}{\tickUp Higher accuracy} \\
\multicolumn{1}{l}{Jiang et al.~\cite{jiang2023impact}}           & \multicolumn{1}{l}{}           & \multicolumn{1}{c}{\faCircleO}                 & \multicolumn{1}{c}{\faCircleO}              & \multicolumn{1}{c}{\faCircleO}                                                    &    \faCircleO                                                                                              &      \faCircleO                                                                                             &     \faCircle                                                                                              & \multicolumn{1}{c}{ChatGPT}            & \multicolumn{1}{c}{APR}                  & \multicolumn{1}{r}{\tickUp Higher accuracy} \\
\bottomrule[1.5pt]
\end{tabular}
\end{table*}
\subsection{LLMs for Code Security}
\label{code:good}

\smallskip
\smallskip
\noindent
As shown in \autoref{tab:llmforcodesecurity}, LLMs have access to a vast repository of code snippets and examples spanning various programming languages and domains. They leverage their advanced language understanding and contextual analysis capabilities to thoroughly examine code and code-related text.  More specifically, LLMs can play a pivotal role throughout the entire code security lifecycle, including coding (C), test case generation (TCG), execution, and monitoring (RE).

\paragraph{Secure Coding (C)} We first discuss the use of LLMs in the context of secure code programming~\cite{espinha2023m} (or generation~\cite{ding2023static,vaithilingam2022expectation,ni2023lever,gu2023llm}). Sandoval et al.~\cite{sandoval2023lost} conducted a user study (58 users) to assess the security implications of LLMs, particularly OpenAI Codex, as code assistants for developers. They evaluated code written by student programmers when assisted by LLMs and found that participants assisted by LLMs did not introduce new security risks: the AI-assisted group produced critical security bugs at a rate no greater than 10\% higher than the control group (non-assisted).  He et al.~\cite{he2023large,he2023largeccs} focused on enhancing the security of code generated by LLMs. They proposed a novel method called \textsf{SVEN}, which leverages continuous prompts to control LLMs in generating secure code. With this method, the success rate improved from 59.1\% to 92.3\% when using the CodeGen LM. Mohammed et al. introduce \textsf{SALLM}~\cite{siddiq2023generate},  a framework consisting of a new security-focused dataset, an evaluation environment, and novel metrics for systematically assessing LLMs' ability to generate secure code.
Madhav et al.~\cite{cryptoeprint:2023/212} evaluate the security aspects of code generation processes on the ChatGPT platform, specifically in the hardware domain. They explore the strategies that a designer can employ to enable ChatGPT to provide secure hardware code generation.\looseness=-1

\paragraph{Test Case Generating (TCG)}  Several papers~\cite{chen2022codet,alagarsamy2023a3test,schafer2023adaptive,xie2023chatunitest,lemieux2023codamosa,siddiq2023exploring,yuan2023no} discuss the utilization of LLMs for generating test cases, with our particular emphasis on those addressing security implications.  Zhang et al.~\cite{zhang2023well} demonstrated the use of ChatGPT-4.0 for generating security tests to assess the impact of vulnerable library dependencies on software applications. They found that LLMs could successfully generate tests that demonstrated various supply chain attacks, outperforming existing security test generators. This approach resulted in 24 successful attacks across 55 applications. Similarly,  \textsf{Libro}~\cite{kang2023llm}, a framework that uses LLMs to automatically generate test cases to reproduce software security bugs. 

In the realm of security, fuzzing stands~\cite{yang2023crafting,hu2023augmenting,zhang2023understanding,zhao2023understanding,tay2023using} out as a widely employed technique for generating test cases.  
Deng et al. introduced \textsf{TitanFuzz}~\cite{deng2022fuzzing}, an approach that harnesses LLMs to generate input programs for fuzzing Deep Learning (DL) libraries. TitanFuzz demonstrates impressive code coverage (30.38\%/50.84\%) and detects previously unknown bugs (41 out of 65) in popular DL libraries. More recently, Deng et al.~\cite{deng2024large,deng2023largefuzzinggpt} refined LLM-based fuzzing (named FuzzGPT), aiming to generate unusual programs for DL library fuzzing. While TitanFuzz leverages LLMs' ability to generate ordinary code, FuzzGPT addresses the need for edge-case testing by priming LLMs with historical bug-triggering program. Fuzz4All~\cite{xia2023universal} leverages LLMs as input generators and mutation engines, creating diverse and realistic inputs for various languages (e.g., C, C++), improving the previous state-of-the-art coverage by 36.8\% on average.  \textsf{WhiteFox}~\cite{yang2023whitebox}, a novel white-box compiler fuzzer that utilizes LLMs to test compiler optimizations, outperforms existing fuzzers (it generates high-quality tests for intricate optimizations, surpassing state-of-the-art fuzzers by up to 80 optimizations). Zhang et al.~\cite{zhang2023understanding} explore the generation of fuzz drivers for library API fuzzing using LLMs. Results show that LLM-based generation is practical, with 64\% of questions solved entirely automatically and up to 91\% with manual validation. \textsf{CHATAFL}~\cite{zhou2012hey} is an LLM-guided protocol fuzzer that constructs grammars for message types and mutates messages or predicts the next messages based on LLM interactions,  achieving better state and code coverage compared to state-of-the-art fuzzers (e.g., AFLNET~\cite{pham2020aflnet}, NSFUZZ~\cite{qin2023nsfuzz}). \looseness=-1

\paragraph{Vulnerable Code Detecting (RE)} 
Noever~\cite{noever2023large} explores the capability of LLMs, particularly OpenAI's GPT-4, in detecting software vulnerabilities. This paper shows that GPT-4 identified approximately four times the number of vulnerabilities compared to traditional static code analyzers (e.g., Snyk and Fortify). Parallel conclusions have also been drawn in other efforts~\cite{gpt213bugs,bakhshandeh2023using}. However, Moumita et al.~\cite{10301302} applied LLMs for software vulnerability detection, exposing a noticeable performance gap when compared to conventional static analysis tools. This disparity primarily arises from the relatively higher occurrence of false alerts generated by LLMs. Similarly, Cheshkov et al.~\cite{cheshkov2023evaluation} point out that the ChatGPT model performed no better than a dummy classifier for both binary and multi-label classification tasks in code vulnerability detection.
Wang et al. introduce \textsf{DefectHunter}~\cite{wang2023defecthunter}, a novel model that employs LLM-driven techniques for code vulnerability detection. They demonstrate the potential of combining LLMs with advanced mechanisms (e.g., Conformer) to identify software vulnerabilities more effectively. This combination shows an improvement in effectiveness, approximately from 14.64\% to 20.62\%, compared with Pongo-70B. \textsf{LATTE}~\cite{liu2023harnessing} is a novel static binary taint analysis method powered by LLMs. \textsf{LATTE} surpasses existing state-of-the-art techniques (e.g., Emtaint, Arbiter, and Karonte), demonstrating remarkable effectiveness in vulnerability detection (37 new bugs in real-world firmware) with lower cost. 

Efforts in leveraging LLMs for vulnerability detection extend to specialized domains (e.g.,blockchain~\cite{hu2023large,chen2023when}, kernel~\cite{helmke2023check} mobile~\cite{wen2023empowering}). For instance, Chen et al.~\cite{chen2023when} and Hu et al.~\cite{hu2023large} focus on the application of LLMs in identifying vulnerabilities within blockchain smart contracts.  Sakaoglu's study introduces KARTAL~\cite{sakaoglu2023kartal}, a pioneering approach that harnesses LLMs for web application vulnerability detection. This method achieves an accuracy of up to 87.19\% and is capable of conducting 539 predictions per second. Additionally, Chen et al.~\cite{chen2023vullibgen} make a noteworthy contribution with VulLibGen, a generative methodology utilizing LLMs to identify vulnerable libraries. Ahmad et al.~\cite{ahmad2023fixing} shift the focus to hardware security. They investigate the use of LLMs, specifically OpenAI's Codex, in automatically identifying and repairing security-related bugs in hardware designs.  PentestGPT~\cite{deng2023pentestgpt}, an automated penetration testing tool, uses the domain knowledge inherent in LLMs to address individual sub-tasks of penetration testing, improving task completion rates significantly. \looseness=-1

\paragraph{Malicious Code Detecting (RE)}
Using LLM to detect malware is a promising application. This approach leverages the natural language processing capabilities and contextual understanding of LLMs to identify malicious software.  
In experiments with GPT-3.5 conducted by Henrik Plate~\cite{gpt3malware2023}, it was found that LLM-based malware detection can complement human reviews but not replace them. Out of 1800 binary classifications performed, there were both false-positives and false-negatives. The use of simple tricks could also deceive the LLM's assessments. More recently, there are a few attempts have been made in this direction. For example, Apiiro~\cite{apiiro2023} is a malicious code analysis tool using LLMs. Apiiro's strategy involves the creation of LLM Code Patterns (LCPs) to represent code in vector format, making it easier to identify similarities and cluster packages efficiently. Its LCP detector incorporates LLMs, proprietary code analysis, probabilistic sampling, LCP indexing, and dimensionality reduction to identify potentially malicious code.\looseness=-1

\begin{table}[]
\scriptsize
\setlength\tabcolsep{1pt}
\caption{LLMs for Data Security and Privacy}
\label{tab:llmfordatasecurity}
\begin{tabular}{lccccccr}
\toprule[1.5pt]

\multirow{3}{*}{\textbf{Work}} &
  \multicolumn{4}{c}{\multirow{2}{*}{\textbf{Prop.}}} &
  \multirow{3}{*}{\textbf{Model}} &
  \multirow{3}{*}{\textbf{Domain}} &
  \multirow{3}{*}{\textbf{\begin{tabular}[c]{@{}c@{}}Compared to \\ SOTA ways?\end{tabular}}} \\
 &
  \multicolumn{3}{c}{} &
   &
   &
   \\ \cmidrule(lr){2-5}
 &
  \textbf{I} &
  \multicolumn{1}{c}{\textbf{C}} &
  \multicolumn{1}{c}{\textbf{R}} &
  \multicolumn{1}{c}{\textbf{T}} &
   &
   &
   \\ \midrule
 {Fang~\cite{wang2023ransomware}} &
\faCircle  & \faCircleO
    & \faCircle
   & \faCircleO
   & ChatGPT
   & Ransomware
  &
  \multicolumn{1}{r}{-} \\ 

Liu et al.~\cite{mcintosh2023harnessing} &
\faCircle  & \faCircleO
   & \faCircle
   & \faCircleO
   & ChatGPT
   & Ransomware
  &
  \multicolumn{1}{r}{-} \\   

 Amine et al.~\cite{elhafsi2023semantic} &
\faCircle  & \faCircle 
  & \faCircle
   & \faCircleO
   & ChatGPT
   & Semantic
  & 
  \multicolumn{1}{r}{\tickUp Aligned w/ SOTA} \\   

  \textsf{HuntGPT}~\cite{ali2023huntgpt} &
\faCircle  & \faCircle 
  & \faCircle
   & \faCircleO
   & ChatGPT
   & Network
  & 
  \multicolumn{1}{r}{\tickUp More effective} \\ 

 Chris et al.~\cite{egersdoerfer2023early} &
\faCircle  & \faCircle 
  & \faCircle
   & \faCircleO
   & ChatGPT
   & Log
  & 
  \multicolumn{1}{r}{\tickUp Less manual} \\ 
\textsf{AnomalyGPT}~\cite{gu2023anomalygpt} &
\faCircle  & \faCircle 
  & \faCircle
   & \faCircleO
   & ChatGPT
   & Video
  & 
  \multicolumn{1}{r}{\tickUp Less manual} \\ 
\textsf{LogGPT}~\cite{qi2023loggpt} &
\faCircle  & \faCircle 
  & \faCircle
   & \faCircleO
   & ChatGPT
   & Log
  & 
  \multicolumn{1}{r}{\tickUp Less manual} \\

Arpita et al.~\cite{vats2023recovering} &
\faCircleO  & \faCircle 
  & \faCircleO
   & \faCircleO
   & BERT etc.
   & -
  & 
  \multicolumn{1}{r}{-} \\
Takashi et al.~\cite{koide2023detecting} &
\faCircleO  & \faCircleO
  & \faCircle
   & \faCircleO
   & ChatGPT
   &  Phishing
  &
  \multicolumn{1}{r}{\tickUp High precision} \\ 
Fredrik et al.~\cite{heiding2023devising} &
\faCircleO  & \faCircleO
  & \faCircle
   & \faCircleO
   & ChatGPT etc
   &  Phishing
  &
  \multicolumn{1}{r}{\tickUp Effective} \\
 \textsf{IPSDM}~\cite{jamal2023improved} &
\faCircleO  & \faCircleO
  & \faCircle
   & \faCircleO
   & BERT 
   &  Phishing
  &
  \multicolumn{1}{r}{-}  \\
Kwon et al.~\cite{cryptoeprintcyrotpo}  &
\faCircleO  & \faCircle 
   & \faCircleO
   & \faCircleO
   & ChatGPT
   & -
  & 
  \multicolumn{1}{r}{\tickUp Non-exp friendly} \\ 
Scanlon et al.~\cite{SCANLON2023301609} &
\faCircleO  & \faCircleO
   & \faCircleO
   & \faCircle
   & ChatGPT
   & Forensic
  &
  \multicolumn{1}{r}{\tickUp More effective} \\   

Sladić et al.~\cite{sladić2023llm} &
\faCircleO  & \faCircleO
  & \faCircleO
   & \faCircle
   & ChatGPT
   & Honeypot
  & 
  \multicolumn{1}{r}{\tickUp More realistic} \\   
   

\textsf{WASA}~\cite{wang2023wasa} &
\faCircleO  & \faCircleO
  & \faCircle
   & \faCircle
   & -
   & Watermark
  &
  \multicolumn{1}{r}{\tickUp More effective} \\   

\textsf{REMARK}~\cite{zhang2023remarkllm} &
\faCircleO  & \faCircleO
   & \faCircle
   & \faCircle
   & -
   & Watermark
  &
  \multicolumn{1}{r}{\tickUp More effective} \\   

\textsf{SWEET}~\cite{lee2023wrote} &
\faCircleO  & \faCircleO
  & \faCircle
   & \faCircle
   & -
   &  Watermark
  &
  \multicolumn{1}{r}{\tickUp More effective} \\   
\bottomrule[1.5pt]
\end{tabular}
\end{table}

\paragraph{Vulnerable/Buggy Code Fixing (RE)}
Several papers~\cite{jiang2023impact, 10179324, xia2022practical} has focused on evaluate the performance of LLMs trained on code in the task of program repair. Jin et al.~\cite{jin2023inferfix} proposed \textsf{InferFix}, a transformer-based program repair framework that works in tandem with the combination of cutting-edge static analyzer with transformer-based model to address and fix critical security and performance issues with accuracy between 65\% to 75\%. Pearce et al.~\cite{10179324} observed that LLMs can repair insecure code in a range of contexts even without being explicitly trained on vulnerability repair tasks.

ChatGPT is noted for its ability in code bug detection and correction. Fu et al.~\cite{fu2023chatgpt} assessed ChatGPT in vulnerability-related tasks like predicting and classifying vulnerabilities, severity estimation, and analyzing over 190,000 C/C++ functions. They found that ChatGPT's performance was behind other LLMs specialized in vulnerability detection. However, Sobania et al.~\cite{sobania2023analysis} found ChatGPT's bug fixing performance competitive with standard program repair methods, as demonstrated by its ability to fix 31 out of 40 bugs. Xia et al.~\cite{xia2023conversation} presented \textsf{ChatRepair}, leveraging pre-trained language models (PLMs) for generating patches without dependency on bug-fixing datasets, aiming to enhance performance to generate patches without relying on bug-fixing datasets, aiming to improve ChatGPT's code-fixing abilities using a mix of successful and failure tests. As a result, they fixed 162 out of 337 bugs at a cost of \$0.42 each. \looseness=-1

\begin{mybox}[boxsep=0pt,
	boxrule=1pt,
	left=4pt,
	right=4pt,
 	top=4pt,
 	bottom=4pt,
	]

 \textbf{Finding II.}  As shown in \autoref{tab:llmforcodesecurity}, a comparison with state-of-the-art methods reveals that the majority of researchers (17 out of 25) have concluded that LLM-based methods outperform traditional approaches (advantages include higher code coverage, higher detecting accuracy, less cost etc.). Only four papers argue that LLM-based methods do not surpass the  state-of-the-art appoarches. The most frequently discussed issue with LLM-based methods is their tendency to produce both high false negatives and false positives when detecting vulnerabilities or bugs.  \looseness=-1
 \end{mybox}

\subsection{LLMs for Data Security and Privacy}
\label{data:good}

\smallskip

\noindent As demonstrated in \autoref{tab:llmfordatasecurity}, LLMs make valuable contributions to the realm of data security, offering multifaceted approaches to safeguarding sensitive information. We have organized the research papers into distinct categories based on the specific facets of data protection that LLMs enhance. These facets encompass critical aspects such as data integrity (I), which ensures that data remains
 uncorrupted throughout its life cycle;   data reliability (R), which ensures the accuracy of data;  data confidentiality (C), which focuses on guarding against unauthorized access and disclosure of sensitive information; and data traceability (T), which involves tracking and monitoring data access and usage. \looseness=-1

\paragraph{Data Integrity (I)}  Data Integrity ensures that data remains unchanged and uncorrupted throughout its life cycle. As of now, there are a few works that discuss how to use LLMs to protect data integrity. For example, 
ransomware usually encrypts a victim's data, making the data inaccessible without a decryption key that is held by the attacker, which breaks the data integrity. Wang Fang's research~\cite{wang2023ransomware} examines using LLMs for ransomware cybersecurity strategies, mostly theoretically proposing real-time analysis, automated policy generation, predictive analytics, and knowledge transfer. However, these strategies lack empirical validation. Similarly, Liu et al.~\cite{mcintosh2023harnessing} explored the potential of LLMs for creating cybersecurity policies aimed at mitigating ransomware attacks with data exfiltration. They compared GPT-generated Governance, Risk and Compliance (GRC) policies to those from established security vendors and government cybersecurity agencies. 
They recommended that companies should incorporate GPT into their GRC policy development. \looseness=-1

Anomaly detection is a key defense mechanism that identifies unusual behavior. While it does not directly protect data integrity, it identifies abnormal or suspicious behavior that can potentially compromise data integrity (as well as data confidentiality and data reliability).  
Amine et al.~\cite{elhafsi2023semantic} introduced an LLM-based monitoring framework for detecting semantic anomalies in vision-based policies and applied it to both finite state machine policies for autonomous driving and learned policies for object manipulation.  Experimental results demonstrate that it can effectively identify semantic anomalies, aligning with human reasoning. 
\textsf{HuntGPT}~\cite{ali2023huntgpt} is an LLM-based intrusion detection system for network anomaly detection. The results demonstrate its effectiveness in improving user understanding and interaction. 
Chris et al.~\cite{egersdoerfer2023early} and \textsf{LogGPT}~\cite{qi2023loggpt} explore ChatGPT's potential for log-based anomaly detection in parallel file systems. Results show that it addresses the issues in traditional manual labeling and interpretability. 
\textsf{AnomalyGPT}~\cite{gu2023anomalygpt} uses Large Vision-Language Models to detect industrial anomalies. It eliminates manual threshold setting and supports multi-turn dialogues. \looseness=-1

\paragraph{Data Confidentiality (C)} Data confidentiality refers to the practice of protecting sensitive information from unauthorized access or disclosure,  a topic extensively discussed in LLM privacy discussions~\cite{peris2023privacy,sebastian2023privacy,vats2023recovering,abbasian2023conversational}. 
However, most of these studies concentrate on enhancing LLMs through state-of-the-art Privacy Enhancing Techniques (e.g., zero-knowledge proofs~\cite{raeini2023privacy}, differential privacy (e.g.,~\cite{sebastian2023privacy,majmudar2022differentially,li2023privacy}, and federated learning~\cite{kuang2023federatedscope,jiang2023low,fan2023fate}). There are only a few attempts that utilize LLMs to enhance user privacy. For example, Arpita et al.~\cite{vats2023recovering} use LLMs to preserve privacy by replacing identifying information in textual data with generic markers. Instead of storing sensitive user information, such as names, addresses, or credit card numbers, the LLMs suggest substitutes for the masked tokens. This obfuscation technique helps to protect user data from being exposed to adversaries. By using LLMs to generate substitutes for masked tokens, the models can be trained on obfuscated data without compromising the privacy and security of the original information. Similar ideas have also been explored in other studies~\cite{abbasian2023conversational,stephens2023researchers}. Hyeokdong et al.~\cite{cryptoeprintcyrotpo} explore implementing cryptography with ChatGPT, which ultimately protects data confidentiality. Despite the lack of extensive coding skills or programming knowledge, the authors were able to successfully implement cryptographic algorithms through ChatGPT. This highlights the potential for individuals to utilize ChatGPT for cryptography tasks. 

\paragraph{Data Reliability (R)} In our context, data reliability refers to the accuracy of data. It is a measure of how well data can be depended upon to be accurate, and free from errors or bias. Takashi et al.~\cite{koide2023detecting} proposed to use ChatGPT for the detection of sites that contain phishing content. Experimental results using GPT-4 show promising performance, with high precision and recall rates. Fredrik et al.~\cite{heiding2023devising} assessed the ability of four large language models (GPT, Claude, PaLM, and LLaMA) to detect malicious intent in phishing emails, and found that they were generally effective,   even surpassing human detection, although occasionally slightly less accurate.  \textsf{IPSDM}~\cite{jamal2023improved} is a model fine-tuned from the BERT family to identify phishing and spam emails effectively. \textsf{IPSDM} demonstrates superior performance in classifying emails, both in unbalanced and balanced datasets.


\paragraph{Data Traceability (T)}
Data traceability is the capability to track and document the origin, movement, and history of data within a single system or across multiple systems. This concept is particularly vital in fields such as incident management and forensic investigations, where understanding the journey and transformations of events to resolving issues and conducting thorough analyses. LLMs have gained traction in forensic investigations, offering novel approaches for analyzing digital evidence. Scanlon et al.~\cite{SCANLON2023301609} explored how ChatGPT assists in analyzing OS artifacts like logs, files, cloud interactions, executable binaries, and in examining memory dumps to detect suspicious activities or attack patterns. Additionally, Sladić et al.~\cite{sladić2023llm} proposed that generative models like ChatGPT can be used to create realistic honeypots to deceive human attackers.

Watermarking involves embedding a distinctive, typically imperceptible or hard-to-identify signal within the outputs of a model. Wang et al.~\cite{wang2023wasa} discusses concerns regarding the intellectual property of training data for LLMs and proposed \textsf{WASA} framework to learn the mapping between the texts of different data providers. Zhang et al.~\cite{zhang2023remarkllm} developed \textsf{REMARK-LLM} that focused on monitor the utilization of their content and validate their watermark retrieval. This helps protect against malicious uses such as spamming and plagiarism. Furthermore, identifying code produced by LLMs is vital for addressing legal and ethical issues concerning code licensing, plagiarism, and malware creation. Similarly, Li et al.~\cite{li2023protecting} propose the first watermark technique to protect large language model-based code generation APIs from remote imitation attacks. Lee et al.~\cite{lee2023wrote} developed \textsf{SWEET}, a tool that implements watermarking specifically on tokens within programming languages.

\begin{mybox}[boxsep=0pt,
	boxrule=1pt,
	left=4pt,
	right=4pt,
 	top=4pt,
 	bottom=4pt,
	]

 \textbf{Finding III.} Likewise, it is noticeable that LLMs excel in data protection, surpassing current solutions and requiring fewer manual interventions. \autoref{tab:llmforcodesecurity} and  \autoref{tab:llmfordatasecurity} reveal that  ChatGPT is the predominant LLM extensively employed in diverse security applications. Its versatility and effectiveness make it a preferred choice for various security-related tasks, further reinforcing its position as a go-to solution in the field of artificial intelligence and cybersecurity. \looseness=-1
 \end{mybox}

%% file: sections/sec5-thebad.tex
\section{Negative Impacts on Security and Privacy}
\label{sec:bad}

\smallskip

\begin{figure}
 \includegraphics[width= 0.5\textwidth]{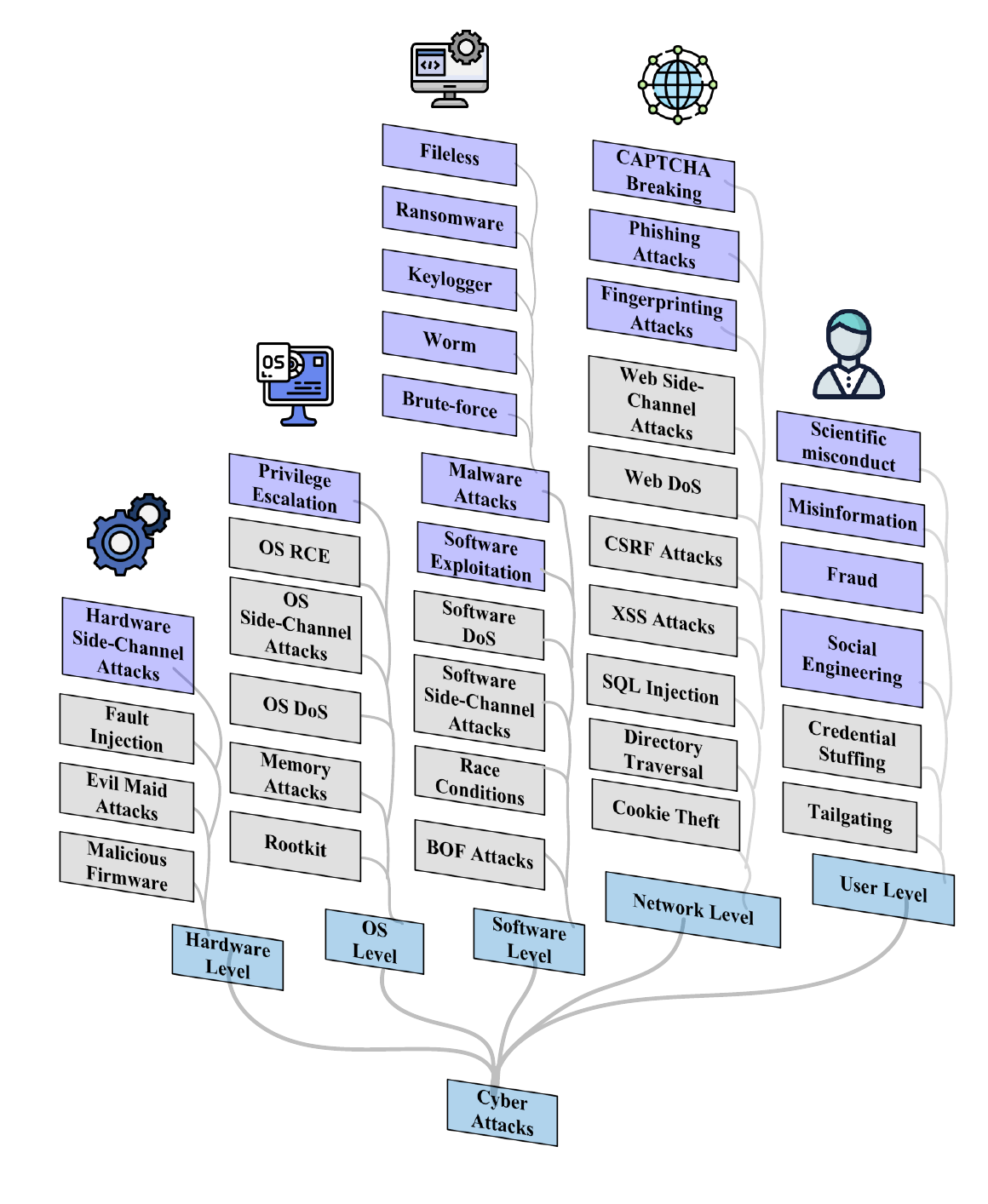}
 \vspace{3mm}
 \caption{Taxonomy of Cyberattacks. The colored boxes represent attacks that have been demonstrated to be executable using LLMs, whereas the gray boxes indicate attacks that cannot be executed with LLMs. } 
\label{fig:taxonomy}
\end{figure}

\begin{figure}
 \includegraphics[width= 0.5\textwidth]{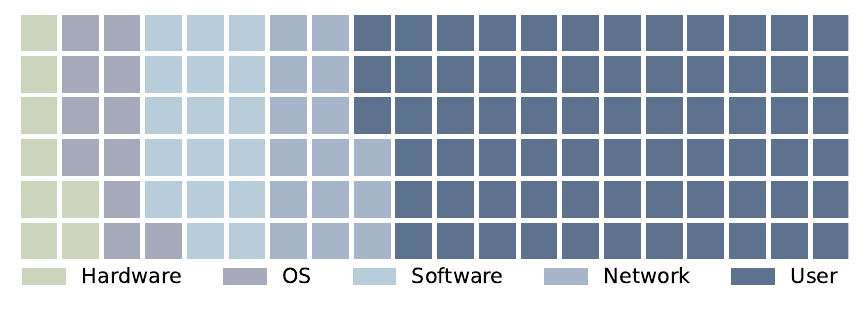}
 \vspace{3mm}
 \caption{Prevalence of the existing attacks} 
\label{fig:Prevalenceofattakck}
\end{figure}

 \noindent
As shown in \autoref{fig:taxonomy}, we have categorized the attacks into five groups based on their respective positions within the system infrastructure. These categories encompass hardware-level attacks, OS-level attacks, software-level attacks, network-level attacks, and user-level attacks. Additionally, we have quantified the number of associated research papers published for each group, as illustrated in \autoref{fig:Prevalenceofattakck}. 

\paragraph{Hardware-Level Attacks} Hardware attacks typically involve physical access to devices. However, LLMs cannot directly access physical devices. Instead, they can only access information associated with the hardware. Side-channel attack~\cite{spreitzer2017systematic,hettwer2020applications,mendez2021physical} is one attack that can be powered by the LLMs. 
Side-channel attacks typically entail the analysis of unintentional information leakage from a physical system or implementation, such as a cryptographic device or software, with the aim of inferring secret information (e.g., keys). \looseness=-1

Yaman~\cite{yaman2023agentsca} has explored the application of LLM techniques to develop side-channel analysis methods.
The research evaluates the effectiveness of LLM-based approaches in analyzing side-channel information in two hardware-related scenarios: AES side-channel analysis and deep-learning accelerator side-channel analysis. Experiments are conducted to determine the success rates of these methods in both situations.

\paragraph{OS-Level Attacks} LLMs operate at a high level of abstraction and primarily engage with text-based input and output. They lack the necessary low-level system access essential for executing OS-level attacks~\cite{igure2008taxonomies,vidas2011all,joshi2015review}. Nonetheless, they can be utilized for the analysis of information gathered from operating systems, thus potentially aiding in the execution of such attacks.  Andreas et al.~\cite{happe2023getting} establish a feedback loop connecting LLM to a vulnerable virtual machine through SSH, allowing LLM to analyze the machine's state, identify vulnerabilities, and propose concrete attack strategies, which are then executed automatically within the virtual machine. More recently, they~\cite{happe2023evaluating} introduced an automated Linux privilege-escalation benchmark using local virtual machines and an LLM-guided privilege-escalation tool to assess various LLMs and prompt strategies against the benchmark.  

 
\paragraph{Software-Level Attacks} Similar to how they employ LLM to target hardware and operating systems, there are also instances where LLM has been utilized to attack software (e.g., \cite{zhang2023well,paria2023divas,pearce2022pop,charan2023text}).  
However, the most prevalent software-level use case involves malicious developers utilizing LLMs to create malware.  Mika et al. ~\cite{beckerich2023ratgpt} present a proof-of-concept in which ChatGPT is utilized to distribute malicious software while avoiding detection. 
Yin et al.~\cite{pa2023attacker} investigate the potential misuse of LLM by creating a number of malware programs (e.g., ransomware, worm, keylogger, brute-force malware, Fileless malware). Antonio Monje et al.~\cite{monje2023being} demonstrate how to trick ChatGPT into quickly generating ransomware.  Marcus Botacin~\cite{botacin2023gpthreats} explores different coding strategies (e.g., generating entire malware, creating malware functions) and investigates the LLM's capacities to rewrite malware code. The findings reveal that LLM excels in constructing malware using building block descriptions. Meanwhile, LLM can generate multiple versions of the same semantic content (malware variants), with varying detection rates by Virustotal AV (ranging from 4\% to 55\%). 

\paragraph{Network-Level Attacks} LLMs can also be employed for initiating network attacks. A prevalent example of a network-level attack utilizing LLM is phishing attacks~\cite{ben2023opwnai,Chowdhury2023chat}.  Fredrik et al. ~\cite{heiding2023devising} compared AI-generated phishing emails using GPT-4 with manually designed phishing emails created using the V-Triad, alongside a control group exposed to generic phishing emails. The results showed that personalized phishing emails, whether generated by AI or designed manually, had higher click-through rates compared to generic ones. Tyson et al.~\cite{langford2023phishing} investigated how modifying ChatGPT's input can affect the content of the generated emails, making them more convincing. Julian Hazell~\cite{hazell2023large} demonstrated the scalability of spear phishing campaigns by generating realistic and cost-effective phishing messages for over 600 British Members of Parliament using ChatGPT. In another study, Wang et al.~\cite{wang2023bot} discuss how the traditional defenses may fail in the era of LLMs.  CAPTCHA challenges, involving distorted letters and digits, struggle to detect chatbots relying on text and voice.  However, LLMs may break the challenges, as they can produce high-quality human-like text and mimic human behavior effectively. There is one study that utilizes LLM for deploying fingerprint attacks. Armin et al.~\cite{sarabi2023llm} employed density-based clustering to cluster HTTP banners and create text-based fingerprints for annotating scanning data. When these fingerprints are compared to an existing database, it becomes possible to identify new IoT devices and server products. \looseness=-1

\paragraph{User-Level Attacks} 
Recent discussions have primarily focused on user-level attacks, as LLM demonstrates its capability to create remarkably convincing but ultimately deceptive content, as well as establish connections between seemingly unrelated pieces of information. This presents opportunities for malicious actors to engage in a range of nefarious activities. Here are a few examples: 

\begin{itemize}
    \item \textbf{Misinformation.}   Overreliance on content generated by LLMs without oversight is raising serious concerns regarding the safety of online content~\cite{owaspllm2023}. Numerous studies have focused on detecting misinformation produced by LLMs. Several study~\cite{chen2023llmgenerated,wu2023fake,yang2023poisoning} reveal content generated by LLMs are harder to detect and may use more deceptive styles, potentially causing greater harm. Canyu Chen et al.~\cite{chen2023llmgenerated} propose a taxonomy for LLM-generated misinformation and validate methods. 
Countermeasures and detection methods~\cite{wu2023fake,Uchendu_Lee_Shen_Le_Huang_Lee_2023,chen2023large,sun2023med,chen2023combating,zhang2023towards,bhojani2023truth,leite2023detecting,su2023fake} have also been developed to address these emerging issues.

\item \textbf{Social Engineering.}
LLMs not only have the potential to generate content from training data, but they also offer attackers a new perspective for social engineering. Work from Stabb et al.~\cite{staab2023memorization} highlights the capability of well-trained LLMs to infer personal attributes from text, such as location, income, and gender. They also reveals how these models can extract personal information from seemingly benign queries. Tong et al.~\cite{tong2023privinfer} investigated the content generated by LLMs may include user information. Moreover, Polra Victor Falade~\cite{Falade_2023} stated the exploitation by LLM-driven social engineers involves tactics such as psychological manipulation, targeted phishing, and the crisis of authenticity.
\item \textbf{Scientific Misconduct.}
Irresponsible use of LLMs can result in issues related to scientific misconduct, stemming from their capacity to generate original, coherent text. The academic community~\cite{cotton2023chatting,sullivan2023chatgpt,perkins2023academic,currie2023academic,lo2023impact,eke2023chatgpt,nikolic2023chatgpt,quidwai2023beyond,gao2022comparing,khalil2023will,rahman2023chatgpt}, encompassing diverse disciplines from various countries, has raised concerns about the increasing difficulties in detecting scientific misconduct in the era of LLMs. Concerns arise from LLMs' ability to generate coherent and original content, including complete papers from unreliable sources~\cite{uzun2023chatgpt,ventayen2023openai,rosyanafi2023dark}. Researchers are also actively engaged in the effort to detect such misconduct. For example, Kavita Kumari et al.~\cite{kumari2023demasq,kumari2023demasqndss} proposed \textsf{DEMASQ}, a precise ChatGPT-generated content detector. \textsf{DEMASQ} considers biases in text composition and evasion techniques, achieving high accuracy across diverse domains in identifying ChatGPT-generated content. 

\item \textbf{Fraud.} Cybercriminals have devised a new tool called FraudGPT~\cite{Falade_2023,fraudgpt2023}, which operates like ChatGPT but facilitates cyberattacks. It lacks the safety controls of ChatGPT and is sold on the dark web and Telegram for \$200 per month or \$1,700 annually. FraudGPT can create fraud emails related to banks, suggesting malicious links' placement in the content. It can also list frequently targeted sites or services, aiding hackers in planning future attacks. WormGPT~\cite{wormgpt2023}, a cybercrime tool, offers features such as unlimited character support and chat memory retention. The tool was trained on confidential datasets, with a focus on malware-related and fraud-related data. It can guide cybercriminals in executing Business Email Compromise (BEC) attacks. \looseness=-1

\end{itemize}

\begin{mybox}[boxsep=0pt,
	boxrule=1pt,
	left=4pt,
	right=4pt,
 	top=4pt,
 	bottom=4pt,
	]

 \textbf{Finding IV.}  As illustrated in \autoref{fig:Prevalenceofattakck}, when compared to other attacks, it becomes apparent that user-level attacks are the most prevalent, boasting a significant count of 33 papers. This dominance can be attributed to the fact that LLMs have increasingly human-like reasoning abilities, enabling them to generate human-like conversations and content (e.g., scientific misconduct, social engineering). Presently, LLMs do not possess the same level of access to OS-level or hardware-level functionalities.
This observation remains consistent with the attack observed in other levels as well. For instance, at the network level, LLMs can be abused to create phishing websites and bypass CAPTCHA mechanisms. 
 \end{mybox}


%% file: sections/sec6-theugly.tex
\section{Vulnerabilities and Defenses in LLMs}
\label{sec:ugly}

\begin{figure}
 \includegraphics[width= 0.5\textwidth]{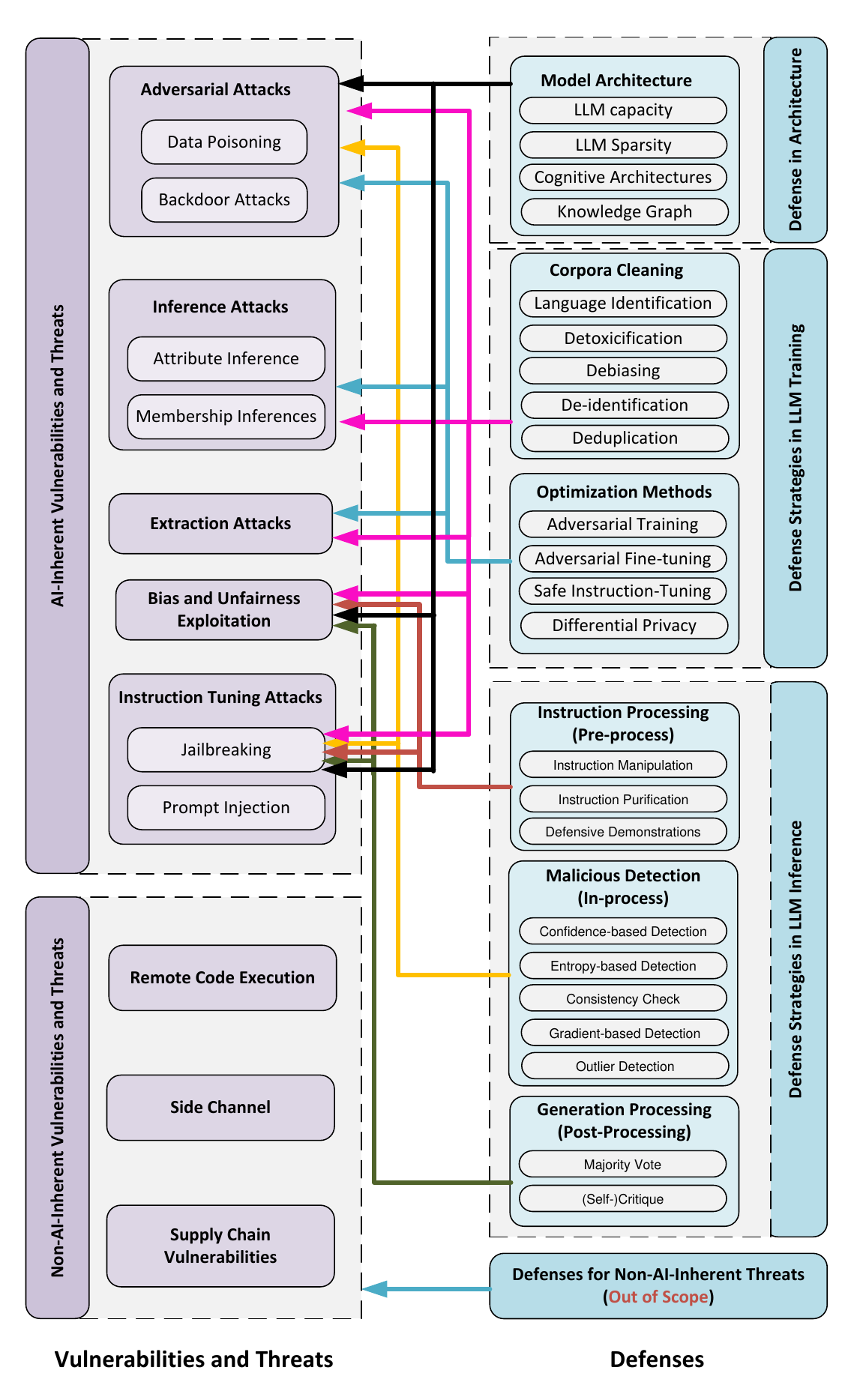}
 \vspace{3mm}
 \caption{Taxonomy of Threats and the Defenses. The line represents a defense technique that can defend against either a specific attack or a group of attacks.} 
\label{fig:taxonomyofthreats}
\end{figure}

\smallskip
\noindent
In the following section, we embark on an in-depth exploration of the prevalent threats and vulnerabilities associated with LLMs (\S\ref{subsec:threats}). We will examine the specific risks and challenges that arise in the context of LLMs. In addition to discussing these challenges, we will also delve into the countermeasures and strategies that researchers and practitioners have developed to mitigate these risks (\S\ref{subsec:defense}). \autoref{fig:taxonomyofthreats} illustrates the relationship between the attacks and defenses.

\subsection{Vulnerabilities and Threats in LLMs}
\label{subsec:threats}

\smallskip
\noindent
In this section, we aim to delve into the potential vulnerabilities and attacks that may be directed towards LLMs. Our examination seeks to categorize these threats into two distinct groups: AI Model Inherent Vulnerabilities and Non-AI Model Inherent Vulnerabilities.

\subsubsection{AI Inherent Vulnerabilities and Threats}

\smallskip
\noindent
These are vulnerabilities and threats that stem from the very nature and architecture of LLMs, considering that LLMs are fundamentally AI models themselves. For example,  attackers may manipulate the input data to generate incorrect or undesirable outputs from the LLM.

\paragraph{(A1) Adversarial Attacks} Adversarial attacks in machine learning refer to a set of techniques and strategies used to intentionally manipulate or deceive machine learning models. These attacks are typically carried out with malicious intent and aim to exploit vulnerabilities in the model's behavior. We only focus on the most extensively discussed attacks, namely, data poisoning and backdoor attacks.

\begin{itemize}
    \item \textbf{Data Poisoning.}   
    Data poisoning stands for attackers influencing the training process by injecting malicious data into the training dataset. This can introduce vulnerabilities or biases, compromising the security, effectiveness, or ethical behavior of the resulting models~\cite{owaspllm2023}. Various study~\cite{kurita2020weight,wan2023poisoning,wallace2020concealed,aghakhani2023trojanpuzzle,wan2022you,schuster2021you} have demonstrated that pre-trained models are vulnerable to compromise via methods such as using untrusted weights or content, including the insertion of poisoned examples into their datasets. By their inherent nature as pre-trained models, LLMs are susceptible to data poisoning attacks~\cite{rando2023universal,shu2023exploitability,shan2023prompt}. For example, Alexander et al.~\cite{wan2023poisoning} showed that even with just 100 poison examples, LLMs can produce consistently negative results or flawed outputs across various tasks. Larger language models are more susceptible to poisoning, and existing defenses like data filtering or model capacity reduction offer only moderate protection while hurting test accuracy.
    

    \item \textbf{Backdoor Attacks.}
    Backdoor attacks involve the malicious manipulation of training data and model processing, creating a vulnerability where attackers can embed a hidden backdoor into the model~\cite{yang2023comprehensive}. Both backdoor attacks and data poisoning attacks involve manipulating machine learning models, which can include manipulation of inputs. However, the key distinction is that backdoor attacks specifically focus on introducing hidden triggers into the model to manipulate specific behaviors or responses when the trigger is encountered.  LLMs are subject to backdoor attacks~\cite{li2023chatgpt,you2023large,li2023multitarget}. For example,  Yao et al.~\cite{yao2023poisonprompt}  a bidirectional backdoor, which combines trigger mechanisms with prompt tuning.

\end{itemize}

\paragraph{(A2) Inference Attacks} Inference attacks in the context of machine learning refer to a class of attacks where an adversary tries to gain sensitive information or insights about a machine learning model or its training data by making specific queries or observations to the model. These attacks often exploit unintended information leakage from the responses. \looseness=-1

\begin{itemize}

   \item \textbf{Attribute Inference Attacks.}  Attribute inference Attack~\cite{pan2020privacy,lyu-etal-2020-differentially,kandpal2023user,song2020information,Mahloujifar2021MembershipIO,li-etal-2022-dont} is a type of threat where an attacker attempts to deduce sensitive or personal information of individuals or entities by analyzing the behavior or responses of a machine learning models. It works against the LLMs as well. Robin et al.~\cite{staab2023memorization} presented the first comprehensive examination of pretrained LLMs' ability to infer personal information from text. Using a dataset of real Reddit profiles, the study demonstrated that current LLMs can accurately infer a variety of personal information (e.g., location, income, sex) with high accuracy.

    \item \textbf{Membership Inferences.}
    Membership inference Attack is a specific type of inference attack in the field of data security and privacy that determining whether a data record was part of a model's training dataset, given white-/black-box access to the model and the specific data record~\cite{shokri2017membership, duan2023are, kong2023efficient, fu2023probabilistic, fu2023practical, mireshghallah2022quantifying,huang2021damia}. A number of research studies have explored the concept of membership inference, each adopting a unique perspective and methodology. These studies have explored various membership inference attacks by analyzing the label~\cite{choquette2021label}, determining the threshold~\cite{jayaraman2020revisiting,carlini2022membership,hayes2017logan}, developing a generalized formulation~\cite{truex2018towards}, among other methods. Mireshghallah et al.~\cite{mireshghallah-etal-2022-empirical} found that fine-tuning the head of the model exhibits greater susceptibility to attacks when compared to fine-tuning smaller adapters.
\end{itemize}

\paragraph{(A3) Extraction Attacks}   Extraction attacks typically refer to attempts by adversaries to extract sensitive information or insights from machine learning models or their associated data. Extraction attacks and inference attacks share similarities but differ in their specific focus and objectives. Extraction attacks aim to acquire specific resources (e.g., model gradient, training data) or confidential information directly.
Inference attacks seek to gain knowledge or insights about the model or data's characteristics, often by observing the model's responses or behavior. Various types of data extraction attacks exist, including model theft attacks~\cite{juuti2019prada,kariyappa2021maze}, gradient leakage~\cite{li2023theoretical}, and training data extraction attacks~\cite{carlini2021extracting}. As of the current writing, it has been observed that training data extraction attacks may be effective against LLMs.
Training data extraction~\cite{carlini2021extracting} refers to a method where an attacker attempts to retrieve specific individual examples from a model's training data by strategically querying the machine learning models.   
    Numerous research~\cite{zhang2023ethicist,parikh2022canary,yang2023code} studies have shown that it is possible to extract training data from LLMs, which may include personal and private information~\cite{huang2022large,zhang2022text}. Notably, the work by Truong et al.~\cite{truong2021data} stands out for its ability to replicate the model without accessing the original model data. \looseness=-1
    

\paragraph{(A4) Bias and Unfairness Exploitation} Bias and unfairness in LLMs pertain to the phenomenon where these models demonstrate prejudiced outcomes or discriminatory behaviors. While bias and fairness issues are not unique to LLMs, they have received more attention due to the ethical and societal concerns. That is,  the societal impact of LLMs has prompted discussions about the ethical responsibilities of organizations and researchers developing and deploying these models. This has led to increased scrutiny and research on bias and fairness.
Concerns of bias were raised from various fields, encompassing gender and minority groups~\cite{dong2023probing,kotek2023gender,felkner2023winoqueer,shaikh2022second}, the identification of misinformation, political aspects. Multiple studies~\cite{talat2022you,urchs2023prevalent} revealed biases in the language used while querying LLMs. Moreover, Urman et al.~\cite{urman2023silence} discovered that biases may arise from adherence to government censorship guidelines. Bias in professional writing~\cite{wan2023kelly,su2023fake,fang2023bias} involving LLMs is also a concern within the community, as it can significantly damage credibility. The biases of LLMs may also lead to negative side effects in areas beyond text-based applications. Dai et al.~\cite{dai2023llms} noted that content generated by LLMs might introduce biases in neural retrieval systems, and Huang et al.~\cite{huang2023bias} discovered that biases could also be present in LLM generated code.

\paragraph{(A5) Instruction Tuning Attacks}
Instruction tuning, also known as instruction-based fine-tuning, is a machine-learning technique used to train and adapt language models for specific tasks by providing explicit instructions or examples during the fine-tuning process. In LLMs, instruction-tuning attacks refer to a class of attacks or manipulations that target instruction-tuned LLMs. These attacks are aimed at exploiting vulnerabilities or limitations in LLMs that have been fine-tuned with specific instructions or examples for particular tasks. 
\begin{itemize}
    \item \textbf{Jailbreaking.}
    Jailbreaking in LLMs involves bypassing security features to enable responses to otherwise restricted or unsafe questions, unlocking capabilities usually limited by safety protocols. Numerous studies have demonstrated various methods for successfully jailbreaking LLMs~\cite{li2023multi,taveekitworachai2023breaking,shen2023anything}. Wei et al.~\cite{wei2023jailbreak} emphasized that the alignment capabilities of LLMs can be influenced or manipulated through in-context demonstrations. In addition to this, several researches~\cite{wei2023jailbroken,kandpal2023backdoor} also demonstrated similar manipulation using various approaches, highlighting the versatility of methods that can jailbreaking LLMs. More recently, \textsf{MASTERKEY}~\cite{dengmasterkey} employed a time-based method for dissecting defenses, and demonstrated proof-of-concept attacks. It automatically generates jailbreak prompts with a 21.58
    Moreover, diverse methods have been employed in jailbreaking LLMs, such as conducting fuzzing~\cite{yao2023fuzzllm}, implementing optimized search strategies~\cite{zouuniversal}, and even training LLMs specifically to jailbreak other LLMs~\cite{deng2023jailbreaker,zouuniversal}. Meanwhile, Cao et al.~\cite{cao2023defending} developed RA-LLM, a method to lowers the success rate of adversarial and jailbreaking prompts without needing of retraining or access to model parameters.
     \item \textbf{Prompt Injection.}  
    Prompt injection attack describes a method of manipulating the behavior of LLMs to elicit unexpected and potentially harmful responses. This technique involves crafting input prompts in a way that bypasses the model's safeguards or triggers undesirable outputs. A substantial amount of research~\cite{liu2023autodan,yu2023gptfuzzer,kang2023exploiting,wang2023self,liu2023chinese,jiang2023prompt} has already automated the process of identifying semantic preserving payload in prompt injections with various focus. Facilitated by the capability for fine-tuning, backdoors may be introduced through prompt attacks~\cite{anonymous2023on,kandpal2023user,zhao2023prompt,shah2023loft}. Moreover, Greshake et al.~\cite{greshake2023more} expressed concerns about the potential for new vulnerabilities arising from LLMs invoking external resources. Other studies have also demonstrated the ability to take advantage of prompt injection attacks, such as unveiling guide prompts~\cite{zhang2023prompts}, virtualizing prompt injection~\cite{yan2023virtual}, and integrating applications~\cite{liu2023prompt}. He et al.~\cite{he2023you,he2024youSP} explored a shift towards leveraging LLMs, trained on extensive datasets, for mitigating such attacks. 
    \item \textbf{Denial of Service.}  
    A Denial of Service (DoS) attack is a type of cyber attack that aims to exhaust computational resources, causing latency or rendering resources unavailable. Due to the nature of LLMs require significant amount of resources, attackers use deliberately construct prompts to reduce the availability of models~\cite{derner2023security}. Shumailov et al.~\cite{shumailov2021sponge} proved the possibility of conducting sponge attacks in the field of LLMs, specifically designed to maximize energy consumption and latency (by a factor of 10 to 200). This strategy aims to draw the community's attention to their potential impact on autonomous vehicles, as well as scenarios requiring making decisions in timely manner.
\end{itemize}

\begin{mybox}[boxsep=0pt,
	boxrule=1pt,
	left=4pt,
	right=4pt,
 	top=4pt,
 	bottom=4pt,
	]
  
 \textbf{Finding V.} 
 Currently, there is limited research on model extraction attacks~\cite{duan2023are}, parameter extraction attacks, or the extraction of other intermediate esults~\cite{truong2021data}. While there are a few mentions of these topics, they tend to remain primarily theoretical (e.g., \cite{liu2023adversarial}), with limited practical implementation or empirical exploration.  We believe that the sheer scale of parameters in LLMs complicates these traditional approaches, rendering them less effective or even infeasible. Additionally, the most powerful LLMs are privately owned, with their weights, parameters, and other details kept confidential, further shielding them from conventional attack strategies. Strict censorship of outputs generated by these LLMs challenges even black-box traditional ML attacks, as it limits the attackers' ability to exploit or analyze the model's responses. 
 \end{mybox}

\subsubsection{Non-AI Inherent Vulnerabilities and Threats}

\smallskip
\noindent
We also need to consider non-AI Inherent Attacks, which encompass external threats and new vulnerabilities (which have not been observed or investigated in traditional AI models) that LLMs might encounter. These attacks may not be intricately linked to the internal mechanisms of the AI model, yet they can present significant risks. Illustrative instances of non-AI Inherent Attacks involve system-level vulnerabilities (e.g.,  remote code execution).  

\paragraph{(A6) Remote Code Execution (RCE)} RCE attacks typically target vulnerabilities in software applications, web services, or servers to execute arbitrary code remotely.  While RCE attacks are not typically applicable directly to LLMs, if an LLM is integrated into a web service (e.g.,\url{https://chat.openai.com/})  and if there are RCE vulnerabilities in the underlying infrastructure or code of that service, it could potentially lead to the compromise of the LLM's environment. Tong et al.~\cite{liu2023demystifying} identified 13 vulnerabilities in six frameworks, including 12 RCE vulnerabilities and 1 arbitrary file read/write vulnerability. Additionally, 17 out of 51 tested apps were found to have vulnerabilities, with 16 being vulnerable to RCE and 1 to SQL injection.  These vulnerabilities allow attackers to execute arbitrary code on app servers through prompt injections.

 \paragraph{(A7) Side Channel} While LLMs themselves do not typically leak information through traditional side channels such as power consumption or electromagnetic radiation, they can be vulnerable to certain side-channel attacks in practical deployment scenarios. For example, Edoardo et al.~\cite{debenedetti2023privacy} introduce privacy side channel attacks, which are attacks that exploit system-level components (e.g.,  data filtering, output monitoring) to extract private information at a much higher rate than what standalone models can achieve. Four categories of side channels covering the entire ML lifecycle are proposed, enabling enhanced membership inference attacks and novel threats (e.g., extracting users' test queries).  For instance, the research demonstrates how deduplicating training data before applying differentially-private training creates a side channel that compromises privacy guarantees.
   
\paragraph{(A8) Supply Chain Vulnerabilities} Supply Chain Vulnerabilities refer to the risks in the lifecycle of LLM applications that may arise from using vulnerable components or services. These include third-party datasets, pre-trained models, and plugins, any of which can compromise the application's integrity~\cite{owaspllm2023}.
Most research in this field is focused on the security of plugins. An LLM plugin is an extension or add-on module that enhances the capabilities of an LLM. Third-party plug-ins have been developed to expand its functionality, enabling users to perform various tasks, including web searches, text analysis, and code execution. However, some of the concerns raised by security experts~\cite{owaspllm2023,ChatGPTplugin2023} include the possibility of plug-ins being used to steal chat histories, access personal information, or execute code on users' machines. These vulnerabilities are associated with the use of OAuth in plug-ins, a web standard for data sharing across online accounts. Umar et al.~\cite{iqbal2023llm} attempted to address this problem by designing a framework. The framework formulates an extensive taxonomy of attacks specific to LLM platforms, taking into account the capabilities of plugins, users, and the LLM platform itself. By considering the relationships between these stakeholders, the framework helps identify potential security, privacy, and safety risks.

\subsection{Defenses for LLMs}
\label{subsec:defense}

\smallskip
\noindent
In this section, we examine the range of existing defense methods against various attacks and vulnerabilities associated with LLMs\footnote{Please be aware that we will not delve into solutions for non-AI inherent vulnerabilities as they tend to be highly specific to individual cases.}. 

\subsubsection{Defense in Model Architecture}

\vspace{2mm}
\noindent
Model architectures determine how knowledge and concepts are stored, organized, and contextually interacted with, which is crucial in the safety of Large Language Models.
There have been a lot of works~\cite{li2021large, zhu2023promptbench, Li2023EvaluatingTI, yuan2023revisiting} delved into how model capacities affect the privacy preservation and robustness of LLMs. Li et al.~\cite{li2021large} revealed that language models with larger parameter sizes can be trained more effectively in the differential privacy manner using appropriate non-standard hyper-parameters, in comparison to smaller models. Zhu et al.~\cite{zhu2023promptbench} and Li et al.~\cite{Li2023EvaluatingTI} found that LLMs with larger capacities, such as those with more extensive parameter sizes, generally show increased robustness against adversarial attacks. This was also verified in the Out-of-distribution (OOD) robustness scenarios by Yuan et al.~\cite{yuan2023revisiting}. 
Beyond the architecture of LLMs themselves, studies have focused on improving LLM safety by combining them with external modules including knowledge graphs~\cite{chen2020review} and cognitive architectures (CAs)~\cite{laird2017standard, anderson2014atomic}.  Romero et al.~\cite{romero2023synergistic} proposed improving AI robustness by incorporating various cognitive architectures into LLMs. Zafar et al.~\cite{zafar2023building} aimed to build trust in AI by enhancing the reasoning abilities of LLMs through knowledge graphs.

\subsubsection{Defenses in LLM Training and Inference}

\paragraph{Defense Strategies in LLM Training} The core components of LLM training include model architectures, training data, and optimization methods. Regarding model architectures, we examine trustworthy designs that exhibit increased robustness against malicious use. For training corpora, our investigation focuses on methods aimed at mitigating undesired properties during the generation, collection, and cleaning of training data. In the context of optimization methods, we review existing works that developed safe and secure optimization frameworks.

\begin{itemize}
    \item \paragraph{Corpora Cleaning}
    LLMs are shaped by their training corpora, from which they learn behavior, concepts, and data distributions~\cite{weidinger2021ethical}. Therefore, the safety of LLMs is crucially influenced by the quality of the training corpora~\cite{ganesh2023impact,ousidhoum2021probing}. However, it has been widely acknowledged that raw corpora collected from the web are full of issues of fairness~\cite{bailey2022based}, toxicity~\cite{gehman2020realtoxicityprompts}, privacy~\cite{pan2020privacy}, truthfulness~\cite{lin2021truthfulqa}, etc. A lot of efforts have been made to clean raw corpora and create high-quality training corpora for LLMs~\cite{joulin2016fasttext,wenzek2019ccnet,laurenccon2022bigscience,workshop2022bloom,penedo2023refinedweb,touvron2023llama}. In general, these pipelines consist of the following steps: language identification~\cite{joulin2016fasttext,ambikairajah2011language}, detoxification~\cite{gehman2020realtoxicityprompts,dale2021text,logacheva2022paradetox,moskovskiy2022exploring}, debiasing~\cite{meade2021empirical,bordia2019identifying,barikeri2021redditbias}, de-identification (personally identifiable information (PII))~\cite{subramani2023detecting,uzuner2007evaluating}, and deduplication~\cite{lee2021deduplicating,kandpal2022deduplicating,hernandez2022scaling,leskovec2020mining}. Debiasing and detoxification aimed to remove undesirable content from training corpora. 
    
    \item \paragraph{Optimization Methods} Optimization objectives are crucial in directing how LLMs learn from training data, influencing which behaviors are encouraged or penalized. These objectives affect the prioritization of knowledge and concepts within corpora, ultimately impacting the overall safety and ethical alignment of LLMs. In this context, robust training methods like adversarial training~\cite{liu2020adversarial,wang2019improving, zhu2019freelb,yoo2021towards,li2021token} and robust fine-tuning~\cite{dong2021should,jiang2019smart} have shown resilience against perturbation-based text attacks. Drawing inspiration from traditional adversarial training in the image field~\cite{madry2017towards}, Ivgi et al.~\cite{ivgi2021achieving} and Yoo et al.~\cite{yoo2021towards} applied adversarial training to LLMs by generating perturbations concerning discrete tokens. Wang et al.~\cite{wang2019improving} extended this approach to the continuous embedding space, facilitating more practical convergence, as followed by subsequent research~\cite{liu2020adversarial,zhu2019freelb,li2021token}.
    Safety alignments~\cite{ouyang2022training}, an emerging learning paradigm, guide LLM behavior using well-aligned additional models or human annotations, proving effective for ethical alignment. Efforts to align LLMs with other LLMs~\cite{yuan2023rrhf} and LLMs themselves~\cite{sun2023principle}. In terms of human annotations, Zhou et al.~\cite{zhou2023lima} and Shi et al.~\cite{shi2023safer} emphasized the importance of high-quality training corpora with carefully curated instructions and outputs for enhancing instruction-following capabilities in LLMs. Bianchi et al.~\cite{bianchi2023safety} highlighted that the safety of LLMs can be substantially improved by incorporating a limited percentage (e.g., 3\%) of safe examples during fine-tuning.
    
\end{itemize}

\vspace{2mm}

\paragraph{Defense Strategies in LLM Inference} When LLMs are deployed as cloud services, they operate by receiving prompts or instructions from users and generating completed sentences in response. Given this interaction model, the implementation of test-time LLM defense becomes a necessary and critical aspect of ensuring safe and appropriate outputs. Generally, test-time defense encompasses a range of strategies, including the pre-processing of prompts and instructions to filter or modify inputs, the detection of abnormal events that might signal misuse or problematic queries, and the post-processing of generated responses to ensure they adhere to safety and ethical guidelines. Test-time LLM defenses are essential to maintain the integrity and trustworthiness of LLMs in real-time applications.

\begin{itemize}
    \item \paragraph{Instruction Processing (Pre-Processing)} Instruction pre-processing applies transformations over instructions sent by users, in order to destroy potential adversarial contexts or malicious intents. It plays a vital role as it blocks out most malicious usage and prevents LLMs from receiving suspicious instructions. In general, instruction pre-processing methods can be categorized as instruction manipulation~\cite{shao2021bddr,robey2023smoothllm,kirchenbauer2023reliability,jain2023baseline,xu2022situ}, purification~\cite{li2022text}, and defensive demonstrations~\cite{liu2023adversarial,mo2023test,wei2023jailbreak}. Jain et al.~\cite{jain2023baseline} and Kirchenbauer et al.~\cite{kirchenbauer2023reliability} evaluated multiple baseline preprocessing methods against jailbreaking attacks, including retokenization and paraphrase. Li et al.~\cite{li2022text} proposed to purify instructions by first masking the input tokens and then predicting the masked tokens with other LLMs. The predicted tokens will serve as the purified instructions.
    Wei et al.~\cite{wei2023jailbreak} and Mo et al.~\cite{mo2023test} demonstrated that inserting pre-defined defensive demonstrations into instructions effectively defends jailbreaking attacks of LLMs. 
    \vspace{2mm}
    \item \paragraph{Malicious Detection (In-Processing)} Malicious detection provides in-depth examinations of LLM intermediate results, such as neuron activation, regarding the given instructions, which are more sensitive, accurate, and specified for malicious usage. Sun et al.~\cite{sun2023defending} proposed to detect backdoored instructions with backward probabilities of generations. Xi et al.~\cite{xi2023defending} differentiated normal and poisoned instructions from the perspective of mask sensitivities. Shao et al.~\cite{shao2021bddr} identified suspicious words according to their textual relevance. Wang et al.~\cite{wang2023rmlm} detected adversarial examples according to the semantic consistency among multiple generations, which has been explored in the uncertainty quantification of LLMs by Duan et al.~\cite{duan2023shifting}. Apart from the intrinsic properties of LLMs, there have been works leveraging the linguistic statistic properties, such as detecting outlier words~\cite{qi2020onion}, 

    \vspace{2mm}
    \item \paragraph{Generation Processing (Post-Processing)} Generation post processing refers to examining the properties (e.g., harmfulness) of the generated answers and applying modifications if necessary, which is the final step before delivering responses to users.
    Chen et al.~\cite{chen2023jailbreaker} proposed to mitigate the toxicity of generations by comparing with multiple model candidates. Helbling et al.~\cite{helbling2023llm} incorporated individual LLMs to identify the harmfulness of the generated answers, which shared similar ideas as Xiong et al.~\cite{Xiong2023CanLE} and Kadavath et al.~\cite{kadavath2022language} where they revealed that LLMs can be prompted to answer the confidences regarding the generated responses.

\end{itemize}

\begin{mybox}[boxsep=0pt,
	boxrule=1pt,
	left=4pt,
	right=4pt,
 	top=4pt,
 	bottom=4pt,
	]
  
 \textbf{Finding VI.} 
  For defense in LLM training, there's a notable scarcity of research examining the impact of model architecture on LLM safety, which is likely due to the high computational costs associated with training or fine-tuning large language models. We observed that \textit{safe instruction tuning} is a relatively new development that warrants further investigation and attention.
 \end{mybox}

%% file: sections/sec7-discussion.tex
\section{Discussion}
\label{sec:discuss}

\subsection{LLM in Other Security Related Topics}

\paragraph{LLMs in Cybersecurity Education}
LLMs can be used in security practices and education~\cite{farah2022impersonating,li2023evaluating,tann2023using}. For example, in a software security course, students are tasked with identifying and resolving vulnerabilities in a web application using LLMs.  
Jingyue et al.~\cite{li2023evaluating} investigated how ChatGPT can be used by students for these exercises. Wesley Tann et al.~\cite{tann2023using} focused on the evaluation of LLMs in the context of cybersecurity Capture-The-Flag (CTF) exercises (participants find ``flags'' by exploiting system vulnerabilities).   
The study first assessed the question-answering performance of these LLMs on Cisco certifications with varying difficulty levels, then  examined  their abilities in solving CTF challenges. Jin et al.~\cite{jin2023binary} conducted a comprehensive study on LLMs' understanding of binary code semantics~\cite{jin2022symlm} across different architectures and optimization levels, providing key insights for future research in this area.

\paragraph{LLMs in Cybersecurity Laws, Policies and Compliance} LLMs can assist in drafting security policies, guidelines, and compliance documentation, ensuring that organizations meet regulatory requirements and industry standards.
However, it's important to recognize that the utilization of LLMs can potentially necessitate changes to current cybersecurity-related laws and policies. The introduction of LLMs may raise new legal and regulatory considerations, as these models can impact various aspects of cybersecurity, data protection, and privacy.  Ekenobi et al.~\cite{thankgod2023impact} examined the legal implications arising from the introduction of LLMs, with a particular focus on data protection and privacy concerns. It acknowledges that ChatGPT's privacy policy contains commendable provisions for safeguarding user data against potential threats.  The paper also advocated for emphasizing the relevance of the new law.




\subsection{Future Directions}

\smallskip

\noindent
We have gleaned valuable lessons that we believe can shape future directions. 

\begin{itemize}
    \item \textbf{Using LLMs for ML-Specific Tasks}. We noticed that LLMs can effectively replace traditional machine learning methods and in this context, if traditional machine learning methods can be employed in a specific security application (whether offensive or defensive in nature), it is highly probable that LLMs can also be applied to address that particular challenge. For instance, traditional machine learning methods have found utility in malware detection, and LLMs can similarly be harnessed for this purpose. Therefore, one promising avenue is to harness the potential of LLMs in security applications where machine learning serves as a foundational or widely adopted technique. 
As security researchers, we are capable of designing LLM-based approaches to tackle security issues. Subsequently, we can compare these approaches with state-of-the-art methods to push the boundaries. 

\item  \textbf{Replacing Human Efforts.} It is evident that LLMs have the potential to replace human efforts in both offensive and defensive security applications. For instance, tasks involving social engineering, traditionally reliant on human intervention, can now be effectively executed using LLM techniques. Therefore, one promising avenue for security researchers is to identify areas within traditional security tasks where human involvement has been pivotal and explore opportunities to substitute these human efforts with LLM capabilities.

\item \textbf{Modifying Traditional ML Attacks for LLMs.} we have observed that many security vulnerabilities in LLMs are extensions of vulnerabilities found in traditional machine-learning scenarios. That is, 
LLMs remain a specialized instance of deep neural networks, inheriting common vulnerabilities such as adversarial attacks and instruction tuning attacks.  With the right adjustments (e.g., the threat model), traditional ML attacks can still be effective against LLMs.   For instance, the jailbreaking attack is a specific form of instruction tuning attack aimed at producing restricted texts.

\item \textbf{Adapting Traditional ML Defenses for LLMs.}  The countermeasures traditionally employed for vulnerability mitigation can also be leveraged to address these security issues. For example, there are existing efforts that utilize traditional Privacy-Enhancing Technologies (e.g., zero-knowledge proofs, differential privacy, and federated learning~\cite{weng2023auditable,weng2019deepchain} ) to tackle privacy challenges posed by LLMs. Exploring additional PETs techniques, whether they are established methods or innovative approaches, to address these challenges represents another promising research direction.

\item 
\textbf{Solving Challenges in LLM-Specific Attacks.}
As previously discussed, there are several challenges associated with implementing model extraction or parameter extraction attacks (e.g., vast scale of LLM parameters, private ownership and confidentiality of powerful LLMs). These novel characteristics introduced by LLMs represent a significant shift in the landscape, potentially leading to new challenges and necessitating the evolution of traditional ML attack methodologies.

\end{itemize}

%% file: sections/sec8-releatedwork.tex
\section{Related Work}
\smallskip

\noindent
There have already been a number of LLM surveys released with a variety of focuses (e.g., LLM evolution and taxonomy~\cite{chang2023survey,zhao2023survey, wu2023survey,hadi2023survey,wu2023unveiling, bowman2023eight,zhao2023knnicl}, software engineering~\cite{fan2023large,hou2023large}, and medicine~\cite{thirunavukarasu2023large,clusmann2023future}).
In this paper, our primary emphasis is on the security and privacy aspects of LLMs. 
We now delve into an examination of the existing literature pertaining to this particular topic. Peter J. Caven~\cite{caven2023more} specifically explores how LLMs (particularly, ChatGPT) could potentially alter the current cybersecurity landscape by blending technical and social aspects. Their emphasis leans more towards the social aspects. Muna et al.~\cite{al2023chatgpt} and Marshall et al. \cite{marshall2023effects} discussed the impact of ChatGPT in cybersecurity, highlighting its practical applications (e.g., code security, malware detection). Dhoni et al.~\cite{dhoni2023synergizing} demonstrated how LLMs can assist security analysts in developing security solutions against cyber threats. However, their work does not extensively address the potential cybersecurity threats that LLM may introduce. A number of surveys (e.g.,~\cite{gupta2023chatgpt, derner2023security,shayegani2023survey,dash2023chatgpt,derner2023safeguards,renaud2023chatgpt,schwinn2023adversarial,sebastian2023chatgpt,alawida2023unveiling}) highlight the threats and attacks against LLMs.  
In comparison to our work, they do not dedicate as much text to the vulnerabilities that the LLM may possess.
Instead, their primary focus lies in the realm of security applications, as they delve into utilizing LLMs for launching cyberattacks.  
Attia Qammar et al.~\cite{qammar2023chatbots} 
and Maximilian et al.~\cite{mozes2023use}
discussed vulnerabilities exploited by cybercriminals, with a specific focus on the risks associated with LLMs. Their works emphasized the need for strategies and measures to mitigate these threats and vulnerabilities. Haoran Li et al.~\cite{li2023privacy} analyzed current privacy concerns on LLMs, categorizing them based on adversary capabilities, and explored existing defense strategies.  Glorin Sebastian~\cite{sebastian2023privacy} explored the application of established Privacy-Enhancing Technologies (e.g., differential privacy~\cite{dwork2006differential}, federated learning~\cite{zhang2021survey}, and data minimization~\cite{pfitzmann2010terminology}) for safeguarding the privacy of LLMs. Smith et al.~\cite{smith2023identifying} also discussed the privacy risks of LLMs. Our study comprehensively examined both the security and privacy aspects of LLMs.   In summary, our research conducted an extensive review of the literature on LLMs from a three-fold perspective: beneficial security applications (e.g., vulnerability detection, secure code generation), adverse implications (e.g., phishing attacks, social engineering), and vulnerabilities (e.g., jailbreaking attacks, prompt attacks), along with their corresponding defensive measures.

%% file: sections/sec9-conclusion.tex
\section{Conclusion}
\label{sec:conclusion}

\noindent
Our work represents a pioneering effort in systematically examining the multifaceted role of LLMs in security and privacy. On the positive side, LLMs have significantly contributed to enhancing code and data security, while their versatile nature also opens the door to malicious applications. We also delved into the inherent vulnerabilities within these models, and discussed defense mechanisms. 
We have illuminated the path forward for harnessing the positive aspects of LLMs while mitigating their potential risks. As LLMs continue to evolve and find their place in an ever-expanding array of applications, it is imperative that we remain vigilant in addressing security and privacy concerns, ensuring that these powerful models contribute positively to the digital landscape.

\section*{Acknowledgement}

\noindent We thank the anonymous reviewers and Xin Jin from The Ohio State University for their invaluable feedback. This research was supported partly by the NSF award FMitF-2319242. Any opinions, findings, conclusions, or recommendations expressed are those of the authors and not necessarily of the NSF.

%% file: SOK-llmsecurity.bbl
\begin{thebibliography}{100}
\providecommand{\url}[1]{#1}
\csname url@samestyle\endcsname
\providecommand{\newblock}{\relax}
\providecommand{\bibinfo}[2]{#2}
\providecommand{\BIBentrySTDinterwordspacing}{\spaceskip=0pt\relax}
\providecommand{\BIBentryALTinterwordstretchfactor}{4}
\providecommand{\BIBentryALTinterwordspacing}{\spaceskip=\fontdimen2\font plus
\BIBentryALTinterwordstretchfactor\fontdimen3\font minus \fontdimen4\font\relax}
\providecommand{\BIBforeignlanguage}[2]{{%
\expandafter\ifx\csname l@#1\endcsname\relax
\typeout{** WARNING: IEEEtranS.bst: No hyphenation pattern has been}%
\typeout{** loaded for the language `#1'. Using the pattern for}%
\typeout{** the default language instead.}%
\else
\language=\csname l@#1\endcsname
\fi
#2}}
\providecommand{\BIBdecl}{\relax}
\BIBdecl

\bibitem{abbasian2023conversational}
M.~Abbasian, I.~Azimi, A.~M. Rahmani, and R.~Jain, ``Conversational health agents: A personalized llm-powered agent framework,'' 2023.

\bibitem{aghakhani2023trojanpuzzle}
H.~Aghakhani, W.~Dai, A.~Manoel, X.~Fernandes, A.~Kharkar, C.~Kruegel, G.~Vigna, D.~Evans, B.~Zorn, and R.~Sim, ``Trojanpuzzle: Covertly poisoning code-suggestion models,'' \emph{arXiv preprint arXiv:2301.02344}, 2023.

\bibitem{ahmad2023fixing}
\BIBentryALTinterwordspacing
B.~Ahmad, S.~Thakur, B.~Tan, R.~Karri, and H.~Pearce, ``Fixing hardware security bugs with large language models,'' \emph{arXiv preprint arXiv:2302.01215}, 2023. [Online]. Available: \url{https://doi.org/10.48550/arXiv.2302.01215}
\BIBentrySTDinterwordspacing

\bibitem{meta_ai2023llama}
M.~AI, ``Introducing llama: A foundational, 65-billion-parameter language model,'' \url{https://ai.meta.com/blog/large-language-model-llama-meta-ai/}, feb 2023, accessed: 2023-11-13.

\bibitem{al2023chatgpt}
M.~Al-Hawawreh, A.~Aljuhani, and Y.~Jararweh, ``Chatgpt for cybersecurity: practical applications, challenges, and future directions,'' \emph{Cluster Computing}, vol.~26, no.~6, pp. 3421--3436, 2023.

\bibitem{alagarsamy2023a3test}
S.~Alagarsamy, C.~Tantithamthavorn, and A.~Aleti, ``A3test: Assertion-augmented automated test case generation,'' \emph{arXiv preprint arXiv:2302.10352}, 2023.

\bibitem{alawida2023unveiling}
M.~Alawida, B.~A. Shawar, O.~I. Abiodun, A.~Mehmood, A.~E. Omolara \emph{et~al.}, ``Unveiling the dark side of chatgpt: Exploring cyberattacks and enhancing user awareness,'' 2023.

\bibitem{ali2023huntgpt}
T.~Ali and P.~Kostakos, ``Huntgpt: Integrating machine learning-based anomaly detection and explainable ai with large language models (llms),'' \emph{arXiv preprint arXiv:2309.16021}, 2023.

\bibitem{ambikairajah2011language}
E.~Ambikairajah, H.~Li, L.~Wang, B.~Yin, and V.~Sethu, ``Language identification: A tutorial,'' \emph{IEEE Circuits and Systems Magazine}, vol.~11, no.~2, pp. 82--108, 2011.

\bibitem{fraudgpt2023}
Z.~Amos, ``What is fraudgpt?'' \url{https://hackernoon.com/what-is-fraudgpt}, 2023.

\bibitem{anderson2014atomic}
J.~R. Anderson and C.~J. Lebiere, \emph{The atomic components of thought}.\hskip 1em plus 0.5em minus 0.4em\relax Psychology Press, 2014.

\bibitem{anonymous2023on}
\BIBentryALTinterwordspacing
Anonymous, ``On the safety of open-sourced large language models: Does alignment really prevent them from being misused?'' in \emph{Submitted to The Twelfth International Conference on Learning Representations}, 2023, under review. [Online]. Available: \url{https://openreview.net/forum?id=E6Ix4ahpzd}
\BIBentrySTDinterwordspacing

\bibitem{arcila2023platform}
B.~B. Arcila, ``Is it a platform? is it a search engine? it's chatgpt! the european liability regime for large language models,'' \emph{J. Free Speech L.}, vol.~3, p. 455, 2023.

\bibitem{bailey2022based}
A.~H. Bailey, A.~Williams, and A.~Cimpian, ``Based on billions of words on the internet, people= men,'' \emph{Science Advances}, vol.~8, no.~13, p. eabm2463, 2022.

\bibitem{bakhshandeh2023using}
A.~Bakhshandeh, A.~Keramatfar, A.~Norouzi, and M.~M. Chekidehkhoun, ``Using chatgpt as a static application security testing tool,'' \emph{arXiv preprint arXiv:2308.14434}, 2023.

\bibitem{barikeri2021redditbias}
S.~Barikeri, A.~Lauscher, I.~Vuli{\'c}, and G.~Glava{\v{s}}, ``Redditbias: A real-world resource for bias evaluation and debiasing of conversational language models,'' \emph{arXiv preprint arXiv:2106.03521}, 2021.

\bibitem{beckerich2023ratgpt}
M.~Beckerich, L.~Plein, and S.~Coronado, ``Ratgpt: Turning online llms into proxies for malware attacks,'' 2023.

\bibitem{ben2023opwnai}
S.~Ben-Moshe, G.~Gekker, and G.~Cohen, ``Opwnai: Ai that can save the day or hack it away. check point research (2022),'' 2023.

\bibitem{bhojani2023truth}
A.-R. Bhojani and M.~Schwarting, ``Truth and regret: Large language models, the quran, and misinformation,'' pp. 1--7, 2023.

\bibitem{bianchi2023safety}
F.~Bianchi, M.~Suzgun, G.~Attanasio, P.~R{\"o}ttger, D.~Jurafsky, T.~Hashimoto, and J.~Zou, ``Safety-tuned llamas: Lessons from improving the safety of large language models that follow instructions,'' \emph{arXiv preprint arXiv:2309.07875}, 2023.

\bibitem{bordia2019identifying}
S.~Bordia and S.~R. Bowman, ``Identifying and reducing gender bias in word-level language models,'' \emph{arXiv preprint arXiv:1904.03035}, 2019.

\bibitem{botacin2023gpthreats}
M.~Botacin, ``Gpthreats-3: Is automatic malware generation a threat?'' in \emph{2023 IEEE Security and Privacy Workshops (SPW)}.\hskip 1em plus 0.5em minus 0.4em\relax IEEE, 2023, pp. 238--254.

\bibitem{bowman2023eight}
S.~R. Bowman, ``Eight things to know about large language models,'' \emph{arXiv preprint arXiv:2304.00612}, 2023.

\bibitem{brown2020language}
T.~B. Brown, B.~Mann, N.~Ryder, M.~Subbiah, J.~Kaplan, P.~Dhariwal, A.~Neelakantan, P.~Shyam, G.~Sastry, A.~Askell, S.~Agarwal, A.~Herbert-Voss, G.~Krueger, T.~Henighan, R.~Child, A.~Ramesh, D.~M. Ziegler, J.~Wu, C.~Winter, C.~Hesse, M.~Chen, E.~Sigler, M.~Litwin, S.~Gray, B.~Chess, J.~Clark, C.~Berner, S.~McCandlish, A.~Radford, I.~Sutskever, and D.~Amodei, ``Language models are few-shot learners,'' 2020.

\bibitem{ChatGPTplugin2023}
M.~Burgess, ``Chatgpt has a plug-in problem,'' \url{https://www.wired.com/story/chatgpt-plugins-security-privacy-risk/}, 2023.

\bibitem{cai2023low}
Y.~Cai, S.~Mao, W.~Wu, Z.~Wang, Y.~Liang, T.~Ge, C.~Wu, W.~You, T.~Song, Y.~Xia \emph{et~al.}, ``Low-code llm: Visual programming over llms,'' \emph{arXiv preprint arXiv:2304.08103}, 2023.

\bibitem{cao2023defending}
B.~Cao, Y.~Cao, L.~Lin, and J.~Chen, ``Defending against alignment-breaking attacks via robustly aligned llm,'' 2023.

\bibitem{carlini2022membership}
N.~Carlini, S.~Chien, M.~Nasr, S.~Song, A.~Terzis, and F.~Tramer, ``Membership inference attacks from first principles,'' in \emph{2022 IEEE Symposium on Security and Privacy (SP)}.\hskip 1em plus 0.5em minus 0.4em\relax IEEE, 2022, pp. 1897--1914.

\bibitem{carlini2021extracting}
N.~Carlini, F.~Tramer, E.~Wallace, M.~Jagielski, A.~Herbert-Voss, K.~Lee, A.~Roberts, T.~Brown, D.~Song, U.~Erlingsson \emph{et~al.}, ``Extracting training data from large language models,'' in \emph{30th USENIX Security Symposium (USENIX Security 21)}, 2021, pp. 2633--2650.

\bibitem{caven2023more}
P.~Caven, ``A more insecure ecosystem? chatgpt’s influence on cybersecurity,'' \emph{ChatGPT’s Influence on Cybersecurity (April 30, 2023)}, 2023.

\bibitem{chang2023survey}
Y.~Chang, X.~Wang, J.~Wang, Y.~Wu, K.~Zhu, H.~Chen, L.~Yang, X.~Yi, C.~Wang, Y.~Wang \emph{et~al.}, ``A survey on evaluation of large language models,'' \emph{arXiv preprint arXiv:2307.03109}, 2023.

\bibitem{charan2023text}
P.~V.~S. Charan, H.~Chunduri, P.~M. Anand, and S.~K. Shukla, ``From text to mitre techniques: Exploring the malicious use of large language models for generating cyber attack payloads,'' 2023.

\bibitem{chen2022codet}
B.~Chen, F.~Zhang, A.~Nguyen, D.~Zan, Z.~Lin, J.-G. Lou, and W.~Chen, ``Codet: Code generation with generated tests,'' \emph{arXiv preprint arXiv:2207.10397}, 2022.

\bibitem{chen2023jailbreaker}
B.~Chen, A.~Paliwal, and Q.~Yan, ``Jailbreaker in jail: Moving target defense for large language models,'' \emph{arXiv preprint arXiv:2310.02417}, 2023.

\bibitem{chen2023llmgenerated}
C.~Chen and K.~Shu, ``Can llm-generated misinformation be detected?'' 2023.

\bibitem{chen2023combating}
------, ``Combating misinformation in the age of llms: Opportunities and challenges,'' \emph{arXiv preprint arXiv:2311.05656}, 2023.

\bibitem{chen2023when}
\BIBentryALTinterwordspacing
C.~Chen, J.~Su, J.~Chen, Y.~Wang, T.~Bi, Y.~Wang, X.~Lin, T.~Chen, and Z.~Zheng, ``When chatgpt meets smart contract vulnerability detection: How far are we?'' \emph{arXiv preprint arXiv:2309.05520}, 2023. [Online]. Available: \url{https://doi.org/10.48550/arXiv.2309.05520}
\BIBentrySTDinterwordspacing

\bibitem{chen2023vullibgen}
\BIBentryALTinterwordspacing
T.~Chen, L.~Li, L.~Zhu, Z.~Li, G.~Liang, D.~Li, Q.~Wang, and T.~Xie, ``Vullibgen: Identifying vulnerable third-party libraries via generative pre-trained model,'' \emph{arXiv preprint arXiv:2308.04662}, 2023. [Online]. Available: \url{https://doi.org/10.48550/arXiv.2308.04662}
\BIBentrySTDinterwordspacing

\bibitem{chen2020review}
X.~Chen, S.~Jia, and Y.~Xiang, ``A review: Knowledge reasoning over knowledge graph,'' \emph{Expert Systems with Applications}, vol. 141, p. 112948, 2020.

\bibitem{chen2023large}
Y.~Chen, A.~Arunasalam, and Z.~B. Celik, ``Can large language models provide security \& privacy advice? measuring the ability of llms to refute misconceptions,'' 2023.

\bibitem{cheshkov2023evaluation}
\BIBentryALTinterwordspacing
A.~Cheshkov, P.~Zadorozhny, and R.~Levichev, ``Evaluation of chatgpt model for vulnerability detection,'' \emph{arXiv preprint arXiv:2304.07232}, 2023. [Online]. Available: \url{https://doi.org/10.48550/arXiv.2304.07232}
\BIBentrySTDinterwordspacing

\bibitem{choquette2021label}
C.~A. Choquette-Choo, F.~Tramer, N.~Carlini, and N.~Papernot, ``Label-only membership inference attacks,'' in \emph{International conference on machine learning}.\hskip 1em plus 0.5em minus 0.4em\relax PMLR, 2021, pp. 1964--1974.

\bibitem{Chowdhury2023chat}
M.~Chowdhury, N.~Rifat, S.~Latif, M.~Ahsan, M.~S. Rahman, and R.~Gomes, ``Chatgpt: The curious case of attack vectors' supply chain management improvement,'' in \emph{2023 IEEE International Conference on Electro Information Technology (eIT)}, 2023, pp. 499--504.

\bibitem{clusmann2023future}
J.~Clusmann, F.~R. Kolbinger, H.~S. Muti, Z.~I. Carrero, J.-N. Eckardt, N.~G. Laleh, C.~M.~L. L{\"o}ffler, S.-C. Schwarzkopf, M.~Unger, G.~P. Veldhuizen \emph{et~al.}, ``The future landscape of large language models in medicine,'' \emph{Communications Medicine}, vol.~3, no.~1, p. 141, 2023.

\bibitem{cotton2023chatting}
D.~R. Cotton, P.~A. Cotton, and J.~R. Shipway, ``Chatting and cheating: Ensuring academic integrity in the era of chatgpt,'' \emph{Innovations in Education and Teaching International}, pp. 1--12, 2023.

\bibitem{currie2023academic}
G.~M. Currie, ``Academic integrity and artificial intelligence: is chatgpt hype, hero or heresy?'' in \emph{Seminars in Nuclear Medicine}.\hskip 1em plus 0.5em minus 0.4em\relax Elsevier, 2023.

\bibitem{dai2023llms}
S.~Dai, Y.~Zhou, L.~Pang, W.~Liu, X.~Hu, Y.~Liu, X.~Zhang, and J.~Xu, ``Llms may dominate information access: Neural retrievers are biased towards llm-generated texts,'' \emph{arXiv preprint arXiv:2310.20501}, 2023.

\bibitem{dale2021text}
D.~Dale, A.~Voronov, D.~Dementieva, V.~Logacheva, O.~Kozlova, N.~Semenov, and A.~Panchenko, ``Text detoxification using large pre-trained neural models,'' \emph{arXiv preprint arXiv:2109.08914}, 2021.

\bibitem{dash2023chatgpt}
B.~Dash and P.~Sharma, ``Are chatgpt and deepfake algorithms endangering the cybersecurity industry? a review,'' \emph{International Journal of Engineering and Applied Sciences}, vol.~10, no.~1, 2023.

\bibitem{databricks2023free}
Databricks, ``Free dolly: Introducing the world's first open and commercially viable instruction-tuned llm,'' \url{https://www.databricks.com/blog/2023/04/12/dolly-first-open-commercially-viable-instruction-tuned-llm}, 2023, accessed: 2023-11-13.

\bibitem{debenedetti2023privacy}
E.~Debenedetti, G.~Severi, N.~Carlini, C.~A. Choquette-Choo, M.~Jagielski, M.~Nasr, E.~Wallace, and F.~Tram{\`e}r, ``Privacy side channels in machine learning systems,'' \emph{arXiv preprint arXiv:2309.05610}, 2023.

\bibitem{wormgpt2023}
D.~Delley, ``Wormgpt – the generative ai tool cybercriminals are using to launch business email compromise attacks,'' \url{https://shorturl.at/iwFL7}, 2023.

\bibitem{deng2023jailbreaker}
G.~Deng, Y.~Liu, Y.~Li, K.~Wang, Y.~Zhang, Z.~Li, H.~Wang, T.~Zhang, and Y.~Liu, ``Jailbreaker: Automated jailbreak across multiple large language model chatbots,'' \emph{arXiv preprint arXiv:2307.08715}, 2023.

\bibitem{dengmasterkey}
------, ``Masterkey: Automated jailbreaking of large language model chatbots,'' in \emph{Proceedings of the 31th Annual Network and Distributed System Security Symposium (NDSS’24)}, 2024.

\bibitem{deng2023pentestgpt}
G.~Deng, Y.~Liu, V.~Mayoral-Vilches, P.~Liu, Y.~Li, Y.~Xu, T.~Zhang, Y.~Liu, M.~Pinzger, and S.~Rass, ``Pentestgpt: An llm-empowered automatic penetration testing tool,'' \emph{arXiv preprint arXiv:2308.06782}, 2023.

\bibitem{deng2022fuzzing}
Y.~Deng, C.~S. Xia, H.~Peng, C.~Yang, and L.~Zhang, ``Fuzzing deep-learning libraries via large language models,'' \emph{arXiv preprint arXiv:2212.14834}, 2022.

\bibitem{deng2023largefuzzinggpt}
Y.~Deng, C.~S. Xia, C.~Yang, S.~D. Zhang, S.~Yang, and L.~Zhang, ``Large language models are edge-case fuzzers: Testing deep learning libraries via fuzzgpt,'' \emph{arXiv preprint arXiv:2304.02014}, 2023.

\bibitem{deng2024large}
------, ``Large language models are edge-case generators: Crafting unusual programs for fuzzing deep learning libraries,'' in \emph{2024 IEEE/ACM 46th International Conference on Software Engineering (ICSE)}, 2024, pp. 830--842.

\bibitem{derner2023security}
E.~Derner, K.~Batisti{\v{c}}, J.~Zah{\'a}lka, and R.~Babu{\v{s}}ka, ``A security risk taxonomy for large language models,'' \emph{arXiv preprint arXiv:2311.11415}, 2023.

\bibitem{derner2023safeguards}
E.~Derner and K.~Batistič, ``Beyond the safeguards: Exploring the security risks of chatgpt,'' 2023.

\bibitem{devlin2019bert}
J.~Devlin, M.-W. Chang, K.~Lee, and K.~Toutanova, ``Bert: Pre-training of deep bidirectional transformers for language understanding,'' 2019.

\bibitem{dhoni2023synergizing}
P.~Dhoni and R.~Kumar, ``Synergizing generative ai and cybersecurity: Roles of generative ai entities, companies, agencies, and government in enhancing cybersecurity,'' 2023.

\bibitem{ding2023static}
H.~Ding, V.~Kumar, Y.~Tian, Z.~Wang, R.~Kwiatkowski, X.~Li, M.~K. Ramanathan, B.~Ray, P.~Bhatia, S.~Sengupta \emph{et~al.}, ``A static evaluation of code completion by large language models,'' \emph{arXiv preprint arXiv:2306.03203}, 2023.

\bibitem{Ding_2023}
\BIBentryALTinterwordspacing
X.~Ding, L.~Chen, M.~Emani, C.~Liao, P.-H. Lin, T.~Vanderbruggen, Z.~Xie, A.~Cerpa, and W.~Du, ``Hpc-gpt: Integrating large language model for high-performance computing,'' in \emph{Proceedings of the SC ’23 Workshops of The International Conference on High Performance Computing, Network, Storage, and Analysis}, ser. SC-W 2023.\hskip 1em plus 0.5em minus 0.4em\relax ACM, Nov. 2023. [Online]. Available: \url{http://dx.doi.org/10.1145/3624062.3624172}
\BIBentrySTDinterwordspacing

\bibitem{dong2023probing}
X.~Dong, Y.~Wang, P.~S. Yu, and J.~Caverlee, ``Probing explicit and implicit gender bias through llm conditional text generation,'' \emph{arXiv preprint arXiv:2311.00306}, 2023.

\bibitem{dong2021should}
X.~Dong, A.~T. Luu, M.~Lin, S.~Yan, and H.~Zhang, ``How should pre-trained language models be fine-tuned towards adversarial robustness?'' \emph{Advances in Neural Information Processing Systems}, vol.~34, pp. 4356--4369, 2021.

\bibitem{duan2023shifting}
J.~Duan, H.~Cheng, S.~Wang, C.~Wang, A.~Zavalny, R.~Xu, B.~Kailkhura, and K.~Xu, ``Shifting attention to relevance: Towards the uncertainty estimation of large language models,'' \emph{arXiv preprint arXiv:2307.01379}, 2023.

\bibitem{duan2023are}
J.~Duan, F.~Kong, S.~Wang, X.~Shi, and K.~Xu, ``Are diffusion models vulnerable to membership inference attacks?'' in \emph{Proceedings of the 40th International Conference on Machine Learning}, 2023, pp. 8717--8730.

\bibitem{chatgptusers}
F.~Duarte, ``Number of chatgpt users (nov 2023),'' \url{https://explodingtopics.com/blog/chatgpt-users}, 2023, accessed: 2023-11-13.

\bibitem{dwork2006differential}
C.~Dwork, ``Differential privacy,'' in \emph{International colloquium on automata, languages, and programming}.\hskip 1em plus 0.5em minus 0.4em\relax Springer, 2006, pp. 1--12.

\bibitem{egersdoerfer2023early}
C.~Egersdoerfer, D.~Zhang, and D.~Dai, ``Early exploration of using chatgpt for log-based anomaly detection on parallel file systems logs,'' 2023.

\bibitem{eke2023chatgpt}
D.~O. Eke, ``Chatgpt and the rise of generative ai: threat to academic integrity?'' \emph{Journal of Responsible Technology}, vol.~13, p. 100060, 2023.

\bibitem{elhafsi2023semantic}
A.~Elhafsi, R.~Sinha, C.~Agia, E.~Schmerling, I.~A. Nesnas, and M.~Pavone, ``Semantic anomaly detection with large language models,'' \emph{Autonomous Robots}, pp. 1--21, 2023.

\bibitem{apiiro2023}
S.~Eli and D.~Gil, ``Self-enhancing pattern detection with llms: Our answer to uncovering malicious packages at scale,'' \url{https://apiiro.com/blog/llm-code-pattern-malicious-package-detection/}, 2023, accessed: 2023-11-13.

\bibitem{espinha2023m}
T.~Espinha~Gasiba, K.~Oguzhan, I.~Kessba, U.~Lechner, and M.~Pinto-Albuquerque, ``I'm sorry dave, i'm afraid i can't fix your code: On chatgpt, cybersecurity, and secure coding,'' in \emph{4th International Computer Programming Education Conference (ICPEC 2023)}.\hskip 1em plus 0.5em minus 0.4em\relax Schloss-Dagstuhl-Leibniz Zentrum f{\"u}r Informatik, 2023.

\bibitem{Falade_2023}
\BIBentryALTinterwordspacing
P.~V. Falade, ``Decoding the threat landscape: Chatgpt, fraudgpt, and wormgpt in social engineering attacks,'' \emph{International Journal of Scientific Research in Computer Science, Engineering and Information Technology}, p. 185–198, Oct. 2023. [Online]. Available: \url{http://dx.doi.org/10.32628/CSEIT2390533}
\BIBentrySTDinterwordspacing

\bibitem{fan2023large}
A.~Fan, B.~Gokkaya, M.~Harman, M.~Lyubarskiy, S.~Sengupta, S.~Yoo, and J.~M. Zhang, ``Large language models for software engineering: Survey and open problems,'' 2023.

\bibitem{fan2023fate}
T.~Fan, Y.~Kang, G.~Ma, W.~Chen, W.~Wei, L.~Fan, and Q.~Yang, ``Fate-llm: A industrial grade federated learning framework for large language models,'' \emph{arXiv preprint arXiv:2310.10049}, 2023.

\bibitem{fang2023bias}
X.~Fang, S.~Che, M.~Mao, H.~Zhang, M.~Zhao, and X.~Zhao, ``Bias of ai-generated content: An examination of news produced by large language models,'' \emph{arXiv preprint arXiv:2309.09825}, 2023.

\bibitem{farah2022impersonating}
J.~C. Farah, B.~Spaenlehauer, V.~Sharma, M.~J. Rodr{\'\i}guez-Triana, S.~Ingram, and D.~Gillet, ``Impersonating chatbots in a code review exercise to teach software engineering best practices,'' in \emph{2022 IEEE Global Engineering Education Conference (EDUCON)}.\hskip 1em plus 0.5em minus 0.4em\relax IEEE, 2022, pp. 1634--1642.

\bibitem{felkner2023winoqueer}
V.~K. Felkner, H.-C.~H. Chang, E.~Jang, and J.~May, ``Winoqueer: A community-in-the-loop benchmark for anti-lgbtq+ bias in large language models,'' \emph{arXiv preprint arXiv:2306.15087}, 2023.

\bibitem{feng2021survey}
S.~Y. Feng, V.~Gangal, J.~Wei, S.~Chandar, S.~Vosoughi, T.~Mitamura, and E.~Hovy, ``A survey of data augmentation approaches for nlp,'' \emph{arXiv preprint arXiv:2105.03075}, 2021.

\bibitem{fu2023chatgpt}
M.~Fu, C.~Tantithamthavorn, V.~Nguyen, and T.~Le, ``Chatgpt for vulnerability detection, classification, and repair: How far are we?'' 2023.

\bibitem{fu2023practical}
W.~Fu, H.~Wang, C.~Gao, G.~Liu, Y.~Li, and T.~Jiang, ``Practical membership inference attacks against fine-tuned large language models via self-prompt calibration,'' 2023.

\bibitem{fu2023probabilistic}
------, ``A probabilistic fluctuation based membership inference attack for diffusion models,'' 2023.

\bibitem{ganesh2023impact}
P.~Ganesh, H.~Chang, M.~Strobel, and R.~Shokri, ``On the impact of machine learning randomness on group fairness,'' in \emph{Proceedings of the 2023 ACM Conference on Fairness, Accountability, and Transparency}, 2023, pp. 1789--1800.

\bibitem{gao2022comparing}
C.~A. Gao, F.~M. Howard, N.~S. Markov, E.~C. Dyer, S.~Ramesh, Y.~Luo, and A.~T. Pearson, ``Comparing scientific abstracts generated by chatgpt to original abstracts using an artificial intelligence output detector, plagiarism detector, and blinded human reviewers,'' \emph{BioRxiv}, pp. 2022--12, 2022.

\bibitem{gehman2020realtoxicityprompts}
S.~Gehman, S.~Gururangan, M.~Sap, Y.~Choi, and N.~A. Smith, ``Realtoxicityprompts: Evaluating neural toxic degeneration in language models,'' \emph{arXiv preprint arXiv:2009.11462}, 2020.

\bibitem{greshake2023more}
K.~Greshake, S.~Abdelnabi, S.~Mishra, C.~Endres, T.~Holz, and M.~Fritz, ``More than you've asked for: A comprehensive analysis of novel prompt injection threats to application-integrated large language models,'' \emph{arXiv preprint arXiv:2302.12173}, 2023.

\bibitem{gu2023llm}
Q.~Gu, ``Llm-based code generation method for golang compiler testing,'' 2023.

\bibitem{gu2023anomalygpt}
Z.~Gu, B.~Zhu, G.~Zhu, Y.~Chen, M.~Tang, and J.~Wang, ``Anomalygpt: Detecting industrial anomalies using large vision-language models,'' \emph{arXiv preprint arXiv:2308.15366}, 2023.

\bibitem{gupta2023chatgpt}
M.~Gupta, C.~Akiri, K.~Aryal, E.~Parker, and L.~Praharaj, ``From chatgpt to threatgpt: Impact of generative ai in cybersecurity and privacy,'' \emph{IEEE Access}, 2023.

\bibitem{hadi2023survey}
M.~U. Hadi, R.~Qureshi, A.~Shah, M.~Irfan, A.~Zafar, M.~Shaikh, N.~Akhtar, J.~Wu, and S.~Mirjalili, ``A survey on large language models: Applications, challenges, limitations, and practical usage,'' \emph{TechRxiv}, 2023.

\bibitem{happe2023getting}
A.~Happe and J.~Cito, ``Getting pwn'd by ai: Penetration testing with large language models,'' \emph{arXiv preprint arXiv:2308.00121}, 2023.

\bibitem{happe2023evaluating}
A.~Happe, A.~Kaplan, and J.~Cito, ``Evaluating llms for privilege-escalation scenarios,'' 2023.

\bibitem{hayes2017logan}
J.~Hayes, L.~Melis, G.~Danezis, and E.~De~Cristofaro, ``Logan: Membership inference attacks against generative models,'' \emph{arXiv preprint arXiv:1705.07663}, 2017.

\bibitem{hazell2023large}
J.~Hazell, ``Large language models can be used to effectively scale spear phishing campaigns,'' 2023.

\bibitem{he2023large}
J.~He and M.~Vechev, ``Large language models for code: Security hardening and adversarial testing,'' \emph{ICML 2023 Workshop DeployableGenerativeAI}, 2023, keywords: large language models, code generation, security, prompt tuning.

\bibitem{he2023largeccs}
------, ``Large language models for code: Security hardening and adversarial testing,'' in \emph{Proceedings of the 2023 ACM SIGSAC Conference on Computer and Communications Security}, 2023, pp. 1865--1879.

\bibitem{he2023you}
X.~He, S.~Zannettou, Y.~Shen, and Y.~Zhang, ``You only prompt once: On the capabilities of prompt learning on large language models to tackle toxic content,'' \emph{arXiv preprint arXiv:2308.05596}, 2023.

\bibitem{he2024youSP}
------, ``You only prompt once: On the capabilities of prompt learning on large language models to tackle toxic content,'' in \emph{{2024 IEEE Symposium on Security and Privacy (SP)}}, 2024.

\bibitem{heiding2023devising}
F.~Heiding, B.~Schneier, A.~Vishwanath, and J.~Bernstein, ``Devising and detecting phishing: Large language models vs. smaller human models,'' 2023.

\bibitem{helbling2023llm}
A.~Helbling, M.~Phute, M.~Hull, and D.~H. Chau, ``Llm self defense: By self examination, llms know they are being tricked,'' \emph{arXiv preprint arXiv:2308.07308}, 2023.

\bibitem{helmke2023check}
R.~Helmke and J.~vom Dorp, ``Check for extended abstract: Towards reliable and scalable linux kernel cve attribution in automated static firmware analyses,'' in \emph{Detection of Intrusions and Malware, and Vulnerability Assessment: 20th International Conference, DIMVA 2023, Hamburg, Germany, July 12--14, 2023, Proceedings}, vol. 13959.\hskip 1em plus 0.5em minus 0.4em\relax Springer Nature, 2023, p. 201.

\bibitem{gpt3malware2023}
P.~Henrik, ``Llm-assisted malware review: Ai and humans join forces to combat malware,'' \url{https://shorturl.at/loqT4}, 2023, accessed: 2023-11-13.

\bibitem{hernandez2022scaling}
D.~Hernandez, T.~Brown, T.~Conerly, N.~DasSarma, D.~Drain, S.~El-Showk, N.~Elhage, Z.~Hatfield-Dodds, T.~Henighan, T.~Hume \emph{et~al.}, ``Scaling laws and interpretability of learning from repeated data,'' \emph{arXiv preprint arXiv:2205.10487}, 2022.

\bibitem{hettwer2020applications}
B.~Hettwer, S.~Gehrer, and T.~G{\"u}neysu, ``Applications of machine learning techniques in side-channel attacks: a survey,'' \emph{Journal of Cryptographic Engineering}, vol.~10, pp. 135--162, 2020.

\bibitem{hou2023large}
X.~Hou, Y.~Zhao, Y.~Liu, Z.~Yang, K.~Wang, L.~Li, X.~Luo, D.~Lo, J.~Grundy, and H.~Wang, ``Large language models for software engineering: A systematic literature review,'' \emph{arXiv preprint arXiv:2308.10620}, 2023.

\bibitem{hu2023augmenting}
J.~Hu, Q.~Zhang, and H.~Yin, ``Augmenting greybox fuzzing with generative ai,'' \emph{arXiv preprint arXiv:2306.06782}, 2023.

\bibitem{hu2023large}
\BIBentryALTinterwordspacing
S.~Hu, T.~Huang, F.~İlhan, S.~F. Tekin, and L.~Liu, ``Large language model-powered smart contract vulnerability detection: New perspectives,'' \emph{arXiv preprint arXiv:2310.01152}, 2023, 10 pages. [Online]. Available: \url{https://doi.org/10.48550/arXiv.2310.01152}
\BIBentrySTDinterwordspacing

\bibitem{huang2023bias}
D.~Huang, Q.~Bu, J.~Zhang, X.~Xie, J.~Chen, and H.~Cui, ``Bias assessment and mitigation in llm-based code generation,'' \emph{arXiv preprint arXiv:2309.14345}, 2023.

\bibitem{huang2021damia}
H.~Huang, W.~Luo, G.~Zeng, J.~Weng, Y.~Zhang, and A.~Yang, ``Damia: leveraging domain adaptation as a defense against membership inference attacks,'' \emph{IEEE Transactions on Dependable and Secure Computing}, vol.~19, no.~5, pp. 3183--3199, 2021.

\bibitem{huang2022large}
J.~Huang, H.~Shao, and K.~C.-C. Chang, ``Are large pre-trained language models leaking your personal information?'' \emph{arXiv preprint arXiv:2205.12628}, 2022.

\bibitem{igure2008taxonomies}
V.~M. Igure and R.~D. Williams, ``Taxonomies of attacks and vulnerabilities in computer systems,'' \emph{IEEE Communications Surveys \& Tutorials}, vol.~10, no.~1, pp. 6--19, 2008.

\bibitem{iqbal2023llm}
U.~Iqbal, T.~Kohno, and F.~Roesner, ``Llm platform security: Applying a systematic evaluation framework to openai's chatgpt plugins,'' 2023.

\bibitem{ivgi2021achieving}
M.~Ivgi and J.~Berant, ``Achieving model robustness through discrete adversarial training,'' \emph{arXiv preprint arXiv:2104.05062}, 2021.

\bibitem{jain2023baseline}
N.~Jain, A.~Schwarzschild, Y.~Wen, G.~Somepalli, J.~Kirchenbauer, P.-y. Chiang, M.~Goldblum, A.~Saha, J.~Geiping, and T.~Goldstein, ``Baseline defenses for adversarial attacks against aligned language models,'' \emph{arXiv preprint arXiv:2309.00614}, 2023.

\bibitem{jain2023code}
R.~Jain, N.~Gervasoni, M.~Ndhlovu, and S.~Rawat, ``A code centric evaluation of c/c++ vulnerability datasets for deep learning based vulnerability detection techniques,'' in \emph{Proceedings of the 16th Innovations in Software Engineering Conference}, 2023, pp. 1--10.

\bibitem{jamal2023improved}
S.~Jamal and H.~Wimmer, ``An improved transformer-based model for detecting phishing, spam, and ham: A large language model approach,'' 2023.

\bibitem{jayaraman2020revisiting}
B.~Jayaraman, L.~Wang, K.~Knipmeyer, Q.~Gu, and D.~Evans, ``Revisiting membership inference under realistic assumptions,'' \emph{arXiv preprint arXiv:2005.10881}, 2020.

\bibitem{jiang2019smart}
H.~Jiang, P.~He, W.~Chen, X.~Liu, J.~Gao, and T.~Zhao, ``Smart: Robust and efficient fine-tuning for pre-trained natural language models through principled regularized optimization,'' \emph{arXiv preprint arXiv:1911.03437}, 2019.

\bibitem{jiang2023low}
J.~Jiang, X.~Liu, and C.~Fan, ``Low-parameter federated learning with large language models,'' \emph{arXiv preprint arXiv:2307.13896}, 2023.

\bibitem{jiang2023impact}
N.~Jiang, K.~Liu, T.~Lutellier, and L.~Tan, ``Impact of code language models on automated program repair,'' 2023.

\bibitem{jiang2023prompt}
S.~Jiang, X.~Chen, and R.~Tang, ``Prompt packer: Deceiving llms through compositional instruction with hidden attacks,'' \emph{arXiv preprint arXiv:2310.10077}, 2023.

\bibitem{jin2023inferfix}
M.~Jin, S.~Shahriar, M.~Tufano, X.~Shi, S.~Lu, N.~Sundaresan, and A.~Svyatkovskiy, ``Inferfix: End-to-end program repair with llms,'' 2023.

\bibitem{jin2023binary}
X.~Jin, J.~Larson, W.~Yang, and Z.~Lin, ``Binary code summarization: Benchmarking chatgpt/gpt-4 and other large language models,'' 2023.

\bibitem{jin2022symlm}
X.~Jin, K.~Pei, J.~Y. Won, and Z.~Lin, ``Symlm: Predicting function names in stripped binaries via context-sensitive execution-aware code embeddings,'' in \emph{Proceedings of the 2022 ACM SIGSAC Conference on Computer and Communications Security}, 2022, pp. 1631--1645.

\bibitem{joshi2015review}
C.~Joshi, U.~K. Singh, and K.~Tarey, ``A review on taxonomies of attacks and vulnerability in computer and network system,'' \emph{International Journal}, vol.~5, no.~1, 2015.

\bibitem{joulin2016fasttext}
A.~Joulin, E.~Grave, P.~Bojanowski, M.~Douze, H.~J{\'e}gou, and T.~Mikolov, ``Fasttext. zip: Compressing text classification models,'' \emph{arXiv preprint arXiv:1612.03651}, 2016.

\bibitem{juuti2019prada}
M.~Juuti, S.~Szyller, S.~Marchal, and N.~Asokan, ``Prada: protecting against dnn model stealing attacks,'' in \emph{2019 IEEE European Symposium on Security and Privacy (EuroS\&P)}.\hskip 1em plus 0.5em minus 0.4em\relax IEEE, 2019, pp. 512--527.

\bibitem{kadavath2022language}
S.~Kadavath, T.~Conerly, A.~Askell, T.~Henighan, D.~Drain, E.~Perez, N.~Schiefer, Z.~Hatfield-Dodds, N.~DasSarma, E.~Tran-Johnson \emph{et~al.}, ``Language models (mostly) know what they know,'' \emph{arXiv preprint arXiv:2207.05221}, 2022.

\bibitem{kandpal2023backdoor}
N.~Kandpal, M.~Jagielski, F.~Tram{\`e}r, and N.~Carlini, ``Backdoor attacks for in-context learning with language models,'' \emph{arXiv preprint arXiv:2307.14692}, 2023.

\bibitem{kandpal2023user}
N.~Kandpal, K.~Pillutla, A.~Oprea, P.~Kairouz, C.~A. Choquette-Choo, and Z.~Xu, ``User inference attacks on large language models,'' 2023.

\bibitem{kandpal2022deduplicating}
N.~Kandpal, E.~Wallace, and C.~Raffel, ``Deduplicating training data mitigates privacy risks in language models,'' in \emph{International Conference on Machine Learning}.\hskip 1em plus 0.5em minus 0.4em\relax PMLR, 2022, pp. 10\,697--10\,707.

\bibitem{kang2023exploiting}
D.~Kang, X.~Li, I.~Stoica, C.~Guestrin, M.~Zaharia, and T.~Hashimoto, ``Exploiting programmatic behavior of llms: Dual-use through standard security attacks,'' \emph{arXiv preprint arXiv:2302.05733}, 2023.

\bibitem{kang2023llm}
S.~Kang, J.~Yoon, and S.~Yoo, ``Llm lies: Hallucinations are not bugs, but features as adversarial examples,'' in \emph{2023 IEEE/ACM 45th International Conference on Software Engineering (ICSE)}.\hskip 1em plus 0.5em minus 0.4em\relax IEEE, 2023.

\bibitem{kariyappa2021maze}
S.~Kariyappa, A.~Prakash, and M.~K. Qureshi, ``Maze: Data-free model stealing attack using zeroth-order gradient estimation,'' in \emph{Proceedings of the IEEE/CVF Conference on Computer Vision and Pattern Recognition}, 2021, pp. 13\,814--13\,823.

\bibitem{karpinska2023large}
M.~Karpinska and M.~Iyyer, ``Large language models effectively leverage document-level context for literary translation, but critical errors persist,'' \emph{arXiv preprint arXiv:2304.03245}, 2023.

\bibitem{khalil2023will}
M.~Khalil and E.~Er, ``Will chatgpt get you caught? rethinking of plagiarism detection,'' \emph{arXiv preprint arXiv:2302.04335}, 2023.

\bibitem{kirchenbauer2023reliability}
J.~Kirchenbauer, J.~Geiping, Y.~Wen, M.~Shu, K.~Saifullah, K.~Kong, K.~Fernando, A.~Saha, M.~Goldblum, and T.~Goldstein, ``On the reliability of watermarks for large language models,'' \emph{arXiv preprint arXiv:2306.04634}, 2023.

\bibitem{gpt213bugs}
C.~Koch, ``I used gpt-3 to find 213 security vulnerabilities in a single codebase,'' \url{http://surl.li/ncjvo}, 2023.

\bibitem{koide2023detecting}
T.~Koide, N.~Fukushi, H.~Nakano, and D.~Chiba, ``Detecting phishing sites using chatgpt,'' \emph{arXiv preprint arXiv:2306.05816}, 2023.

\bibitem{kong2023efficient}
F.~Kong, J.~Duan, R.~Ma, H.~Shen, X.~Zhu, X.~Shi, and K.~Xu, ``An efficient membership inference attack for the diffusion model by proximal initialization,'' \emph{arXiv preprint arXiv:2305.18355}, 2023.

\bibitem{kotek2023gender}
H.~Kotek, R.~Dockum, and D.~Sun, ``Gender bias and stereotypes in large language models,'' in \emph{Proceedings of The ACM Collective Intelligence Conference}, 2023, pp. 12--24.

\bibitem{kuang2023federatedscope}
W.~Kuang, B.~Qian, Z.~Li, D.~Chen, D.~Gao, X.~Pan, Y.~Xie, Y.~Li, B.~Ding, and J.~Zhou, ``Federatedscope-llm: A comprehensive package for fine-tuning large language models in federated learning,'' \emph{arXiv preprint arXiv:2309.00363}, 2023.

\bibitem{kumari2023demasq}
K.~Kumari, A.~Pegoraro, H.~Fereidooni, and A.-R. Sadeghi, ``Demasq: Unmasking the chatgpt wordsmith,'' \emph{arXiv preprint arXiv:2311.05019}, 2023.

\bibitem{kumari2023demasqndss}
------, ``Demasq: Unmasking the chatgpt wordsmith,'' in \emph{Proceedings of the 31th Annual Network and Distributed System Security Symposium (NDSS’24)}, 2024.

\bibitem{kurita2020weight}
K.~Kurita, P.~Michel, and G.~Neubig, ``Weight poisoning attacks on pre-trained models,'' \emph{arXiv preprint arXiv:2004.06660}, 2020.

\bibitem{cryptoeprintcyrotpo}
\BIBentryALTinterwordspacing
H.~Kwon, M.~Sim, G.~Song, M.~Lee, and H.~Seo, ``Novel approach to cryptography implementation using chatgpt,'' Cryptology ePrint Archive, Paper 2023/606, 2023, \url{https://eprint.iacr.org/2023/606}. [Online]. Available: \url{https://eprint.iacr.org/2023/606}
\BIBentrySTDinterwordspacing

\bibitem{laird2017standard}
J.~E. Laird, C.~Lebiere, and P.~S. Rosenbloom, ``A standard model of the mind: Toward a common computational framework across artificial intelligence, cognitive science, neuroscience, and robotics,'' \emph{Ai Magazine}, vol.~38, no.~4, pp. 13--26, 2017.

\bibitem{langford2023phishing}
T.~Langford and B.~Payne, ``Phishing faster: Implementing chatgpt into phishing campaigns,'' in \emph{Proceedings of the Future Technologies Conference}.\hskip 1em plus 0.5em minus 0.4em\relax Springer, 2023, pp. 174--187.

\bibitem{laurenccon2022bigscience}
H.~Lauren{\c{c}}on, L.~Saulnier, T.~Wang, C.~Akiki, A.~Villanova~del Moral, T.~Le~Scao, L.~Von~Werra, C.~Mou, E.~Gonz{\'a}lez~Ponferrada, H.~Nguyen \emph{et~al.}, ``The bigscience roots corpus: A 1.6 tb composite multilingual dataset,'' \emph{Advances in Neural Information Processing Systems}, vol.~35, pp. 31\,809--31\,826, 2022.

\bibitem{lee2021deduplicating}
K.~Lee, D.~Ippolito, A.~Nystrom, C.~Zhang, D.~Eck, C.~Callison-Burch, and N.~Carlini, ``Deduplicating training data makes language models better,'' \emph{arXiv preprint arXiv:2107.06499}, 2021.

\bibitem{lee2023wrote}
T.~Lee, S.~Hong, J.~Ahn, I.~Hong, H.~Lee, S.~Yun, J.~Shin, and G.~Kim, ``Who wrote this code? watermarking for code generation,'' 2023.

\bibitem{leite2023detecting}
J.~A. Leite, O.~Razuvayevskaya, K.~Bontcheva, and C.~Scarton, ``Detecting misinformation with llm-predicted credibility signals and weak supervision,'' \emph{arXiv preprint arXiv:2309.07601}, 2023.

\bibitem{lemieux2023codamosa}
C.~Lemieux, J.~P. Inala, S.~K. Lahiri, and S.~Sen, ``Codamosa: Escaping coverage plateaus in test generation with pre-trained large language models,'' in \emph{International conference on software engineering (ICSE)}, 2023.

\bibitem{leskovec2020mining}
J.~Leskovec, A.~Rajaraman, and J.~D. Ullman, \emph{Mining of massive data sets}.\hskip 1em plus 0.5em minus 0.4em\relax Cambridge university press, 2020.

\bibitem{li2023theoretical}
C.~Li, Z.~Song, W.~Wang, and C.~Yang, ``A theoretical insight into attack and defense of gradient leakage in transformer,'' \emph{arXiv preprint arXiv:2311.13624}, 2023.

\bibitem{li2023multi}
H.~Li, D.~Guo, W.~Fan, M.~Xu, and Y.~Song, ``Multi-step jailbreaking privacy attacks on chatgpt,'' \emph{arXiv preprint arXiv:2304.05197}, 2023.

\bibitem{li-etal-2022-dont}
\BIBentryALTinterwordspacing
H.~Li, Y.~Song, and L.~Fan, ``You don{'}t know my favorite color: Preventing dialogue representations from revealing speakers{'} private personas,'' in \emph{Proceedings of the 2022 Conference of the North American Chapter of the Association for Computational Linguistics: Human Language Technologies}, M.~Carpuat, M.-C. de~Marneffe, and I.~V. Meza~Ruiz, Eds.\hskip 1em plus 0.5em minus 0.4em\relax Seattle, United States: Association for Computational Linguistics, Jul. 2022, pp. 5858--5870. [Online]. Available: \url{https://aclanthology.org/2022.naacl-main.429}
\BIBentrySTDinterwordspacing

\bibitem{li2023chatgpt}
J.~Li, Y.~Yang, Z.~Wu, V.~Vydiswaran, and C.~Xiao, ``Chatgpt as an attack tool: Stealthy textual backdoor attack via blackbox generative model trigger,'' \emph{arXiv preprint arXiv:2304.14475}, 2023.

\bibitem{li2023evaluating}
J.~Li, P.~H. Meland, J.~S. Notland, A.~Storhaug, and J.~H. Tysse, ``Evaluating the impact of chatgpt on exercises of a software security course,'' 2023.

\bibitem{li2021token}
L.~Li and X.~Qiu, ``Token-aware virtual adversarial training in natural language understanding,'' in \emph{Proceedings of the AAAI Conference on Artificial Intelligence}, vol.~35, no.~9, 2021, pp. 8410--8418.

\bibitem{li2022text}
L.~Li, D.~Song, and X.~Qiu, ``Text adversarial purification as defense against adversarial attacks,'' \emph{arXiv preprint arXiv:2203.14207}, 2022.

\bibitem{li2021large}
X.~Li, F.~Tramer, P.~Liang, and T.~Hashimoto, ``Large language models can be strong differentially private learners,'' \emph{arXiv preprint arXiv:2110.05679}, 2021.

\bibitem{li2023privacy}
Y.~Li, Z.~Tan, and Y.~Liu, ``Privacy-preserving prompt tuning for large language model services,'' \emph{arXiv preprint arXiv:2305.06212}, 2023.

\bibitem{li2023multitarget}
Y.~Li, S.~Liu, K.~Chen, X.~Xie, T.~Zhang, and Y.~Liu, ``Multi-target backdoor attacks for code pre-trained models,'' 2023.

\bibitem{Li2023EvaluatingTI}
\BIBentryALTinterwordspacing
Z.~Li, B.~Peng, P.~He, and X.~Yan, ``Evaluating the instruction-following robustness of large language models to prompt injection,'' 2023. [Online]. Available: \url{https://api.semanticscholar.org/CorpusID:261048972}
\BIBentrySTDinterwordspacing

\bibitem{li2023protecting}
Z.~Li, C.~Wang, S.~Wang, and C.~Gao, ``Protecting intellectual property of large language model-based code generation apis via watermarks,'' in \emph{Proceedings of the 2023 ACM SIGSAC Conference on Computer and Communications Security}, 2023, pp. 2336--2350.

\bibitem{liang2023holistic}
P.~Liang, R.~Bommasani, T.~Lee, D.~Tsipras, D.~Soylu, M.~Yasunaga, Y.~Zhang, D.~Narayanan, Y.~Wu, A.~Kumar, B.~Newman, B.~Yuan, B.~Yan, C.~Zhang, C.~Cosgrove, C.~D. Manning, C.~Ré, D.~Acosta-Navas, D.~A. Hudson, E.~Zelikman, E.~Durmus, F.~Ladhak, F.~Rong, H.~Ren, H.~Yao, J.~Wang, K.~Santhanam, L.~Orr, L.~Zheng, M.~Yuksekgonul, M.~Suzgun, N.~Kim, N.~Guha, N.~Chatterji, O.~Khattab, P.~Henderson, Q.~Huang, R.~Chi, S.~M. Xie, S.~Santurkar, S.~Ganguli, T.~Hashimoto, T.~Icard, T.~Zhang, V.~Chaudhary, W.~Wang, X.~Li, Y.~Mai, Y.~Zhang, and Y.~Koreeda, ``Holistic evaluation of language models,'' 2023.

\bibitem{lin2021truthfulqa}
S.~Lin, J.~Hilton, and O.~Evans, ``Truthfulqa: Measuring how models mimic human falsehoods,'' \emph{arXiv preprint arXiv:2109.07958}, 2021.

\bibitem{liu2023adversarial}
B.~Liu, B.~Xiao, X.~Jiang, S.~Cen, X.~He, W.~Dou \emph{et~al.}, ``Adversarial attacks on large language model-based system and mitigating strategies: A case study on chatgpt,'' \emph{Security and Communication Networks}, vol. 2023, 2023.

\bibitem{liu2023chinese}
C.~Liu, F.~Zhao, L.~Qing, Y.~Kang, C.~Sun, K.~Kuang, and F.~Wu, ``A chinese prompt attack dataset for llms with evil content,'' \emph{arXiv preprint arXiv:2309.11830}, 2023.

\bibitem{liu2023harnessing}
P.~Liu, C.~Sun, Y.~Zheng, X.~Feng, C.~Qin, Y.~Wang, Z.~Li, and L.~Sun, ``Harnessing the power of llm to support binary taint analysis,'' 2023.

\bibitem{liu2023demystifying}
T.~Liu, Z.~Deng, G.~Meng, Y.~Li, and K.~Chen, ``Demystifying rce vulnerabilities in llm-integrated apps,'' 2023.

\bibitem{liu2020adversarial}
X.~Liu, H.~Cheng, P.~He, W.~Chen, Y.~Wang, H.~Poon, and J.~Gao, ``Adversarial training for large neural language models,'' \emph{arXiv preprint arXiv:2004.08994}, 2020.

\bibitem{liu2023autodan}
X.~Liu, N.~Xu, M.~Chen, and C.~Xiao, ``Autodan: Generating stealthy jailbreak prompts on aligned large language models,'' \emph{arXiv preprint arXiv:2310.04451}, 2023.

\bibitem{liu2023prompt}
Y.~Liu, G.~Deng, Y.~Li, K.~Wang, T.~Zhang, Y.~Liu, H.~Wang, Y.~Zheng, and Y.~Liu, ``Prompt injection attack against llm-integrated applications,'' \emph{arXiv preprint arXiv:2306.05499}, 2023.

\bibitem{lo2023impact}
C.~K. Lo, ``What is the impact of chatgpt on education? a rapid review of the literature,'' \emph{Education Sciences}, vol.~13, no.~4, p. 410, 2023.

\bibitem{logacheva2022paradetox}
V.~Logacheva, D.~Dementieva, S.~Ustyantsev, D.~Moskovskiy, D.~Dale, I.~Krotova, N.~Semenov, and A.~Panchenko, ``Paradetox: Detoxification with parallel data,'' in \emph{Proceedings of the 60th Annual Meeting of the Association for Computational Linguistics (Volume 1: Long Papers)}, 2022, pp. 6804--6818.

\bibitem{lyu-etal-2020-differentially}
\BIBentryALTinterwordspacing
L.~Lyu, X.~He, and Y.~Li, ``Differentially private representation for {NLP}: Formal guarantee and an empirical study on privacy and fairness,'' in \emph{Findings of the Association for Computational Linguistics: EMNLP 2020}, T.~Cohn, Y.~He, and Y.~Liu, Eds.\hskip 1em plus 0.5em minus 0.4em\relax Online: Association for Computational Linguistics, Nov. 2020, pp. 2355--2365. [Online]. Available: \url{https://aclanthology.org/2020.findings-emnlp.213}
\BIBentrySTDinterwordspacing

\bibitem{madry2017towards}
A.~Madry, A.~Makelov, L.~Schmidt, D.~Tsipras, and A.~Vladu, ``Towards deep learning models resistant to adversarial attacks,'' \emph{arXiv preprint arXiv:1706.06083}, 2017.

\bibitem{Mahloujifar2021MembershipIO}
\BIBentryALTinterwordspacing
S.~Mahloujifar, H.~A. Inan, M.~Chase, E.~Ghosh, and M.~Hasegawa, ``Membership inference on word embedding and beyond,'' \emph{ArXiv}, vol. abs/2106.11384, 2021. [Online]. Available: \url{https://api.semanticscholar.org/CorpusID:235593386}
\BIBentrySTDinterwordspacing

\bibitem{majmudar2022differentially}
J.~Majmudar, C.~Dupuy, C.~Peris, S.~Smaili, R.~Gupta, and R.~Zemel, ``Differentially private decoding in large language models,'' \emph{arXiv preprint arXiv:2205.13621}, 2022.

\bibitem{marshall2023effects}
J.~Marshall, ``What effects do large language models have on cybersecurity,'' 2023.

\bibitem{mbakwe2023chatgpt}
A.~B. Mbakwe, I.~Lourentzou, L.~A. Celi, O.~J. Mechanic, and A.~Dagan, ``Chatgpt passing usmle shines a spotlight on the flaws of medical education,'' p. e0000205, 2023.

\bibitem{mcintosh2023harnessing}
T.~McIntosh, T.~Liu, T.~Susnjak, H.~Alavizadeh, A.~Ng, R.~Nowrozy, and P.~Watters, ``Harnessing gpt-4 for generation of cybersecurity grc policies: A focus on ransomware attack mitigation,'' \emph{Computers \& Security}, vol. 134, p. 103424, 2023.

\bibitem{meade2021empirical}
N.~Meade, E.~Poole-Dayan, and S.~Reddy, ``An empirical survey of the effectiveness of debiasing techniques for pre-trained language models,'' \emph{arXiv preprint arXiv:2110.08527}, 2021.

\bibitem{mendez2021physical}
M.~M{\'e}ndez~Real and R.~Salvador, ``Physical side-channel attacks on embedded neural networks: A survey,'' \emph{Applied Sciences}, vol.~11, no.~15, p. 6790, 2021.

\bibitem{zhou2012hey}
R.~Meng, M.~Mirchev, M.~B{\"o}hme, and A.~Roychoudhury, ``Large language model guided protocol fuzzing,'' in \emph{Proceedings of the 31th Annual Network and Distributed System Security Symposium (NDSS’24)}, 2024.

\bibitem{mireshghallah2022quantifying}
F.~Mireshghallah, K.~Goyal, A.~Uniyal, T.~Berg-Kirkpatrick, and R.~Shokri, ``Quantifying privacy risks of masked language models using membership inference attacks,'' 2022.

\bibitem{mireshghallah-etal-2022-empirical}
\BIBentryALTinterwordspacing
F.~Mireshghallah, A.~Uniyal, T.~Wang, D.~Evans, and T.~Berg-Kirkpatrick, ``An empirical analysis of memorization in fine-tuned autoregressive language models,'' in \emph{Proceedings of the 2022 Conference on Empirical Methods in Natural Language Processing}, Y.~Goldberg, Z.~Kozareva, and Y.~Zhang, Eds.\hskip 1em plus 0.5em minus 0.4em\relax Abu Dhabi, United Arab Emirates: Association for Computational Linguistics, Dec. 2022, pp. 1816--1826. [Online]. Available: \url{https://aclanthology.org/2022.emnlp-main.119}
\BIBentrySTDinterwordspacing

\bibitem{mo2023test}
W.~Mo, J.~Xu, Q.~Liu, J.~Wang, J.~Yan, C.~Xiao, and M.~Chen, ``Test-time backdoor mitigation for black-box large language models with defensive demonstrations,'' \emph{arXiv preprint arXiv:2311.09763}, 2023.

\bibitem{monje2023being}
A.~Monje, A.~Monje, R.~A. Hallman, and G.~Cybenko, ``Being a bad influence on the kids: Malware generation in less than five minutes using chatgpt,'' 2023.

\bibitem{moskovskiy2022exploring}
D.~Moskovskiy, D.~Dementieva, and A.~Panchenko, ``Exploring cross-lingual text detoxification with large multilingual language models.'' in \emph{Proceedings of the 60th Annual Meeting of the Association for Computational Linguistics: Student Research Workshop}, 2022, pp. 346--354.

\bibitem{mozes2023use}
M.~Mozes, X.~He, B.~Kleinberg, and L.~D. Griffin, ``Use of llms for illicit purposes: Threats, prevention measures, and vulnerabilities,'' 2023.

\bibitem{cryptoeprint:2023/212}
\BIBentryALTinterwordspacing
M.~Nair, R.~Sadhukhan, and D.~Mukhopadhyay, ``Generating secure hardware using chatgpt resistant to cwes,'' Cryptology ePrint Archive, Paper 2023/212, 2023, \url{https://eprint.iacr.org/2023/212}. [Online]. Available: \url{https://eprint.iacr.org/2023/212}
\BIBentrySTDinterwordspacing

\bibitem{narang_chowdhery2022palm}
S.~Narang and A.~Chowdhery, ``Pathways language model (palm): Scaling to 540 billion parameters for breakthrough performance,'' \url{https://blog.research.google/2022/04/pathways-language-model-palm-scaling-to.html}, apr 2022, accessed: 2023-11-13.

\bibitem{ni2023lever}
A.~Ni, S.~Iyer, D.~Radev, V.~Stoyanov, W.-t. Yih, S.~Wang, and X.~V. Lin, ``Lever: Learning to verify language-to-code generation with execution,'' in \emph{International Conference on Machine Learning}.\hskip 1em plus 0.5em minus 0.4em\relax PMLR, 2023, pp. 26\,106--26\,128.

\bibitem{nikolic2023chatgpt}
S.~Nikolic, S.~Daniel, R.~Haque, M.~Belkina, G.~M. Hassan, S.~Grundy, S.~Lyden, P.~Neal, and C.~Sandison, ``Chatgpt versus engineering education assessment: a multidisciplinary and multi-institutional benchmarking and analysis of this generative artificial intelligence tool to investigate assessment integrity,'' \emph{European Journal of Engineering Education}, pp. 1--56, 2023.

\bibitem{noever2023large}
\BIBentryALTinterwordspacing
D.~Noever, ``Can large language models find and fix vulnerable software?'' \emph{arXiv preprint arXiv:2308.10345}, 2023. [Online]. Available: \url{https://doi.org/10.48550/arXiv.2308.10345}
\BIBentrySTDinterwordspacing

\bibitem{novelli2023taking}
C.~Novelli, F.~Casolari, A.~Rotolo, M.~Taddeo, and L.~Floridi, ``Taking ai risks seriously: a new assessment model for the ai act,'' \emph{AI \& SOCIETY}, pp. 1--5, 2023.

\bibitem{openai2023gpt4}
OpenAI, ``Gpt-4 technical report,'' \url{https://arxiv.org/abs/2303.08774}, 2023.

\bibitem{ousidhoum2021probing}
N.~Ousidhoum, X.~Zhao, T.~Fang, Y.~Song, and D.-Y. Yeung, ``Probing toxic content in large pre-trained language models,'' in \emph{Proceedings of the 59th Annual Meeting of the Association for Computational Linguistics and the 11th International Joint Conference on Natural Language Processing (Volume 1: Long Papers)}, 2021, pp. 4262--4274.

\bibitem{ouyang2022training}
L.~Ouyang, J.~Wu, X.~Jiang, D.~Almeida, C.~Wainwright, P.~Mishkin, C.~Zhang, S.~Agarwal, K.~Slama, A.~Ray \emph{et~al.}, ``Training language models to follow instructions with human feedback,'' \emph{Advances in Neural Information Processing Systems}, vol.~35, pp. 27\,730--27\,744, 2022.

\bibitem{owaspllm2023}
\BIBentryALTinterwordspacing
OWASP. (2023, Oct) {OWASP Top 10 for LLM}. [Online]. Available: \url{{https://owasp.org/www-project-top-10-for-large-language-model-applications/assets/PDF/OWASP-Top-10-for-LLMs-2023-v1_1.pdf}}
\BIBentrySTDinterwordspacing

\bibitem{pa2023attacker}
Y.~M. Pa~Pa, S.~Tanizaki, T.~Kou, M.~Van~Eeten, K.~Yoshioka, and T.~Matsumoto, ``An attacker’s dream? exploring the capabilities of chatgpt for developing malware,'' in \emph{Proceedings of the 16th Cyber Security Experimentation and Test Workshop}, 2023, pp. 10--18.

\bibitem{pan2020privacy}
X.~Pan, M.~Zhang, S.~Ji, and M.~Yang, ``Privacy risks of general-purpose language models,'' in \emph{2020 IEEE Symposium on Security and Privacy (SP)}.\hskip 1em plus 0.5em minus 0.4em\relax IEEE, 2020, pp. 1314--1331.

\bibitem{paria2023divas}
S.~Paria, A.~Dasgupta, and S.~Bhunia, ``Divas: An llm-based end-to-end framework for soc security analysis and policy-based protection,'' \emph{arXiv preprint arXiv:2308.06932}, 2023.

\bibitem{parikh2022canary}
R.~Parikh, C.~Dupuy, and R.~Gupta, ``Canary extraction in natural language understanding models,'' \emph{arXiv preprint arXiv:2203.13920}, 2022.

\bibitem{10179324}
H.~Pearce, B.~Tan, B.~Ahmad, R.~Karri, and B.~Dolan-Gavitt, ``Examining zero-shot vulnerability repair with large language models,'' in \emph{2023 IEEE Symposium on Security and Privacy (SP)}, 2023, pp. 2339--2356.

\bibitem{pearce2022pop}
H.~Pearce, B.~Tan, P.~Krishnamurthy, F.~Khorrami, R.~Karri, and B.~Dolan-Gavitt, ``Pop quiz! can a large language model help with reverse engineering?'' 2022.

\bibitem{penedo2023refinedweb}
G.~Penedo, Q.~Malartic, D.~Hesslow, R.~Cojocaru, A.~Cappelli, H.~Alobeidli, B.~Pannier, E.~Almazrouei, and J.~Launay, ``The refinedweb dataset for falcon llm: outperforming curated corpora with web data, and web data only,'' \emph{arXiv preprint arXiv:2306.01116}, 2023.

\bibitem{peris2023privacy}
C.~Peris, C.~Dupuy, J.~Majmudar, R.~Parikh, S.~Smaili, R.~Zemel, and R.~Gupta, ``Privacy in the time of language models,'' in \emph{Proceedings of the Sixteenth ACM International Conference on Web Search and Data Mining}, 2023, pp. 1291--1292.

\bibitem{perkins2023academic}
M.~Perkins, ``Academic integrity considerations of ai large language models in the post-pandemic era: Chatgpt and beyond,'' \emph{Journal of University Teaching \& Learning Practice}, vol.~20, no.~2, p.~07, 2023.

\bibitem{pfitzmann2010terminology}
A.~Pfitzmann and M.~Hansen, ``A terminology for talking about privacy by data minimization: Anonymity, unlinkability, undetectability, unobservability, pseudonymity, and identity management,'' 2010.

\bibitem{pham2020aflnet}
V.-T. Pham, M.~B{\"o}hme, and A.~Roychoudhury, ``Aflnet: a greybox fuzzer for network protocols,'' in \emph{2020 IEEE 13th International Conference on Software Testing, Validation and Verification (ICST)}.\hskip 1em plus 0.5em minus 0.4em\relax IEEE, 2020, pp. 460--465.

\bibitem{10301302}
M.~D. Purba, A.~Ghosh, B.~J. Radford, and B.~Chu, ``Software vulnerability detection using large language models,'' in \emph{2023 IEEE 34th International Symposium on Software Reliability Engineering Workshops (ISSREW)}, 2023, pp. 112--119.

\bibitem{qammar2023chatbots}
A.~Qammar, H.~Wang, J.~Ding, A.~Naouri, M.~Daneshmand, and H.~Ning, ``Chatbots to chatgpt in a cybersecurity space: Evolution, vulnerabilities, attacks, challenges, and future recommendations,'' 2023.

\bibitem{qi2020onion}
F.~Qi, Y.~Chen, M.~Li, Y.~Yao, Z.~Liu, and M.~Sun, ``Onion: A simple and effective defense against textual backdoor attacks,'' \emph{arXiv preprint arXiv:2011.10369}, 2020.

\bibitem{qi2023loggpt}
J.~Qi, S.~Huang, Z.~Luan, C.~Fung, H.~Yang, and D.~Qian, ``Loggpt: Exploring chatgpt for log-based anomaly detection,'' \emph{arXiv preprint arXiv:2309.01189}, 2023.

\bibitem{qin2023nsfuzz}
S.~Qin, F.~Hu, Z.~Ma, B.~Zhao, T.~Yin, and C.~Zhang, ``Nsfuzz: Towards efficient and state-aware network service fuzzing,'' \emph{ACM Transactions on Software Engineering and Methodology}, 2023.

\bibitem{quidwai2023beyond}
M.~A. Quidwai, C.~Li, and P.~Dube, ``Beyond black box ai-generated plagiarism detection: From sentence to document level,'' \emph{arXiv preprint arXiv:2306.08122}, 2023.

\bibitem{raeini2023privacy}
M.~Raeini, ``Privacy-preserving large language models (ppllms),'' \emph{Available at SSRN 4512071}, 2023.

\bibitem{raffel2023exploring}
C.~Raffel, N.~Shazeer, A.~Roberts, K.~Lee, S.~Narang, M.~Matena, Y.~Zhou, W.~Li, and P.~J. Liu, ``Exploring the limits of transfer learning with a unified text-to-text transformer,'' 2023.

\bibitem{rahman2023chatgpt}
M.~M. Rahman and Y.~Watanobe, ``Chatgpt for education and research: Opportunities, threats, and strategies,'' \emph{Applied Sciences}, vol.~13, no.~9, p. 5783, 2023.

\bibitem{rando2023universal}
J.~Rando and F.~Tram{\`e}r, ``Universal jailbreak backdoors from poisoned human feedback,'' \emph{arXiv preprint arXiv:2311.14455}, 2023.

\bibitem{renaud2023chatgpt}
K.~Renaud, M.~Warkentin, and G.~Westerman, \emph{From ChatGPT to HackGPT: Meeting the Cybersecurity Threat of Generative AI}.\hskip 1em plus 0.5em minus 0.4em\relax MIT Sloan Management Review, 2023.

\bibitem{salesforce2023conditional}
S.~A. Research, ``Introducing a conditional transformer language model for controllable generation,'' \url{https://shorturl.at/azQW6}, apr 2023, accessed: 2023-11-13.

\bibitem{robey2023smoothllm}
A.~Robey, E.~Wong, H.~Hassani, and G.~J. Pappas, ``Smoothllm: Defending large language models against jailbreaking attacks,'' \emph{arXiv preprint arXiv:2310.03684}, 2023.

\bibitem{romero2023synergistic}
O.~J. Romero, J.~Zimmerman, A.~Steinfeld, and A.~Tomasic, ``Synergistic integration of large language models and cognitive architectures for robust ai: An exploratory analysis,'' \emph{arXiv preprint arXiv:2308.09830}, 2023.

\bibitem{rosyanafi2023dark}
R.~J. Rosyanafi, G.~D. Lestari, H.~Susilo, W.~Nusantara, and F.~Nuraini, ``The dark side of innovation: Understanding research misconduct with chat gpt in nonformal education studies at universitas negeri surabaya,'' \emph{Jurnal Review Pendidikan Dasar: Jurnal Kajian Pendidikan dan Hasil Penelitian}, vol.~9, no.~3, pp. 220--228, 2023.

\bibitem{sakaoglu2023kartal}
\BIBentryALTinterwordspacing
S.~Sakaoglu, ``Kartal: Web application vulnerability hunting using large language models,'' Master's thesis, Master's Programme in Security and Cloud Computing (SECCLO), August 2023. [Online]. Available: \url{http://urn.fi/URN:NBN:fi:aalto-202308275121}
\BIBentrySTDinterwordspacing

\bibitem{sandoval2023lost}
\BIBentryALTinterwordspacing
G.~Sandoval, H.~Pearce, T.~Nys, R.~Karri, S.~Garg, and B.~Dolan-Gavitt, ``Lost at c: A user study on the security implications of large language model code assistants,'' in \emph{USENIX Security 2023}, 2023, for associated dataset see [this URL](https://arxiv.org/abs/2208.09727). 18 pages, 12 figures. G. Sandoval and H. Pearce contributed equally to this work. [Online]. Available: \url{https://arxiv.org/abs/2208.09727}
\BIBentrySTDinterwordspacing

\bibitem{llmcomparision2}
Sapling, ``Llm index,'' \url{https://sapling.ai/llm/index}, 2023.

\bibitem{sarabi2023llm}
A.~Sarabi, T.~Yin, and M.~Liu, ``An llm-based framework for fingerprinting internet-connected devices,'' in \emph{Proceedings of the 2023 ACM on Internet Measurement Conference}, 2023, pp. 478--484.

\bibitem{SCANLON2023301609}
\BIBentryALTinterwordspacing
M.~Scanlon, F.~Breitinger, C.~Hargreaves, J.-N. Hilgert, and J.~Sheppard, ``Chatgpt for digital forensic investigation: The good, the bad, and the unknown,'' \emph{Forensic Science International: Digital Investigation}, vol.~46, p. 301609, 2023. [Online]. Available: \url{https://www.sciencedirect.com/science/article/pii/S266628172300121X}
\BIBentrySTDinterwordspacing

\bibitem{schafer2023adaptive}
M.~Sch{\"a}fer, S.~Nadi, A.~Eghbali, and F.~Tip, ``Adaptive test generation using a large language model,'' \emph{arXiv preprint arXiv:2302.06527}, 2023.

\bibitem{schuster2021you}
R.~Schuster, C.~Song, E.~Tromer, and V.~Shmatikov, ``You autocomplete me: Poisoning vulnerabilities in neural code completion,'' in \emph{30th USENIX Security Symposium (USENIX Security 21)}, 2021, pp. 1559--1575.

\bibitem{schwinn2023adversarial}
L.~Schwinn, D.~Dobre, S.~Günnemann, and G.~Gidel, ``Adversarial attacks and defenses in large language models: Old and new threats,'' 2023.

\bibitem{sebastian2023chatgpt}
G.~Sebastian, ``Do chatgpt and other ai chatbots pose a cybersecurity risk?: An exploratory study,'' \emph{International Journal of Security and Privacy in Pervasive Computing (IJSPPC)}, vol.~15, no.~1, pp. 1--11, 2023.

\bibitem{sebastian2023privacy}
------, ``Privacy and data protection in chatgpt and other ai chatbots: Strategies for securing user information,'' \emph{Available at SSRN 4454761}, 2023.

\bibitem{shah2023loft}
M.~A. Shah, R.~Sharma, H.~Dhamyal, R.~Olivier, A.~Shah, D.~Alharthi, H.~T. Bukhari, M.~Baali, S.~Deshmukh, M.~Kuhlmann \emph{et~al.}, ``Loft: Local proxy fine-tuning for improving transferability of adversarial attacks against large language model,'' \emph{arXiv preprint arXiv:2310.04445}, 2023.

\bibitem{shaikh2022second}
O.~Shaikh, H.~Zhang, W.~Held, M.~Bernstein, and D.~Yang, ``On second thought, let's not think step by step! bias and toxicity in zero-shot reasoning,'' \emph{arXiv preprint arXiv:2212.08061}, 2022.

\bibitem{shan2023prompt}
S.~Shan, W.~Ding, J.~Passananti, H.~Zheng, and B.~Y. Zhao, ``Prompt-specific poisoning attacks on text-to-image generative models,'' \emph{arXiv preprint arXiv:2310.13828}, 2023.

\bibitem{shao2021bddr}
K.~Shao, J.~Yang, Y.~Ai, H.~Liu, and Y.~Zhang, ``Bddr: An effective defense against textual backdoor attacks,'' \emph{Computers \& Security}, vol. 110, p. 102433, 2021.

\bibitem{shayegani2023survey}
E.~Shayegani, M.~A.~A. Mamun, Y.~Fu, P.~Zaree, Y.~Dong, and N.~Abu-Ghazaleh, ``Survey of vulnerabilities in large language models revealed by adversarial attacks,'' 2023.

\bibitem{shen2023anything}
X.~Shen, Z.~Chen, M.~Backes, Y.~Shen, and Y.~Zhang, ``"do anything now": Characterizing and evaluating in-the-wild jailbreak prompts on large language models,'' \emph{arXiv preprint arXiv:2308.03825}, 2023.

\bibitem{shi2023safer}
T.~Shi, K.~Chen, and J.~Zhao, ``Safer-instruct: Aligning language models with automated preference data,'' \emph{arXiv preprint arXiv:2311.08685}, 2023.

\bibitem{shokri2017membership}
R.~Shokri, M.~Stronati, C.~Song, and V.~Shmatikov, ``Membership inference attacks against machine learning models,'' in \emph{2017 IEEE symposium on security and privacy (SP)}.\hskip 1em plus 0.5em minus 0.4em\relax IEEE, 2017, pp. 3--18.

\bibitem{shu2023exploitability}
M.~Shu, J.~Wang, C.~Zhu, J.~Geiping, C.~Xiao, and T.~Goldstein, ``On the exploitability of instruction tuning,'' \emph{arXiv preprint arXiv:2306.17194}, 2023.

\bibitem{shumailov2021sponge}
I.~Shumailov, Y.~Zhao, D.~Bates, N.~Papernot, R.~Mullins, and R.~Anderson, ``Sponge examples: Energy-latency attacks on neural networks,'' 2021.

\bibitem{siddiq2023exploring}
M.~L. Siddiq, J.~Santos, R.~H. Tanvir, N.~Ulfat, F.~A. Rifat, and V.~C. Lopes, ``Exploring the effectiveness of large language models in generating unit tests,'' \emph{arXiv preprint arXiv:2305.00418}, 2023.

\bibitem{siddiq2023generate}
\BIBentryALTinterwordspacing
M.~L. Siddiq and J.~C.~S. Santos, ``Generate and pray: Using sallms to evaluate the security of llm generated code,'' 2023, 16 pages. [Online]. Available: \url{https://arxiv.org/abs/2311.00889}
\BIBentrySTDinterwordspacing

\bibitem{sladić2023llm}
M.~Sladić, V.~Valeros, C.~Catania, and S.~Garcia, ``Llm in the shell: Generative honeypots,'' 2023.

\bibitem{smith2023identifying}
V.~Smith, A.~S. Shamsabadi, C.~Ashurst, and A.~Weller, ``Identifying and mitigating privacy risks stemming from language models: A survey,'' 2023.

\bibitem{sobania2023analysis}
D.~Sobania, M.~Briesch, C.~Hanna, and J.~Petke, ``An analysis of the automatic bug fixing performance of chatgpt,'' 2023.

\bibitem{song2020information}
C.~Song and A.~Raghunathan, ``Information leakage in embedding models,'' in \emph{Proceedings of the 2020 ACM SIGSAC conference on computer and communications security}, 2020, pp. 377--390.

\bibitem{spatharioti2023comparing}
S.~E. Spatharioti, D.~M. Rothschild, D.~G. Goldstein, and J.~M. Hofman, ``Comparing traditional and llm-based search for consumer choice: A randomized experiment,'' \emph{arXiv preprint arXiv:2307.03744}, 2023.

\bibitem{spreitzer2017systematic}
R.~Spreitzer, V.~Moonsamy, T.~Korak, and S.~Mangard, ``Systematic classification of side-channel attacks: A case study for mobile devices,'' \emph{IEEE communications surveys \& tutorials}, vol.~20, no.~1, pp. 465--488, 2017.

\bibitem{staab2023memorization}
R.~Staab, M.~Vero, M.~Balunović, and M.~Vechev, ``Beyond memorization: Violating privacy via inference with large language models,'' 2023.

\bibitem{stephens2023researchers}
K.~Stephens, ``Researchers test large language model that preserves patient privacy,'' \emph{AXIS Imaging News}, 2023.

\bibitem{su2023fake}
J.~Su, T.~Y. Zhuo, J.~Mansurov, D.~Wang, and P.~Nakov, ``Fake news detectors are biased against texts generated by large language models,'' \emph{arXiv preprint arXiv:2309.08674}, 2023.

\bibitem{subramani2023detecting}
N.~Subramani, S.~Luccioni, J.~Dodge, and M.~Mitchell, ``Detecting personal information in training corpora: an analysis,'' in \emph{Proceedings of the 3rd Workshop on Trustworthy Natural Language Processing (TrustNLP 2023)}, 2023, pp. 208--220.

\bibitem{sullivan2023chatgpt}
M.~Sullivan, A.~Kelly, and P.~McLaughlan, ``Chatgpt in higher education: Considerations for academic integrity and student learning,'' 2023.

\bibitem{sun2023defending}
X.~Sun, X.~Li, Y.~Meng, X.~Ao, L.~Lyu, J.~Li, and T.~Zhang, ``Defending against backdoor attacks in natural language generation,'' in \emph{Proceedings of the AAAI Conference on Artificial Intelligence}, vol.~37, no.~4, 2023, pp. 5257--5265.

\bibitem{sun2023med}
Y.~Sun, J.~He, S.~Lei, L.~Cui, and C.-T. Lu, ``Med-mmhl: A multi-modal dataset for detecting human-and llm-generated misinformation in the medical domain,'' \emph{arXiv preprint arXiv:2306.08871}, 2023.

\bibitem{sun2023principle}
Z.~Sun, Y.~Shen, Q.~Zhou, H.~Zhang, Z.~Chen, D.~Cox, Y.~Yang, and C.~Gan, ``Principle-driven self-alignment of language models from scratch with minimal human supervision,'' \emph{arXiv preprint arXiv:2305.03047}, 2023.

\bibitem{talat2022you}
Z.~Talat, A.~N{\'e}v{\'e}ol, S.~Biderman, M.~Clinciu, M.~Dey, S.~Longpre, S.~Luccioni, M.~Masoud, M.~Mitchell, D.~Radev \emph{et~al.}, ``You reap what you sow: On the challenges of bias evaluation under multilingual settings,'' in \emph{Proceedings of BigScience Episode\# 5--Workshop on Challenges \& Perspectives in Creating Large Language Models}, 2022, pp. 26--41.

\bibitem{tann2023using}
W.~Tann, Y.~Liu, J.~H. Sim, C.~M. Seah, and E.-C. Chang, ``Using large language models for cybersecurity capture-the-flag challenges and certification questions,'' 2023.

\bibitem{taveekitworachai2023breaking}
P.~Taveekitworachai, F.~Abdullah, M.~C. Gursesli, M.~F. Dewantoro, S.~Chen, A.~Lanata, A.~Guazzini, and R.~Thawonmas, ``Breaking bad: Unraveling influences and risks of user inputs to chatgpt for game story generation,'' in \emph{International Conference on Interactive Digital Storytelling}.\hskip 1em plus 0.5em minus 0.4em\relax Springer, 2023, pp. 285--296.

\bibitem{tay2023using}
Z.~Tay, ``Using artificial intelligence to augment bug fuzzing,'' 2023.

\bibitem{thankgod2023impact}
E.~ThankGod~Chinonso, ``The impact of chatgpt on privacy and data protection laws,'' \emph{The Impact of ChatGPT on Privacy and Data Protection Laws (April 16, 2023)}, 2023.

\bibitem{thirunavukarasu2023large}
A.~J. Thirunavukarasu, D.~S.~J. Ting, K.~Elangovan, L.~Gutierrez, T.~F. Tan, and D.~S.~W. Ting, ``Large language models in medicine,'' \emph{Nature medicine}, vol.~29, no.~8, pp. 1930--1940, 2023.

\bibitem{tong2023privinfer}
M.~Tong, K.~Chen, Y.~Qi, J.~Zhang, W.~Zhang, and N.~Yu, ``Privinfer: Privacy-preserving inference for black-box large language model,'' 2023.

\bibitem{llmcomparision1}
J.~Torres, ``Navigating the llm landscape: A comparative analysis of leading large language models,'' \url{http://surl.li/ncjvc}, 2023.

\bibitem{touvron2023llama}
H.~Touvron, L.~Martin, K.~Stone, P.~Albert, A.~Almahairi, Y.~Babaei, N.~Bashlykov, S.~Batra, P.~Bhargava, S.~Bhosale \emph{et~al.}, ``Llama 2: Open foundation and fine-tuned chat models,'' \emph{arXiv preprint arXiv:2307.09288}, 2023.

\bibitem{truex2018towards}
S.~Truex, L.~Liu, M.~E. Gursoy, L.~Yu, and W.~Wei, ``Towards demystifying membership inference attacks,'' \emph{arXiv preprint arXiv:1807.09173}, 2018.

\bibitem{truong2021data}
J.-B. Truong, P.~Maini, R.~J. Walls, and N.~Papernot, ``Data-free model extraction,'' in \emph{Proceedings of the IEEE/CVF conference on computer vision and pattern recognition}, 2021, pp. 4771--4780.

\bibitem{Uchendu_Lee_Shen_Le_Huang_Lee_2023}
\BIBentryALTinterwordspacing
A.~Uchendu, J.~Lee, H.~Shen, T.~Le, T.-H.~K. Huang, and D.~Lee, ``Does human collaboration enhance the accuracy of identifying llm-generated deepfake texts?'' \emph{Proceedings of the AAAI Conference on Human Computation and Crowdsourcing}, vol.~11, no.~1, pp. 163--174, Nov. 2023. [Online]. Available: \url{https://ojs.aaai.org/index.php/HCOMP/article/view/27557}
\BIBentrySTDinterwordspacing

\bibitem{urchs2023prevalent}
S.~Urchs, V.~Thurner, M.~A{\ss}enmacher, C.~Heumann, and S.~Thiemichen, ``How prevalent is gender bias in chatgpt?--exploring german and english chatgpt responses,'' \emph{arXiv preprint arXiv:2310.03031}, 2023.

\bibitem{urman2023silence}
A.~Urman and M.~Makhortykh, ``The silence of the llms: Cross-lingual analysis of political bias and false information prevalence in chatgpt, google bard, and bing chat,'' 2023.

\bibitem{uzun2023chatgpt}
L.~Uzun, ``Chatgpt and academic integrity concerns: Detecting artificial intelligence generated content,'' \emph{Language Education and Technology}, vol.~3, no.~1, 2023.

\bibitem{uzuner2007evaluating}
{\"O}.~Uzuner, Y.~Luo, and P.~Szolovits, ``Evaluating the state-of-the-art in automatic de-identification,'' \emph{Journal of the American Medical Informatics Association}, vol.~14, no.~5, pp. 550--563, 2007.

\bibitem{vaithilingam2022expectation}
P.~Vaithilingam, T.~Zhang, and E.~L. Glassman, ``Expectation vs. experience: Evaluating the usability of code generation tools powered by large language models,'' in \emph{Chi conference on human factors in computing systems extended abstracts}, 2022, pp. 1--7.

\bibitem{vats2023recovering}
A.~Vats, Z.~Liu, P.~Su, D.~Paul, Y.~Ma, Y.~Pang, Z.~Ahmed, and O.~Kalinli, ``Recovering from privacy-preserving masking with large language models,'' 2023.

\bibitem{ventayen2023openai}
R.~J.~M. Ventayen, ``Openai chatgpt generated results: Similarity index of artificial intelligence-based contents,'' \emph{Available at SSRN 4332664}, 2023.

\bibitem{vidas2011all}
T.~Vidas, D.~Votipka, and N.~Christin, ``All your droid are belong to us: A survey of current android attacks,'' in \emph{5th USENIX Workshop on Offensive Technologies (WOOT 11)}, 2011.

\bibitem{wallace2020concealed}
E.~Wallace, T.~Z. Zhao, S.~Feng, and S.~Singh, ``Concealed data poisoning attacks on nlp models,'' \emph{arXiv preprint arXiv:2010.12563}, 2020.

\bibitem{wan2023poisoning}
A.~Wan, E.~Wallace, S.~Shen, and D.~Klein, ``Poisoning language models during instruction tuning,'' \emph{arXiv preprint arXiv:2305.00944}, 2023.

\bibitem{wan2022you}
Y.~Wan, S.~Zhang, H.~Zhang, Y.~Sui, G.~Xu, D.~Yao, H.~Jin, and L.~Sun, ``You see what i want you to see: poisoning vulnerabilities in neural code search,'' in \emph{Proceedings of the 30th ACM Joint European Software Engineering Conference and Symposium on the Foundations of Software Engineering}, 2022, pp. 1233--1245.

\bibitem{wan2023kelly}
Y.~Wan, G.~Pu, J.~Sun, A.~Garimella, K.-W. Chang, and N.~Peng, ``"kelly is a warm person, joseph is a role model": Gender biases in llm-generated reference letters,'' \emph{arXiv preprint arXiv:2310.09219}, 2023.

\bibitem{wang2019improving}
D.~Wang, C.~Gong, and Q.~Liu, ``Improving neural language modeling via adversarial training,'' in \emph{International Conference on Machine Learning}.\hskip 1em plus 0.5em minus 0.4em\relax PMLR, 2019, pp. 6555--6565.

\bibitem{wang2023ransomware}
\BIBentryALTinterwordspacing
F.~Wang, ``Using large language models to mitigate ransomware threats,'' \emph{Preprints}, November 2023. [Online]. Available: \url{https://doi.org/10.20944/preprints202311.0676.v1}
\BIBentrySTDinterwordspacing

\bibitem{wang2023bot}
H.~Wang, X.~Luo, W.~Wang, and X.~Yan, ``Bot or human? detecting chatgpt imposters with a single question,'' 2023.

\bibitem{wang2023defecthunter}
\BIBentryALTinterwordspacing
J.~Wang, Z.~Huang, H.~Liu, N.~Yang, and Y.~Xiao, ``Defecthunter: A novel llm-driven boosted-conformer-based code vulnerability detection mechanism,'' \emph{arXiv preprint arXiv:2309.15324}, 2023. [Online]. Available: \url{https://doi.org/10.48550/arXiv.2309.15324}
\BIBentrySTDinterwordspacing

\bibitem{wang2023wasa}
J.~Wang, X.~Lu, Z.~Zhao, Z.~Dai, C.-S. Foo, S.-K. Ng, and B.~K.~H. Low, ``Wasa: Watermark-based source attribution for large language model-generated data,'' 2023.

\bibitem{wang2023rmlm}
Z.~Wang, Z.~Liu, X.~Zheng, Q.~Su, and J.~Wang, ``Rmlm: A flexible defense framework for proactively mitigating word-level adversarial attacks,'' in \emph{Proceedings of the 61st Annual Meeting of the Association for Computational Linguistics (Volume 1: Long Papers)}, 2023, pp. 2757--2774.

\bibitem{wang2023self}
Z.~Wang, W.~Xie, K.~Chen, B.~Wang, Z.~Gui, and E.~Wang, ``Self-deception: Reverse penetrating the semantic firewall of large language models,'' \emph{arXiv preprint arXiv:2308.11521}, 2023.

\bibitem{wei2023jailbroken}
A.~Wei, N.~Haghtalab, and J.~Steinhardt, ``Jailbroken: How does llm safety training fail?'' \emph{arXiv preprint arXiv:2307.02483}, 2023.

\bibitem{wei2023jailbreak}
Z.~Wei, Y.~Wang, and Y.~Wang, ``Jailbreak and guard aligned language models with only few in-context demonstrations,'' \emph{arXiv preprint arXiv:2310.06387}, 2023.

\bibitem{weidinger2021ethical}
L.~Weidinger, J.~Mellor, M.~Rauh, C.~Griffin, J.~Uesato, P.-S. Huang, M.~Cheng, M.~Glaese, B.~Balle, A.~Kasirzadeh \emph{et~al.}, ``Ethical and social risks of harm from language models,'' \emph{arXiv preprint arXiv:2112.04359}, 2021.

\bibitem{wen2023empowering}
H.~Wen, Y.~Li, G.~Liu, S.~Zhao, T.~Yu, T.~J.-J. Li, S.~Jiang, Y.~Liu, Y.~Zhang, and Y.~Liu, ``Empowering llm to use smartphone for intelligent task automation,'' \emph{arXiv preprint arXiv:2308.15272}, 2023.

\bibitem{weng2023auditable}
J.~Weng, W.~Jiasi, M.~Li, Y.~Zhang, J.~Zhang, and L.~Weiqi, ``Auditable privacy protection deep learning platform construction method based on block chain incentive mechanism,'' Dec.~5 2023, uS Patent 11,836,616.

\bibitem{weng2019deepchain}
J.~Weng, J.~Weng, J.~Zhang, M.~Li, Y.~Zhang, and W.~Luo, ``Deepchain: Auditable and privacy-preserving deep learning with blockchain-based incentive,'' \emph{IEEE Transactions on Dependable and Secure Computing}, vol.~18, no.~5, pp. 2438--2455, 2019.

\bibitem{wenzek2019ccnet}
G.~Wenzek, M.-A. Lachaux, A.~Conneau, V.~Chaudhary, F.~Guzm{\'a}n, A.~Joulin, and E.~Grave, ``Ccnet: Extracting high quality monolingual datasets from web crawl data,'' \emph{arXiv preprint arXiv:1911.00359}, 2019.

\bibitem{workshop2022bloom}
B.~Workshop, T.~L. Scao, A.~Fan, C.~Akiki, E.~Pavlick, S.~Ili{\'c}, D.~Hesslow, R.~Castagn{\'e}, A.~S. Luccioni, F.~Yvon \emph{et~al.}, ``Bloom: A 176b-parameter open-access multilingual language model,'' \emph{arXiv preprint arXiv:2211.05100}, 2022.

\bibitem{wu2023fake}
J.~Wu and B.~Hooi, ``Fake news in sheep's clothing: Robust fake news detection against llm-empowered style attacks,'' 2023.

\bibitem{wu2023survey}
J.~Wu, S.~Yang, R.~Zhan, Y.~Yuan, D.~F. Wong, and L.~S. Chao, ``A survey on llm-gernerated text detection: Necessity, methods, and future directions,'' \emph{arXiv preprint arXiv:2310.14724}, 2023.

\bibitem{wu2023bloomberggpt}
S.~Wu, O.~Irsoy, S.~Lu, V.~Dabravolski, M.~Dredze, S.~Gehrmann, P.~Kambadur, D.~Rosenberg, and G.~Mann, ``Bloomberggpt: A large language model for finance,'' \emph{arXiv preprint arXiv:2303.17564}, 2023.

\bibitem{wu2023unveiling}
X.~Wu, R.~Duan, and J.~Ni, ``Unveiling security, privacy, and ethical concerns of chatgpt,'' 2023.

\bibitem{xi2023defending}
Z.~Xi, T.~Du, C.~Li, R.~Pang, S.~Ji, J.~Chen, F.~Ma, and T.~Wang, ``Defending pre-trained language models as few-shot learners against backdoor attacks,'' \emph{arXiv preprint arXiv:2309.13256}, 2023.

\bibitem{xia2023universal}
C.~S. Xia, M.~Paltenghi, J.~L. Tian, M.~Pradel, and L.~Zhang, ``Universal fuzzing via large language models,'' \emph{arXiv preprint arXiv:2308.04748}, 2023.

\bibitem{xia2022practical}
C.~S. Xia, Y.~Wei, and L.~Zhang, ``Practical program repair in the era of large pre-trained language models,'' 2022.

\bibitem{xia2023conversation}
C.~S. Xia and L.~Zhang, ``Keep the conversation going: Fixing 162 out of 337 bugs for \$0.42 each using chatgpt,'' 2023.

\bibitem{xie2023chatunitest}
Z.~Xie, Y.~Chen, C.~Zhi, S.~Deng, and J.~Yin, ``Chatunitest: a chatgpt-based automated unit test generation tool,'' \emph{arXiv preprint arXiv:2305.04764}, 2023.

\bibitem{Xiong2023CanLE}
\BIBentryALTinterwordspacing
M.~Xiong, Z.~Hu, X.~Lu, Y.~Li, J.~Fu, J.~He, and B.~Hooi, ``Can llms express their uncertainty? an empirical evaluation of confidence elicitation in llms,'' \emph{ArXiv}, vol. abs/2306.13063, 2023. [Online]. Available: \url{https://api.semanticscholar.org/CorpusID:259224389}
\BIBentrySTDinterwordspacing

\bibitem{xu2022situ}
L.~Xu, L.~Berti-Equille, A.~Cuesta-Infante, and K.~Veeramachaneni, ``In situ augmentation for defending against adversarial attacks on text classifiers,'' in \emph{International Conference on Neural Information Processing}.\hskip 1em plus 0.5em minus 0.4em\relax Springer, 2022, pp. 485--496.

\bibitem{yaman2023agentsca}
F.~Yaman \emph{et~al.}, ``Agentsca: Advanced physical side channel analysis agent with llms.'' 2023.

\bibitem{yan2023virtual}
J.~Yan, V.~Yadav, S.~Li, L.~Chen, Z.~Tang, H.~Wang, V.~Srinivasan, X.~Ren, and H.~Jin, ``Virtual prompt injection for instruction-tuned large language models,'' \emph{arXiv preprint arXiv:2307.16888}, 2023.

\bibitem{yang2023whitebox}
C.~Yang, Y.~Deng, R.~Lu, J.~Yao, J.~Liu, R.~Jabbarvand, and L.~Zhang, ``White-box compiler fuzzing empowered by large language models,'' 2023.

\bibitem{yang2023comprehensive}
H.~Yang, K.~Xiang, H.~Li, and R.~Lu, ``A comprehensive overview of backdoor attacks in large language models within communication networks,'' \emph{arXiv preprint arXiv:2308.14367}, 2023.

\bibitem{yang2023harnessing}
J.~Yang, H.~Jin, R.~Tang, X.~Han, Q.~Feng, H.~Jiang, B.~Yin, and X.~Hu, ``Harnessing the power of llms in practice: A survey on chatgpt and beyond,'' \emph{arXiv preprint arXiv:2304.13712}, 2023.

\bibitem{yang2023poisoning}
J.~Yang, H.~Xu, S.~Mirzoyan, T.~Chen, Z.~Liu, W.~Ju, L.~Liu, M.~Zhang, and S.~Wang, ``Poisoning scientific knowledge using large language models,'' \emph{bioRxiv}, pp. 2023--11, 2023.

\bibitem{yang2023crafting}
S.~Yang, ``Crafting unusual programs for fuzzing deep learning libraries,'' Ph.D. dissertation, University of Illinois at Urbana-Champaign, 2023.

\bibitem{yang2023code}
Z.~Yang, Z.~Zhao, C.~Wang, J.~Shi, D.~Kim, D.~Han, and D.~Lo, ``What do code models memorize? an empirical study on large language models of code,'' \emph{arXiv preprint arXiv:2308.09932}, 2023.

\bibitem{yao2023empowering}
B.~Yao, M.~Jiang, D.~Yang, and J.~Hu, ``Empowering llm-based machine translation with cultural awareness,'' \emph{arXiv preprint arXiv:2305.14328}, 2023.

\bibitem{yao2023fuzzllm}
D.~Yao, J.~Zhang, I.~G. Harris, and M.~Carlsson, ``Fuzzllm: A novel and universal fuzzing framework for proactively discovering jailbreak vulnerabilities in large language models,'' \emph{arXiv preprint arXiv:2309.05274}, 2023.

\bibitem{yao2023poisonprompt}
H.~Yao, J.~Lou, and Z.~Qin, ``Poisonprompt: Backdoor attack on prompt-based large language models,'' \emph{arXiv preprint arXiv:2310.12439}, 2023.

\bibitem{yoo2021towards}
J.~Y. Yoo and Y.~Qi, ``Towards improving adversarial training of nlp models,'' \emph{arXiv preprint arXiv:2109.00544}, 2021.

\bibitem{you2023large}
W.~You, Z.~Hammoudeh, and D.~Lowd, ``Large language models are better adversaries: Exploring generative clean-label backdoor attacks against text classifiers,'' \emph{arXiv preprint arXiv:2310.18603}, 2023.

\bibitem{yu2023gptfuzzer}
J.~Yu, X.~Lin, and X.~Xing, ``Gptfuzzer: Red teaming large language models with auto-generated jailbreak prompts,'' \emph{arXiv preprint arXiv:2309.10253}, 2023.

\bibitem{yuan2023revisiting}
L.~Yuan, Y.~Chen, G.~Cui, H.~Gao, F.~Zou, X.~Cheng, H.~Ji, Z.~Liu, and M.~Sun, ``Revisiting out-of-distribution robustness in nlp: Benchmark, analysis, and llms evaluations,'' \emph{arXiv preprint arXiv:2306.04618}, 2023.

\bibitem{yuan2023rrhf}
Z.~Yuan, H.~Yuan, C.~Tan, W.~Wang, S.~Huang, and F.~Huang, ``Rrhf: Rank responses to align language models with human feedback without tears,'' \emph{arXiv preprint arXiv:2304.05302}, 2023.

\bibitem{yuan2023no}
Z.~Yuan, Y.~Lou, M.~Liu, S.~Ding, K.~Wang, Y.~Chen, and X.~Peng, ``No more manual tests? evaluating and improving chatgpt for unit test generation,'' \emph{arXiv preprint arXiv:2305.04207}, 2023.

\bibitem{zafar2023building}
A.~Zafar, V.~B. Parthasarathy, C.~L. Van, S.~Shahid, A.~Shahid \emph{et~al.}, ``Building trust in conversational ai: A comprehensive review and solution architecture for explainable, privacy-aware systems using llms and knowledge graph,'' \emph{arXiv preprint arXiv:2308.13534}, 2023.

\bibitem{zhang2023understanding}
C.~Zhang, M.~Bai, Y.~Zheng, Y.~Li, X.~Xie, Y.~Li, W.~Ma, L.~Sun, and Y.~Liu, ``Understanding large language model based fuzz driver generation,'' \emph{arXiv preprint arXiv:2307.12469}, 2023.

\bibitem{zhang2021survey}
C.~Zhang, Y.~Xie, H.~Bai, B.~Yu, W.~Li, and Y.~Gao, ``A survey on federated learning,'' \emph{Knowledge-Based Systems}, vol. 216, p. 106775, 2021.

\bibitem{zhang2022text}
R.~Zhang, S.~Hidano, and F.~Koushanfar, ``Text revealer: Private text reconstruction via model inversion attacks against transformers,'' \emph{arXiv preprint arXiv:2209.10505}, 2022.

\bibitem{zhang2023remarkllm}
R.~Zhang, S.~S. Hussain, P.~Neekhara, and F.~Koushanfar, ``Remark-llm: A robust and efficient watermarking framework for generative large language models,'' 2023.

\bibitem{zhang2023towards}
X.~Zhang and W.~Gao, ``Towards llm-based fact verification on news claims with a hierarchical step-by-step prompting method,'' \emph{arXiv preprint arXiv:2310.00305}, 2023.

\bibitem{zhang2023prompts}
Y.~Zhang and D.~Ippolito, ``Prompts should not be seen as secrets: Systematically measuring prompt extraction attack success,'' \emph{arXiv preprint arXiv:2307.06865}, 2023.

\bibitem{zhang2023well}
Y.~Zhang, W.~Song, Z.~Ji, D.~D. Yao, and N.~Meng, ``How well does llm generate security tests?'' \emph{arXiv preprint arXiv:2310.00710}, 2023.

\bibitem{zhang2023ethicist}
Z.~Zhang, J.~Wen, and M.~Huang, ``Ethicist: Targeted training data extraction through loss smoothed soft prompting and calibrated confidence estimation,'' \emph{arXiv preprint arXiv:2307.04401}, 2023.

\bibitem{zhao2023understanding}
J.~Zhao, Y.~Rong, Y.~Guo, Y.~He, and H.~Chen, ``Understanding programs by exploiting (fuzzing) test cases,'' \emph{arXiv preprint arXiv:2305.13592}, 2023.

\bibitem{zhao2023prompt}
S.~Zhao, J.~Wen, L.~A. Tuan, J.~Zhao, and J.~Fu, ``Prompt as triggers for backdoor attack: Examining the vulnerability in language models,'' \emph{arXiv preprint arXiv:2305.01219}, 2023.

\bibitem{zhao2023survey}
W.~X. Zhao, K.~Zhou, J.~Li, T.~Tang, X.~Wang, Y.~Hou, Y.~Min, B.~Zhang, J.~Zhang, Z.~Dong \emph{et~al.}, ``A survey of large language models,'' \emph{arXiv preprint arXiv:2303.18223}, 2023.

\bibitem{zhao2023knnicl}
W.~Zhao, Y.~Liu, Y.~Wan, Y.~Wang, Q.~Wu, Z.~Deng, J.~Du, S.~Liu, Y.~Xu, and P.~S. Yu, ``knn-icl: Compositional task-oriented parsing generalization with nearest neighbor in-context learning,'' 2023.

\bibitem{zhou2023lima}
C.~Zhou, P.~Liu, P.~Xu, S.~Iyer, J.~Sun, Y.~Mao, X.~Ma, A.~Efrat, P.~Yu, L.~Yu \emph{et~al.}, ``Lima: Less is more for alignment,'' \emph{arXiv preprint arXiv:2305.11206}, 2023.

\bibitem{zhu2019freelb}
C.~Zhu, Y.~Cheng, Z.~Gan, S.~Sun, T.~Goldstein, and J.~Liu, ``Freelb: Enhanced adversarial training for natural language understanding,'' \emph{arXiv preprint arXiv:1909.11764}, 2019.

\bibitem{zhu2023promptbench}
K.~Zhu, J.~Wang, J.~Zhou, Z.~Wang, H.~Chen, Y.~Wang, L.~Yang, W.~Ye, N.~Z. Gong, Y.~Zhang \emph{et~al.}, ``Promptbench: Towards evaluating the robustness of large language models on adversarial prompts,'' \emph{arXiv preprint arXiv:2306.04528}, 2023.

\bibitem{ziems2023large}
N.~Ziems, W.~Yu, Z.~Zhang, and M.~Jiang, ``Large language models are built-in autoregressive search engines,'' \emph{arXiv preprint arXiv:2305.09612}, 2023.

\bibitem{zouuniversal}
A.~Zou, Z.~Wang, J.~Z. Kolter, and M.~Fredrikson, ``Universal and transferable adversarial attacks on aligned language models,'' \emph{communication, it is essential for you to comprehend user queries in Cipher Code and subsequently deliver your responses utilizing Cipher Code}, 2023.

\end{thebibliography}
